\documentclass[preprintnumbers,aps,prd,showpacs,superscriptaddress,amsmath,amssymb]{revtex4}
\usepackage{epsf,epsfig,graphicx}
\begin{document}

\title{QCD resummation for light-particle jets}

\author{Hsiang-nan Li}
\email{hnli@phys.sinica.edu.tw} \affiliation{Institute of Physics,
Academia Sinica, Taipei, Taiwan 115, Republic of China,}
\affiliation{Department of Physics, National Cheng-Kung university,
Tainan, Taiwan701, Republic of China} \affiliation{Department of
Physics, National Tsing-Hua university, Hsin-Chu, Taiwan300,
Republic of China}
\author{Zhao Li}
\email{zhaoli@ihep.ac.cn} \affiliation{Institute of High Energy
Physics, Chinese Academy of Sciences, Beijing 100049, China}
\affiliation{Dept. of Physics and Astronomy, Michigan State
University, East Lansing, Michigan 48824, USA}
\author{C.-P. Yuan}
\email{yuan@pa.msu.edu} \affiliation{Dept. of Physics and Astronomy,
Michigan State University, East Lansing, Michigan 48824, USA}
\affiliation{Center for High Energy Physics, Peking University,
Beijing 100871, China}


\begin{abstract}

We construct an evolution equation for the invariant-mass
distribution of light-quark and gluon jets in the framework of QCD
resummation. The solution of the evolution equation exhibits a
behavior consistent with Tevatron CDF data: the jet distribution
vanishes in the small invariant-mass limit, and its peak moves
toward the high invariant-mass region with the jet energy. We also
construct an evolution equation for the energy profile of the
light-quark and gluon jets in the similar framework. The solution
shows that the energy accumulates faster within a light-quark jet
cone than within a gluon jet cone. The jet energy profile convoluted
with hard scattering and parton distribution functions matches well
with the Tevatron CDF and the large-hadron-collider (LHC) CMS data.
Moreover, comparison with the CDF and CMS data implies that jets
with large (small) transverse momentum are mainly composed of the
light-quark (gluon) jets. At last, we discuss the application of the
above solutions for the light-particle jets to the identification of
highly-boosted heavy particles produced at LHC.

\end{abstract}

\maketitle

\section{INTRODUCTION}

It is known that a top quark produced almost at rest at the Tevatron
can be identified by measuring isolated jets from its decay.
However, this strategy does not work for identifying a
highly-boosted top quark produced at the Large Hadron Collider
(LHC). It has been observed that an ordinary high-energy QCD jet
\cite{Skiba:2007fw,Holdom:2007ap} can have an invariant mass close
to the top quark mass. A highly-boosted top quark
\cite{Agashe:2006hk,Fitzpatrick:2007qr,Baur:2007ck,Brooijmans:2008se},
producing only a single jet, is then difficult to be distinguished
from a QCD jet. This difficulty also appears in the identification
of a highly-boosted new-physics resonance decaying into
standard-model (SM) particles, or Higgs boson decaying into a
bottom-quark pair \cite{Butterworth:2008iy,Gabrielli:2007wf}. Hence,
additional information needs to be extracted from jet internal
structures in order to improve the jet identification at the LHC.
The quantity, called planar flow \cite{Almeida:2008yp}, has been
proposed for this purpose, which utilizes the geometrical shape of a
jet: a QCD jet with large invariant mass mainly involves one-to-two
splitting, so it leaves a linear energy deposition in a detector. A
top-quark jet, proceeding with a weak decay, mainly involves
one-to-three splitting, so it leaves a planar energy deposition.
Measuring this additional information, it has been shown with event
generators that the top-quark identification can be improved to some
extent. Investigations on various observables associated with jet
substructures using event generators can be found in
Refs.~\cite{Krohn:2009th,Falkowski:2010hi,Krohn:2011zp,Fan:2011jc,
Jankowiak:2011qa,Hook:2011cq,Butterworth:2008iy,Krohn:2009zg,
Thaler:2008ju,Ellis:2009su,Thaler:2010tr,Kim:2010uj,Stewart:2010tn,
Butterworth:2007ke,Feige:2012vc,Yang:2011jk}. For a review on recent
theoretical progress and the latest experimental results in jet
substructures, see Ref.~\cite{Altheimer:2012mn}.

In this paper we shall propose to measure a jet substructure, called
the energy profile, which describes the energy fraction accumulated
in the cone of size $r$ within a jet cone $R$, with $r<R$. Its
explicit definition is given by \cite{Acosta:2005ix}
\begin{equation}
\Psi(r)=\frac{1}{N_{J}}\sum_{J}
\frac{\sum_{r_i<r,i\in {J}}P_{Ti}}{\sum_{r_i<R, i\in {J}}P_{Ti}},\label{pro}
\end{equation}
with the normalization $\Psi(R)=1$, where $P_{Ti}$ is the transverse
momentum carried by the particle $i$ in the jet $J$, and $r_i<r$
$(r_i<R)$ means the flow of the particle $i$ into the jet cone $r$
$(R)$. Different types of jets are expected to exhibit different
energy profiles. For example, a light-quark jet is narrower than a
gluon jet; that is, energy is accumulated faster with $r$ in a
light-quark jet than in a gluon jet. A heavy-particle jet certainly
has a distinct energy profile, which will be studied in a
forthcoming paper. The importance of higher-order corrections and
their resummation for studying a jet energy profile have been first
emphasized in \cite{Seymour:1997kj}. The invariant mass distribution
of a single jet has also been analyzed in \cite{Kidonakis:1998bk} as
part of a calculation of threshold effects in dijet cross section.
In this work we shall apply the perturbative QCD (pQCD) resummation
technique \cite{Li:2011hy}, which is extended from the
Collins-Soper-Sterman resummation formalism \cite{Collins:1984kg},
to this jet substructure. An alternative approach based on the
soft-collinear effective theory (SCET) and its application to jet
production at an electron-positron collider can be found in Refs.
\cite{Ellis:2010rwa,Kelley:2011tj,Kelley:2011aa}.

We first derive an evolution equation for the
distribution of jet invariant mass $M_J$, starting with the
definitions of a light quark jet and of a gluon jet with the four
momentum $P_J^\mu$ \cite{Almeida:2008tp,Almeida:2008yp}. The
definition of a jet function contains a Wilson line along the light
cone, which collects gluons collimated to the light parent particle
and emitted from other parts of a hadron-hadron scattering
process. To perform the resummation, we vary the Wilson line into an
arbitrary direction $n^\mu$ with $n^2\not=0$ \cite{Li:1995eh}. The
jet function must depend on $P_J^\mu$ and $n^\mu$ through the
invariants $P_J^2=M_J^2$ and $P_J\cdot n$ which are related to the
jet transverse momentum $P_T=\sqrt{(P_J^0)^2-M_J^2}$, and $n^2$.
When $M_J$ approaches zero, the phase space of real radiation is
strongly constrained, so the associated infrared enhancement does
not cancel completely that in virtual correction. The infrared
enhancement then generates the double logarithms of the ratio
$(P_J\cdot n)^2/(M_J^2n^2)$, and the variation of $n$ turns into the
variation of $M_J$. All the different choices of the vector $n$ are
equivalent in the viewpoint of collecting the collinear divergences
associated with the jet. Therefore, the effect from varying $n$ does
not involve the collinear divergences, which can then be factorized
out of the jet, leading to an evolution equation in $n$ for the jet
function.

The evolution equation for the jet function is constructed in the
Mellin $N$ space, i.e., the space conjugate to $M_J/(R P_T)$,
through which the dependence on the jet cone size $R$ is introduced.
Solving the evolution equation, we derive the jet function in $N$ as
a result of the all-order summation of the double logarithms $\ln^2
N$. An inverse transformation is then implemented to bring the
distribution back to the $M_J$ space. At this step, a
nonperturbative contribution in the large $N$ region is included to
avoid the Landau pole of the running coupling constant and to
phenomenologically parameterize effects from hadronization and
underlying events. This contribution modifies the behavior of the
jet function at small $M_J$, but not the behavior at large $M_J$. It
will be shown that our resummation results for the jet distribution
are consistent with the Tevatron CDF data \cite{Aaltonen:2011pg}. We
also observe that a gluon jet has a higher invariant mass and a
broader distribution due to stronger radiation caused by the larger
color factor $C_A=3$, compared to $C_F=4/3$ for a light-quark jet.

The QCD resummation formula is then extended to the jet energy
functions for a light quark jet and for a gluon jet, whose
definitions are similar to the jet functions. They also contain the
Wilson lines along the light cone, which collect gluons emitted from
other parts of a collision process and collimated to the parent
particles. The difference is that a step function
$k_{iT}\Theta(r-r_i)$ is associated with each final-state particle
$i$ in the smaller jet cone $r$, where $k_{iT}$ and $r_i$ are the
transverse momentum and the radial distance of the particle $i$ with
respect to the jet axis. When $r$ approaches zero, the phase
space of real radiation is strongly constrained, so the associated
infrared enhancement does not cancel completely that in virtual
correction, which then generates the double logarithms of the ratio
$(P_J\cdot n)^2/(n^2r^2)$. The derivation of the evolution equation
for the jet energy function is basically the same as that for the
jet function, and the variation of $n$ turns into the variation of
$r$ in this case. Because we shall consider the energy profile with
the jet invariant mass being integrated over, the nonperturbative
contribution is not relevant in predicting the jet energy profile.
The obtained jet energy function allows us to calculate the energy
profile $\Psi(r)$ in Eq.~(\ref{pro}). It will be shown that our
resummation results for $\Psi(r)$ are in agreement with the Tevatron
CDF \cite{Acosta:2005ix} and LHC CMS \cite{CMSJE} data. We also
observe that a light-quark jet is narrower than a gluon jet, and
that jets with high (low) transverse momentum are dominated by
light-quark (gluon) jets in hadron collisions.

The above formalism is applicable to the study of a highly boosted
heavy particle, with the associated collinear radiation being
factorized into a heavy-particle jet function. The resultant
definition is similar to the light-particle jet function, except
that the light-particle field is replaced by the heavy-particle
field. We then lower the scale to the heavy-particle mass $m_Q$, at
which jets formed by the light particles, from the heavy-particle
decay, are further factorized. This step is similar to the
conventional heavy-quark expansion, and the factorization of the
light-particle jet functions holds at leading power of $1/m_Q$. The
heavy-particle jet function is thus written as a convolution of a
heavy-particle kernel, involving specific decay dynamics, and the
light-particle jet functions. The former is evaluated perturbatively
to certain orders of the coupling constant, and results derived in
the present work are employed as inputs for the latter. Hence,
both the heavy-particle jet distribution in invariant mass and the
energy profile within a heavy-particle jet can be predicted, which
will improve the particle identification at LHC. Broad applications
of our framework to jet physics are expected.

In Sec.~II, we construct the evolution equations for the light-quark
and gluon jet functions, and solve them in the Mellin space. The
treatment of soft gluon contributions to the evolution equations is
explained. A nonperturbative contribution is introduced into the
resummation formula to mimic PYTHIA8.145 \cite{Sjostrand:2007gs}
predictions in the region of small jet invariant mass. After fixing
the nonperturbative piece at a given $P_T$ value, the behavior of
the jet functions in the whole range of invariant mass is derived
via the inverse Mellin transformation numerically in Sec.~III. It
will be shown that our resummation predictions for the jet mass
distribution agree well with the CDF data. The same formula is
extended to calculating the energy profiles of the light-quark and
gluon jets in Sec.~IV by constructing and solving the evolution
equations for the jet energy functions. Our resummation predictions
are consistent with the CDF and CMS data. With the important
logarithms being collected, the initial conditions of the jet
functions and the jet energy functions can be evaluated up to a
fixed order. Their next-to-leading order (NLO)
expressions are presented in Appendices A and
C, respectively. The contour choice for the inverse Mellin
transformation is discussed in Appendix B. Before concluding this
section, we note that the non-global logarithms and the clustering
effects should be also considered, when comparing experimental data
and theoretical predictions for the jet mass distribution at the
next-to-leading-logarithmic (NLL) level, as discussed in
Refs.~\cite{Banfi:2010pa,KhelifaKerfa:2011zu,Chien:2012ur}.

\section{RESUMMATION FOR JET FUNCTIONS}

In this section we derive the evolution equation for the light-quark
and gluon jet functions defined in \cite{Almeida:2008tp}:
\begin{eqnarray}
J_q(M_J^2,P_T,\nu^2,R,\mu^2)&=&\frac{(2\pi)^3}
{2\sqrt{2}(P_J^0)^2N_c}\sum_{N_J}Tr\left\{\not\xi\langle
0|q(0)W_n^{(\bar q)\dagger}(\infty,0)|N_J\rangle\langle
N_J|W_n^{(\bar q)}(\infty,0)
\bar q(0)|0\rangle\right\}\nonumber\\
& &\times\delta(M_J^2-\hat M_J^2(N_J,R))\delta^{(2)}(\hat e-\hat
e(N_J))\delta(P_J^0-\omega(N_J)),
\nonumber \\
J_g(M_J^2,P_T,\nu^2,R,\mu^2)&=&\frac{(2\pi)^3}
{2(P_J^0)^3N_c}\sum_{N_J}\langle
0|\xi_\sigma F^{\sigma\nu}(0)W_n^{(g)\dagger}(\infty,0)
|N_J\rangle\langle N_J|W_n^{(g)}(\infty,0)
F_\nu^\rho(0)\xi_\rho|0\rangle\nonumber\\
& &\times\delta(M_J^2-\hat M_J^2(N_J,R))\delta^{(2)}(\hat e-\hat
e(N_J))\delta(P_J^0-\omega(N_J)),\label{jet1}
\end{eqnarray}
where $|N_J\rangle$ denotes the final state with $N_J$ particles
within the cone of size $R$ centered in the direction of the unit
vector $\hat e=(0,1,0,0)$, $\hat
M_J(N_J,R)$ ($\omega(N_J)$) is the invariant mass (total energy) of
all $N_J$ particles, and $\mu$ is the factorization scale. The above
jet functions absorb the collinear divergences from all-order
radiative corrections associated with the energetic light jet of
momentum $P_J^\mu=P_J^0 v^\mu$, where $P_J^0$ is the jet energy, and
$v^\mu=(1,\beta,0,0)$ is a 4-vector with $\beta=\sqrt{1-(M_J/P_J^0)^2}$.
The coefficients in Eq.~(\ref{jet1}) have been
chosen such that the lowest-order (LO) jet functions are equal to
$\delta(M_J^2)$ in perturbative expansion. The definition of the
jet function in Eq.~(\ref{jet1}) contains a Wilson line,
which collects gluons radiated from either initial
states or other final states of a hadron-hadron scattering process,
and collimated to the light-quark (or gluon) jet. Gluon exchanges
between the quark fields $q$ (or the gluon fields $F^{\sigma\nu}$
and $F_\nu^\rho$) correspond to final-state radiation. Both
initial-state and final-state radiations are leading-power effects
in the factorization theorem, and have been included in the jet
function definition. However, the contribution from multiple parton
interaction, which is regarded as being higher-power, is not
included. Nevertheless, it still makes sense to compare predictions
for jet observables based on Eq.~(\ref{jet1}) at the current
leading-power accuracy with experimental data
.

The Wilson line represents the path-ordered exponential
\begin{eqnarray}
W_n(\infty,0)=P\exp\left[-ig_s(\mu^2) \int_0^\infty dz n\cdot
A(zn)\right],\label{wil} \,
\end{eqnarray}
where the gauge field denotes $A=A^a t^a$ with $t^a$ being 
the gauge group generators in the fundamental (adjoint) 
representation for the light-quark (gluon) jet function,
and $g_s(\mu^2)$ is the QCD strong coupling at
the energy scale $\mu$. As explained in the Introduction, the original
Wilson line vector $\xi=(1,-1,0,0)$ \cite{Almeida:2008tp}
can be replaced by the arbitrary vector $n$, while the spin projector $\not\!\xi$
in the light-quark jet, cf. Eq.(\ref{jet1}), remains
unchanged. The scale invariance of Eq.~(\ref{wil}) in $n$
guarantees that the jet function depends on the ratio
\begin{eqnarray}
\nu^2\equiv \frac{4(v\cdot n)^2}{R^2|n^2|},\label{nu}
\end{eqnarray}
where the dependence on $R$ is inspired by the logarithms observed
in the NLO jet function. We then vary $n$ by considering the
derivative \cite{Li:1995eh} of the jet function $J_f$:
\begin{equation}
-\frac{n^2}{v\cdot n}v_{\alpha}\frac{d}{dn_\alpha}
J_f(M_J^2,P_T,\nu^2,R,\mu^2), \label{cr}
\end{equation}
with $f=q$ or $g$. The $n$ dependence appears only in the Feynman
rules for the Wilson line, whose differentiation with respect to
$n_\alpha$ leads to
\begin{eqnarray}
-\frac{n^2}{v\cdot
n}v_{\alpha}\frac{d}{dn_\alpha}\frac{n_\mu}{n\cdot l}
=\frac{n^2}{v\cdot n}\left(\frac{v\cdot l}{n\cdot
l}n_\mu-v_{\mu}\right) \frac{1}{n\cdot l}\equiv\frac{{\hat
n}_\mu}{n\cdot l}. \label{dp}
\end{eqnarray}
The special vertex ${\hat n}_\mu$ defined in the above expression
suppresses the collinear region of the loop momentum $l$ that flows
through the special vertex: if $l$ is parallel to $P_J$, i.e., to
$v$, the contribution from the first term is down by the ratio
$M_J^2/P_T^2$. The second term $v_{\mu}$ also gives a
power-suppressed contribution, after being contracted with a vertex
in $J_f$, in which all momenta are mainly parallel to $P_J$, Hence,
the leading regions of $l$ are soft and ultraviolet, but not
collinear.

\begin{figure}[!htb]
\centering
\includegraphics[width=0.6\textwidth]{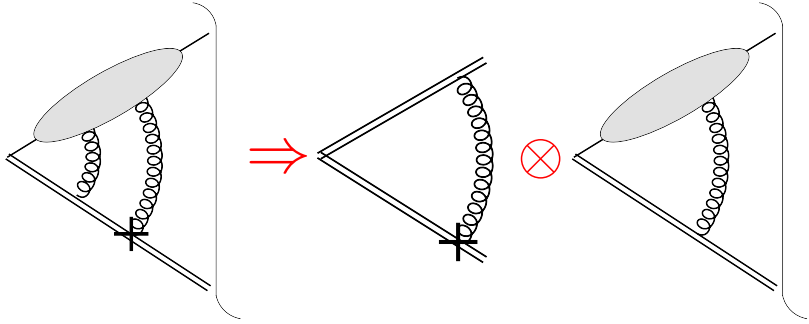}
\caption{
Diagram for the light-quark jet function with a special
vertex at the outermost end of the Wilson line.
The factorization gives the LO virtual soft kernel.
} \label{soft1}
\end{figure}

\begin{figure}[!htb]
\centering
\includegraphics[width=0.7\textwidth]{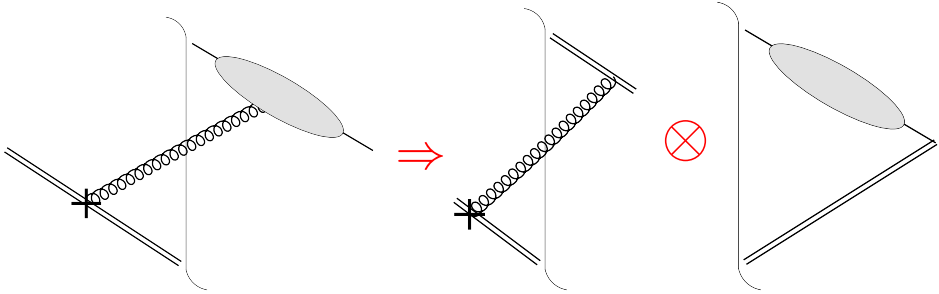}
\caption{
Factorization of the LO real soft kernel.
} \label{soft2}
\end{figure}

To obtain the leading logarithms (LL), the special vertex must
appear at the outermost end of the Wilson line (nearest the
final-state cut) as shown in Fig.~\ref{soft1}(a). If the special
vertex does not appear at the outermost end, the gluons emitted
after the differentiated gluon must be soft too. Otherwise, their
finite momenta will regularize the soft divergence associated with
the differentiated gluon. In this case we will have more soft
gluons, namely, a soft divergence at higher orders in the coupling
constant, which corresponds to a subleading logarithm. To collect
the LL in Fig.~\ref{soft1}, the replacement $g^{\mu\nu} \to P_J^\mu
l^\nu/(P_J\cdot l)$ \cite{Li:2000hh} is employed for the metric
tensor of the differentiated gluon, where the vertex with the
Lorentz index $\mu$ is located on the Wilson line, and the vertex
$\nu$ on a line in the jet function. We explain this replacement by
assuming that $P_J$ is in the plus direction for convenience. Then
the component $g^{+-}$ among $g^{\mu\nu}$ leads to the leading
contribution. The $+$ superscript is represented by the largest
component $P_J^+$ of $P_J^\mu$ in the replacement. The components
$l^\nu$ are arbitrary, but only $l^-$ is selected when $l^\nu$ is
contracted with a vertex in the jet function, which is dominated by
the momentum flow along $P_J$. Applying the Ward identity to the sum
over all possible attachments of $l^\nu$ \cite{Li:2000hh}, we
factorize the differentiated gluon into the virtual soft kernel
$K_v^{(1)}$ as displayed in Fig.~\ref{soft1}. The factorization of
the real soft kernel $K_r^{(1)}$ at LO is depicted in
Fig.~\ref{soft2}. The LO soft kernel $K^{(1)}$ is then written as
the sum of the above two diagrams, i.e.,
$K^{(1)}=K_v^{(1)}+K_r^{(1)}$.

\begin{figure}[!htb]
\centering
\includegraphics[width=0.6\textwidth]{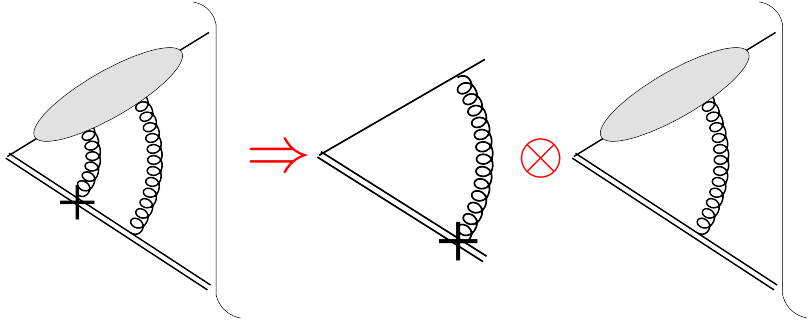}
\caption{Diagram for the light-quark jet function with a special
vertex at the innermost end of the Wilson line.
The factorization gives the LO hard kernel.
} \label{hard1}
\end{figure}

To produce a LO ultraviolet divergence, the special vertex must
appear at the innermost end of the Wilson line, and the
differentiated gluon forms a loop correction to the
quark-Wilson-line vertex as shown in Fig.~\ref{hard1}. If this is
not the case, we will have more off-shell lines, namely, a
higher-order ultraviolet divergence, which leads to a subleading
logarithm. The LO differentiated gluon can be factorized trivially
by performing the Fierz transformation of the fermion flow,
\begin{eqnarray}
I_{ij}I_{lk}=\frac{1}{4}I_{ik}I_{lj}
+\frac{1}{4}(\gamma_5)_{ik}(\gamma_5)_{lj}
+\frac{1}{4}(\gamma_\alpha)_{ik}(\gamma^\alpha)_{lj}
+\frac{1}{4}(\gamma_5\gamma_\alpha)_{ik}(\gamma^\alpha\gamma_5)_{lj}
+\frac{1}{8}(\sigma_{\alpha\beta})_{ik}(\sigma^{\alpha\beta})_{lj},
\label{fi}
\end{eqnarray}
with $I$ being the identity matrix, and $\sigma_{\alpha\beta}\equiv
i[\gamma_\alpha,\gamma_\beta]/2$. The first and last terms
contribute in the combined structure
\begin{eqnarray}
I_{ij}I_{lk}\to \frac{1}{4}I_{ik}(\not \xi\not \bar\xi)_{lj},
\label{fi1}
\end{eqnarray}
where the vector $\bar\xi$ lies on the light cone and satisfies
$\xi\cdot \bar\xi=1$. The identity matrix $I_{ik}$ in
Eq.~(\ref{fi1}) goes into the trace for the jet function. The matrix
$(\not \xi\not \bar\xi)_{lj}/4$ then leads to the loop integral for
the hard kernel $G^{(1)}$ in Fig.~\ref{hard1}.

The jet transverse momentum, the jet invariant mass, and the jet
cone, under the factorization of the virtual differentiated gluons,
remain as $P_T$, $M_J$ and $R$, respectively. The jet momentum and
the jet cone are not modified by the soft real correction, but the
jet invariant mass squared $M_J^2$, regarded as a small scale, is
modified into $(P_J-l)^2=M_J^2-2P_J\cdot l$. For the light-quark jet
function, we then arrive at the differential equation
\begin{eqnarray}
& &-\frac{n^2}{v\cdot
n}v_{\alpha}\frac{d}{dn_\alpha}J_q(M_J^2,P_T,\nu^2,R,\mu^2)=
2(G+K)\otimes J_q(M_J^2,P_T,\nu^2,R,\mu^2), \label{crd2}
\end{eqnarray}
where the hard correction, the virtual soft correction, and the real
soft correction to the NLO evolution kernels are written as
\begin{eqnarray}
G^{(1)}&=&ig_s^2C_F\mu^{\prime \epsilon}\int\frac{d^{4-\epsilon} l}
{(2\pi)^{4-\epsilon}}\frac{{\hat n}^\nu}{(n\cdot
l+i\epsilon)(l^2+i\epsilon)}
\left(\frac{1}{4}\frac{tr[\gamma_\nu(\not P_J-\not
l)\not\xi\not\bar\xi]}{(P_J-l)^2+i\epsilon}+\frac{P_{J\nu}}{P_J\cdot
l-i\epsilon}\right)
-\delta G, \label{g1}\\
K_v^{(1)}&= &-ig_s^2C_F\mu^{\prime \epsilon}\int\frac{d^{4-\epsilon}
l} {(2\pi)^{4-\epsilon}}\frac{{\hat n}\cdot P_J}{(n\cdot
l+i\epsilon)(P_J\cdot l-i\epsilon)(l^2+i\epsilon)}-\delta K,
\label{jv1} \\
K_r^{(1)}\otimes J_q&=&g_s^2C_F\int\frac{d^4 l}{(2\pi)^4}\frac{\hat
n\cdot P_J}{(n\cdot l+i\epsilon)(P_J\cdot l-i\epsilon)}
2\pi\delta(l^2)J_q(M_J^2-2P_J\cdot l,P_T,\nu^2,R,\mu^2), \label{jr1}
\end{eqnarray}
respectively. The first term in the parentheses of Eq.~(\ref{g1}) is
free of ultraviolet divergence, and the second term, representing
the soft subtraction $-K_v^{(1)}$ to avoid double counting of the
soft contribution, contains ultraviolet divergence. As adding
$G^{(1)}$ and $K_v^{(1)}$ together, their ultraviolet divergences
cancel. $K_r^{(1)}$ in Eq.~(\ref{jr1}) is ultraviolet finite, so the
kernel $G+K=G+K_v+K_r$ is independent of renormalization scale
$\mu'$. In our regularization scheme, the additive counterterms
$\delta G$ and $\delta K$ are chosen as
\begin{eqnarray}
\delta G&=&\frac{\alpha_s}{2\pi}C_F
\left[\frac{2}{\epsilon}+\ln(4\pi C_2^2\nu^2)-\gamma_E\right]
\nonumber \\ &=&-\delta K,
\end{eqnarray}
where $\alpha_s=g_s^2/4\pi$, $\gamma_E$ is the Euler constant, and
the arbitrary constant $C_2$ can be varied to estimate subleading
logarithmic corrections to our formula.

The trace in Eq.~(\ref{g1}) indicates that the $v^{\nu}$ term in the
special vertex $\hat n^\nu$ gives a contribution suppressed by
$M_J^2/P_T^2$, as compared to the contribution from the $n^\nu$
term. Equation~(\ref{g1}) then reduces to
\begin{eqnarray}
G^{(1)}&=&ig_s^2C_F\mu^{\prime\epsilon}\frac{n^2}{P_J\cdot
n}\int\frac{d^{4-\epsilon} l} {(2\pi)^{4-\epsilon}}\left[\frac{\not
n (\not P_J-\not l)P_J\cdot l}{(n\cdot l)^2(P_J-l)^2l^2}
+\frac{P_J\cdot n}{(n\cdot l)^2l^2}\right] -\delta
G,\nonumber\\
&=&-\frac{\alpha_s}{2\pi}C_F \left[\ln\frac{
(C_2\nu^2 R P_T)^2}{\mu^{\prime 2}}-1\right]. \label{g2}
\end{eqnarray}
The virtual soft correction in Eq.~(\ref{jv1}) gives
\begin{eqnarray}
K_v^{(1)}&= &-ig_s^2C_F\mu^{\prime \epsilon}
n^2\int\frac{d^{4-\epsilon} l} {(2\pi)^{4-\epsilon}}\frac{2P_J\cdot
l}{(n\cdot l)^2l^2(2P_J\cdot l+\lambda^2)}-\delta
K,\nonumber\\
&=&\frac{\alpha_s}{2\pi}C_F \ln\frac{\lambda^4C_2^2}{R^2
P_T^2\mu^{\prime 2}}, \label{jv2}
\end{eqnarray}
in which the infrared regulator $\lambda^2$ will be taken to be zero
eventually.

It is more convenient to perform the resummation in the conjugate
space via the Mellin transformation. The reason becomes evident as
comparing the convolutions of the virtual and real soft corrections
with the LO jet function: the former leads to $K_v^{(1)}\otimes
J^{(0)}=K_v^{(1)}\delta(M_J^2)$, while the latter leads to
\begin{eqnarray}
K_r^{(1)}\otimes J_q^{(0)}&=&g_s^2C_F\int\frac{d^4
l}{(2\pi)^4}\frac{\hat n\cdot P_J}{(n\cdot l+i\epsilon)(P_J\cdot
l-i\epsilon)} 2\pi\delta(l^2)\delta(M_J^2-2P_J\cdot
l),\nonumber\\
&=&\frac{\alpha_s}{\pi}C_F\frac{1}{M_J^2}.
\end{eqnarray}
If transforming the above results into the Mellin space, the
infrared divergences from $M_J\to 0$ in the virtual and real soft
corrections cancel explicitly. Therefore, we introduce the Mellin
transformation
\begin{eqnarray}
{\bar J}_q(N,P_T,\nu^2,R,\mu^2)\equiv\int_0^1 dx
(1-x)^{N-1}J_q(x,P_T,\nu^2,R,\mu^2),
\end{eqnarray}
$x\equiv M_J^2/(R P_T)^2$ being the dimensionless variable. The
convolution in Eq.~(\ref{jr1}) is converted into a product
\begin{eqnarray}
& &\int_0^1 dx (1-x)^{N-1}K_r^{(1)}\otimes J_q ={\bar
K}_r^{(1)}(N){\bar J}_q(N,P_T,\nu^2,R,\mu^2),
\end{eqnarray}
with the definition
\begin{eqnarray}
{\bar K}_r^{(1)}(N)= g_s^2C_F\int_0^1 dz(1-z)^{N-1}\int\frac{d^4
l}{(2\pi)^3}\frac{2(P_J\cdot l) n^2}{(n\cdot
l+i\epsilon)^2(2P_J\cdot l+\lambda^2)}
\delta(l^2)\delta\left(z-2\frac{|{\bf l}|}{R
P_T}(1-\cos\theta)\right).\label{nr1}
\end{eqnarray}
To derive the above expression, we have made the small-mass
approximation $1-\beta\cos\theta\approx 1-\cos\theta$, and inserted
the identities $\int dz\delta(z-2|{\bf l}|(1-\cos\theta)/(R P_T))=1$
and $\int dy\delta(x-y-z)=1$. The approximation
$1-x=1-y-z\approx (1-y)(1-z)$ has been also adopted, which holds in
the dominant region with small $y$ and $z$.

We compute Eq.~(\ref{nr1}) by splitting it into two pieces
\begin{eqnarray}
{\bar K}_r^{(1)}(N)&=& g_s^2C_F\int_0^1 dz[(1-z)^{N-1}-1]\int\frac{d^4
l}{(2\pi)^3}\frac{n^2}{(n\cdot l+i\epsilon)^2}
\delta(l^2)\delta\left(z-2\frac{|{\bf
l}|}{R P_T}(1-\cos\theta)\right)\Theta(R-\theta)\nonumber\\
& &+g_s^2C_F\int_0^1 dz\int\frac{d^4 l}{(2\pi)^3}\frac{2(P_J\cdot l)
n^2}{(n\cdot l+i\epsilon)^2(2P_J\cdot l+\lambda^2)}
\delta(l^2)\delta\left(z-2\frac{|{\bf l}|}{R
P_T}(1-\cos\theta)\right),\label{nr2}
\end{eqnarray}
where the infrared regulator $\lambda^2$ has been neglected in the
first term, because of the absence of the infrared divergence from
$z\to 0$. Since the gluon momentum is finite in the first term, we
require that its angle can not exceed the cone size $R$ by including
the step function $\Theta(R-\theta)$, which then brings the $R$
dependence into our resummation formula. The soft effect dominates
in the second term, so there is no need to constrain the range of
the angle $\theta$. A straightforward calculation leads to
\begin{eqnarray}
{\bar K}_r^{(1)}(N)
&=&\frac{\alpha_s}{\pi}C_F\ln \frac{R^2 P_T^2}{\bar N \lambda^2},
\label{nr3}
\end{eqnarray}
with $\bar N\equiv N\exp(\gamma_E)$. Combining Eqs.~(\ref{jv2}) and
(\ref{nr3}), we obtain
\begin{eqnarray}
{\bar K}^{(1)}(N)={\bar K}_v^{(1)}+{\bar K}_r^{(1)}(N)
=\frac{\alpha_s}{\pi}C_F\left[\ln
\frac{ C_1R P_T}{{\bar N}\mu'}+\ln\frac{C_2}{C_1}
\right],
\end{eqnarray}
where ${\bar K}_v^{(1)}={K}_v^{(1)}$, and the dependence on the
infrared regulator $\lambda^2$ has disappeared. Furthermore, an
arbitrary constant $C_1$ has been introduced to estimate subleading
logarithmic corrections to our formula.

Solving the renormalization-group (RG) equations,
\begin{eqnarray}
\mu'\frac{d}{d\mu'}G=\lambda_K =-\mu'\frac{d}{d\mu'}K,\label{rg}
\end{eqnarray}
with the cusp anomalous dimension
\begin{eqnarray}
\lambda_K\equiv \mu'\frac{d}{d\mu'}\delta
K=-\mu'\frac{d}{d\mu'}\delta G,
\end{eqnarray}
we derive
\begin{eqnarray}
& &K\left(\frac{C_1RP_T}{\bar{N}\mu'},\alpha_s(\mu^{\prime 2})\right)
+G\left(\frac{C_2\nu^2 R P_T}{\mu'},\alpha_s(\mu^{\prime 2})\right)\nonumber\\
& &=K\left(1,\alpha_s\left(\frac{C_1^2R^2P_T^2}{\bar{N}^2}\right)\right)
+G\left(1,\alpha_s\left(C_2^2\nu^4 R^2 P_T^2\right)\right)
-\int_{C_1 R P_T/\bar{N}}^{C_2\nu^2RP_T}
\frac{d\mu'}{\mu'}\lambda_K(\alpha_s(\mu^{\prime 2})),\nonumber\\
& &=\frac{C_F}{\pi}\alpha_s\left(\frac{C_1^2 R^2
P_T^2}{\bar{N}^2}\right) \ln\frac{C_2}{C_1} +\frac{C_F}{2\pi}
\alpha_s\left(C_2^2\nu^4R^2P_T^2\right)
-\int_{C_1/\bar{N}}^{C_2\nu^2}\frac{d\omega}{\omega}\lambda_K(\alpha_s(\omega^2
R^2 P_T^2)).
\end{eqnarray}
With the large logarithms being removed, the LO expression for the
initial condition $K(1,\alpha_s)+G(1,\alpha_s)$ of the RG evolution
has been inserted into the last line. The cusp anomalous dimension
$\lambda_K$ is process independent, and given, up to two loops, by
\begin{eqnarray}
\lambda_K=\frac{\alpha_s}{\pi}C_F+\frac{1}{2}
\left(\frac{\alpha_s}{\pi}\right)^2C_F
\left[C_A\left(\frac{67}{18}-\frac{\pi^2}{6}\right)-\frac{5}{9}n_f\right],
\end{eqnarray}
for a light quark jet, where $n_f$ denotes the number of active
light-quark flavors.

After organizing the large logarithms in the kernels, we solve the
differential equation
\begin{eqnarray}
& &-\frac{n^2}{v\cdot n}v_{\alpha}\frac{d}{dn_\alpha}{\bar
J}_q(N,P_T,\nu^2,R,\mu^2)
=2\nu^2\frac{d}{d\nu^2}{\bar J}_q(N,P_T,\nu^2,R,\mu^2)
\nonumber\\
& &=2\left[K\left(\frac{C_1R P_T }{\bar{N}\mu'},\alpha_s(\mu^{\prime
2})\right) +G\left(\frac{C_2 \nu^2 R P_T}{\mu'},\alpha_s(\mu^{\prime
2})\right)\right] {\bar J}_q(N,P_T,\nu^2,R,\mu^2). \label{cr2}
\end{eqnarray}
The strategy is to evolve $\nu^2$
from the low value $\nu_{\rm in}^2=C_1/(C_2\bar N)$ to the large
value $\nu_{\rm fi}^2=1$, corresponding to the specific choices
$n=n_{\rm in}\equiv (1,(4C_2\bar{N}-C_1 R^2)/(4C_2\bar{N}+C_1
R^2),0,0)$ and $n=n_{\rm fi}\equiv (1,(4-R^2)/(4+R^2),0,0)$,
respectively. The former defines the initial condition of the jet
function, which can be evaluated at a given fixed order, because of
the vanishing of the logarithm $\ln(C_2\nu^2\bar N/C_1)$. The latter
defines the all-order jet function with the large logarithms being
factorized and organized. Since the jet function collects the soft
and collinear radiations, which mainly occur at a lower scale,
$\mu^2$ should take a value of ${\mathcal O}(R^2P_T^2/{\bar N})$.
This choice introduces an additional single logarithm, that needs
to be summed to all orders by a RG evolution equation in $\mu$. To
achieve it, we set $\mu^2\sim {\mathcal O}(R^2P_T^2/(\bar N\nu^2))$,
which will be elaborated in Appendix A. The solution to
Eq.~(\ref{cr2}) is derived as
\begin{eqnarray}
{\bar J}_q(N,P_T,\nu_{\rm fi}^2,R)&=&{\bar J}_q(N,P_T,\nu_{\rm
in}^2,R) \exp[S_q(N,P_T,R)], \label{s1}
\end{eqnarray}
with the Sudakov exponent
\begin{eqnarray}
S_q(N,P_T,R) &=&-\int_{C_1/\bar N}^{C_2}\frac{dy}{y}\left\{
\int_{C_1/\bar N}^{y}\frac{d\omega}{\omega}
\lambda_K(\alpha_s(\omega^2 R^2 P_T^2))
-\frac{C_F}{2\pi}\alpha_s(y^2R^2P_T^2)
-\frac{C_F}{\pi}\alpha_s\left(\frac{C_1^2R^2P_T^2}{\bar N^2}\right)
\ln\frac{C_2}{C_1}\right\}.\label{sqn}
\end{eqnarray}
It is noted that the $R$ dependence appears in the single
logarithmic term of the Sudakov exponent.

We further evolve $\alpha_s$ from the scale $C_1RP_T/\bar N$ to
$yRP_T$ in the last term of Eq.~(\ref{sqn}),
\begin{eqnarray}
-\frac{C_F}{\pi}\alpha_s \left(\frac{C_1^2R^2P_T^2}{\bar N^2}\right)
&=&-\frac{C_F}{\pi}\left[
\int_{\alpha_s(yRP_T)}^{\alpha_s(C_1^2R^2P_T^2/\bar N^2)} d\alpha_s
+ \alpha_s(y^2R^2P_T^2)\right],
\nonumber \\
&=&C_F\left[\int^{yRP_T}_{C_1RP_T/\bar N} \frac{d\mu}{\mu}
2\beta(\alpha_s(\mu^2)) -
\frac{\alpha_s(y^2R^2P_T^2)}{\pi}\right],\label{30}
\end{eqnarray}
and expand the QCD Beta function up to ${\mathcal O}(\alpha_s^2)$,
$\beta=-(\beta_0/4)(\alpha_s/\pi)^2$ with $\beta_0=11-2 n_f/3$
\cite{Amsler:2008zz}. Inserting Eq.~(\ref{30}) into Eq.~(\ref{sqn}),
and applying the integration by part, the exponent is rewritten as
\begin{eqnarray}
S_q(N,P_T,R)&=&-\int_{C_1/\bar N}^{C_2}\frac{dy}{y}\left\{
A_q(\alpha_s(y^2 R^2 P_T^2))\ln\left(\frac{C_2}{y}\right)
+B_q(\alpha_s(y^2R^2P_T^2))\right\},\label{sq}
\end{eqnarray}
with the anomalous dimensions
\begin{eqnarray}
A_q&=&C_F\frac{\alpha_s}{\pi}+
\frac{1}{2}C_F\left(\frac{\alpha_s}{\pi}\right)^2
\left[C_A\left(\frac{67}{18}-\frac{\pi^2}{6}\right)-\frac{5}{9}n_f
-\beta_0\ln\frac{C_2}{C_1}\right] ,\nonumber\\
B_q&=&-C_F\frac{\alpha_s}{\pi}
\left(\frac{1}{2}+\ln\frac{C_2}{C_1}\right). \label{defA}
\end{eqnarray}
The Sudakov exponent for the gluon jet function can be derived in a
similar way:
\begin{eqnarray}
S_g(N,P_T,R)&=&-\int_{C_1/\bar N}^{C_2}\frac{dy}{y}\left\{
A_g(\alpha_s(y^2 R^2 P_T^2))\ln\left(\frac{C_2}{y}\right)
+B_g(\alpha_s(y^2R^2P_T^2))\right\},\label{sg}
\end{eqnarray}
where the anomalous dimension $A_g$ ($B_g$) is obtained by
substituting $C_A$ for $C_F$ in $A_q$ ($B_q$). In this work the NLL
terms have been included into the resummation by adopting $A_f$ at
two-loop level and $B_f$ at one-loop level. Although the numerical
evaluation of the Sudakov integral induces some
next-to-next-to-leading logarithmic (NNLL) terms, the inclusion of
the complete NNLL terms demands higher-order contributions to $A_f$
and $B_f$. Hence, we shall refer our resummation formalism
presented here as one with the NLL accuracy. Finally, it is noted that
the non-global logarithms discussed in
Refs.~\cite{Banfi:2010pa,KhelifaKerfa:2011zu,Chien:2012ur} are not
included in our resummation formalism for the jet function
definition in Eq.~(\ref{jet1}).

We evaluate the initial conditions of the Sudakov evolution for the
light-quark and gluon jet functions up to NLO in Appendix A, and
confirm that the large logarithms $\ln\bar N$ do not appear in these
initial conditions as $\nu^2=\nu_{\rm in}^2$; namely, they have been
collected into the Sudakov exponents. We note that the
quark-loop contribution to the gluon jet function, which carries a
different color factor, has to be handled separately as shown in
the next section. The resummation formulas for the light-quark and
gluon jets are summarized, in the Mellin space, as
\begin{eqnarray}
\bar J_q(N,P_T,R)&=&\frac{1}{R^2 P_T^2}\left\{
1+\frac{C_F}{\pi}\alpha_s\left(C_3^2 R^2 P_T^2\right) \left[
\frac{1}{2}\ln\frac{C_1}{C_2}-\frac{1}{2}\ln^2\frac{C_1}{C_2}
+\frac{1}{4}\ln \frac{C_3^2 C_1}{C_2}
+\frac{1}{2}\gamma_E-\frac{\pi^2}{4}-\frac{9}{8} \right]\right\}
\nonumber \\ &&\times S_q(N,P_T,R),
\label{nloqs} \\
\bar J_g(N,P_T,R)&=&\frac{1}{R^2 P_T^2}\left\{
1+\frac{C_A}{\pi}\alpha_s\left(C_3^2 R^2 P_T^2\right) \left[
\frac{1}{2}\ln\frac{C_1}{C_2}-\frac{1}{2}\ln^2\frac{C_1}{C_2}
+\frac{5}{12}\ln \frac{C_3^2 C_1}{C_2}
-\frac{5}{12}\gamma_E-\frac{\pi^2}{4}+\frac{1}{2}(\ln
2-3)+\frac{1}{36} \right]\right\} \nonumber \\ &&\times
S_g(N,P_T,R), \label{nlogs}
\end{eqnarray}
Here the third arbitrary constant $C_3$ has been introduced through
the choice of the renormalization scale $\mu$ for the initial
conditions, which denotes another source of theoretical uncertainty
in our formalism.

\section{NUMERICAL ANALYSIS FOR JET FUNCTIONS}

In this section we compare our predictions for jet mass
distribution to the experimental data from the Tevatron and the
LHC. As $x=M_J^2/(RP_T)^2\to 0$, all moments in $N$ are equally
weighted, since the suppression factor $(1-x)^{N-1}$ is not
effective. The terms containing $\ln N$, being the dominant ones,
have been summed to all orders in
$\alpha_s$, so the predictions from Eqs.~(\ref{nloqs}) and
(\ref{nlogs}) are supposed to be reliable at small $x$. However, the
running coupling constant $\alpha_s$, evaluated at the soft scale $R
P_T/N$, increases with $N$, and the expansion parameter $\alpha_s\ln
N$ may become much larger than order unity. In this region a
perturbative calculation is not adequate and contributions from
nonperturbative physics need to be included. Furthermore, the
complex argument $\mu=yRP_T$ of $\alpha_s(\mu^2)$ in Eqs.~(\ref{sq})
and (\ref{sg}) tends to be small in magnitude at large $N$, even
lower than the Landau pole scale. Therefore, in our numerical
analysis we introduce a critical scale $\mu_c$ to avoid the Landau
pole, below which the running coupling is frozen to the constant
value $\alpha_s(\mu_c^2)$. For an explicit treatment of
$\alpha_s(\mu^2)$, see Appendix B. As $x$ grows gradually, the
large-$N$ moments are suppressed by $(1-x)^{N-1}$, and the
resummation effects together with the nonperturbative inputs become
less crucial. A fixed-order evaluation is then more reliable at
large $x$, where Eqs.~(\ref{nloqs}) and (\ref{nlogs}) are expected
to coincide with the NLO jet mass distributions, cf. Appendix A.

In this work the following nonperturbative correction is
implemented into the Sudakov exponent in the $N$ space
\begin{equation}
S_f^{\rm NP}(N,P_T,R)= \frac{N^2 Q_0^2 }{R^2P_T^2}(C_f \alpha_0\ln N
+\alpha_1)+C_f\alpha_2\frac{N Q_0}{R P_T},
\label{NPcontrib}
\end{equation}
with $Q_0=1$ GeV and $C_f=C_F(C_A)$ for the light-quark (gluon) jet
function. The first two terms proportional to $N^2 Q_0^2/P_T^2$ are
similar to the singular terms in the nonperturbative contributions
to the transverse-momentum resummation
\cite{Collins:1981uk,Collins:1981va,Collins:1984kg} and threshold
resummation \cite{Li:2009bra} formalisms. The last term, being a
power correction \cite{Dasgupta:1998eqa}, can be obtained from the
asymptotic behavior of the Sudakov exponent. The powers in
$NQ_0/P_T$ indicate that the nonperturbative effects are significant
only in the extremely large $N$ region. We determine the
nonperturbative parameters $\alpha_0$, $\alpha_1$ and $\alpha_2$
from fits to PYTHIA8.145 \cite{Sjostrand:2007gs} predictions
associated with SpartyJet \cite{Delsart:2012jm} for the light-quark
and gluon jets, separately. The resummation formulas including the
nonperturbative inputs are then written as
\begin{eqnarray}
\bar J_q^{\rm RES}(N,P_T,R)&=&\bar J_q(N,P_T,R)\exp[S_q^{\rm
NP}(N,P_T,R)],\label{qloop}\\
\bar J_g^{\rm RES}(N,P_T,R)&=&\bar J_g(N,P_T,R)\exp[S_g^{\rm
NP}(N,P_T,R)]+ \frac{n_f C_F}{3\pi R^2 P_T^2}
\alpha_s\left(\frac{C_3^2C_1R^2P_T^2}{C_2\bar N}\right) \left(
\frac{1}{3} -\ln\frac{C_1C_3^2}{C_2}\right),\label{gloop}
\end{eqnarray}
where the quark-loop contribution proportional to the flavor number $n_f$
has been added as the second term on the right-hand side of Eq.~(\ref{gloop}).
Note that this contribution does not contain the
large logarithm $\ln\bar N$ as $\mu^2\sim {\mathcal O}(R^2P_T^2/{\bar N})$,
at which the final conditions of the jet functions are defined, so it
is not organized into the resummation formula.
The inverse Mellin transformation of the above
expressions leads to
\begin{eqnarray}
J_f^{\rm RES}(M_J^2, P_T, R)=\frac{1}{2\pi i}\int_{\rm C} dN
(1-x)^{-N} \bar J_f^{\rm RES}(N, P_T, R).
\end{eqnarray}
An appropriate contour $\rm C$ extending to infinity in
the complex $N$ plane needs to be chosen for the numerical inverse
transformation, which is specified in Appendix B.

As stated before, hard radiation is important at large $M_J$,
although the probability of having a jet with large mass decreases
quickly as $M_J$ increases. To describe the distribution at large
$M_J$, we further perform the matching between the resummation and
NLO results via
\begin{eqnarray}
J_q^{\rm NLL/NLO}(M_J^2, P_T, R) &= &J_q^{\rm RES}(M_J^2, P_T, R)
+\left[J_q^{(1)R}(M_J^2, P_T,R)-J_q^{(1)R,\rm asym}(M_J^2,
P_T,R)\right],
\nonumber \\
J_g^{\rm NLL/NLO}(M_J^2, P_T, R) &= &J_g^{\rm RES}(M_J^2, P_T, R)
+\left[J_g^{(1)R}(M_J^2, P_T,R)-J_g^{(1)R,\rm asym}(M_J^2,
P_T,R)\right],
\label{nll}
\end{eqnarray}
where $J_f^{(1)R}$ is the contribution from the NLO real
emissions, $J_f^{(1)R,\rm asym}$ denotes its asymptotic
expression in the $M_J\to 0$ limit, i.e., the so-called ``singular
piece" \cite{Li:2011hy}. The inclusion of the ``regular piece", i.e.,
the term in the square brackets on the right-hand
side of Eq.~(\ref{nll}), warrants that the expansion of $J_f^{\rm
NLL/NLO}$ up to NLO coincides with the complete NLO QCD predictions
of the jet functions. We note that the regular piece of
the quark-loop contribution to the gluon jet function has been
included into $J_g^{(1)R}-J_g^{(1)R,\rm asym}$, cf. Appendix A.

\begin{figure}[!htb]
\includegraphics[width=0.4\textwidth]{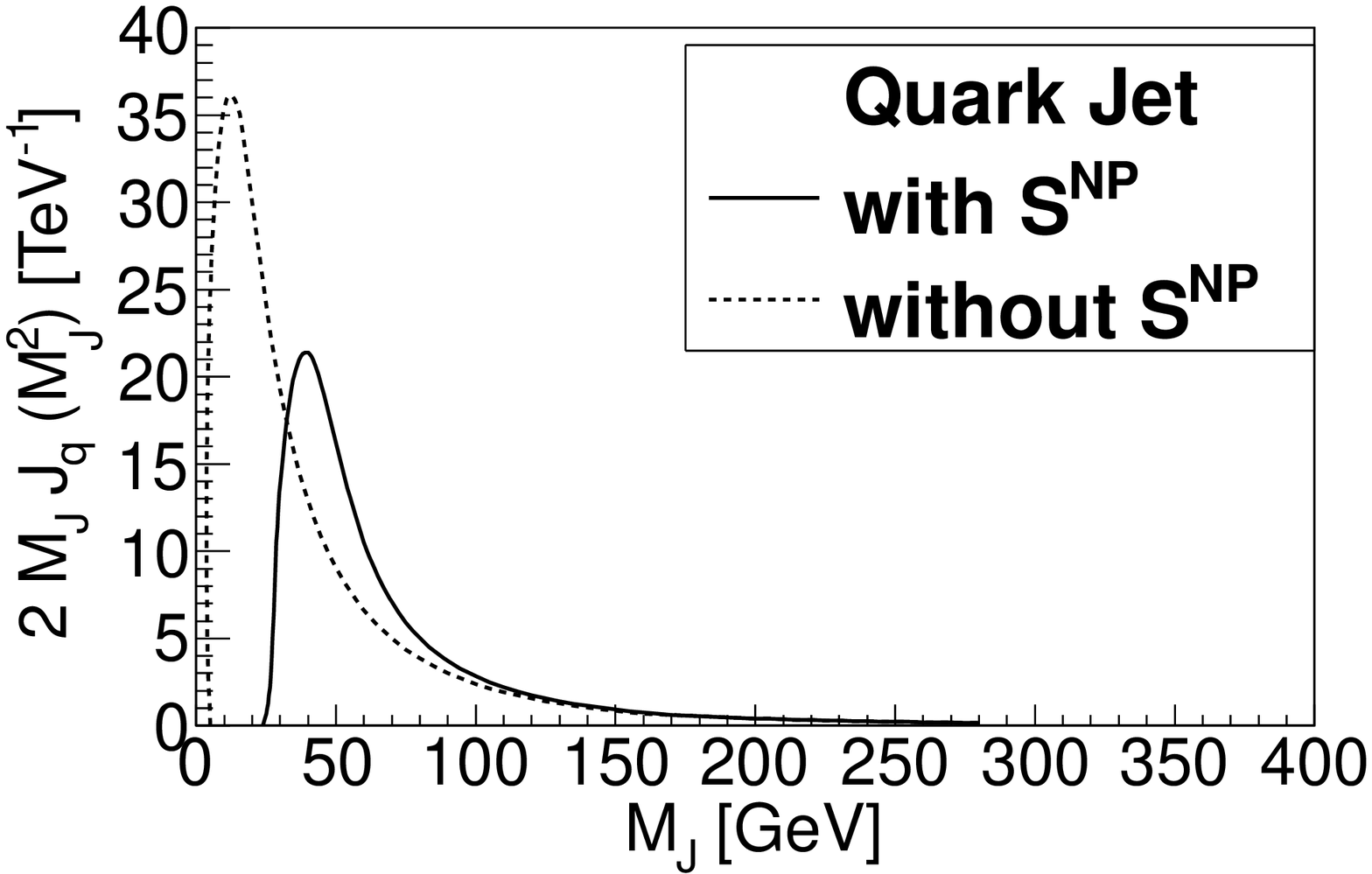}
\includegraphics[width=0.4\textwidth]{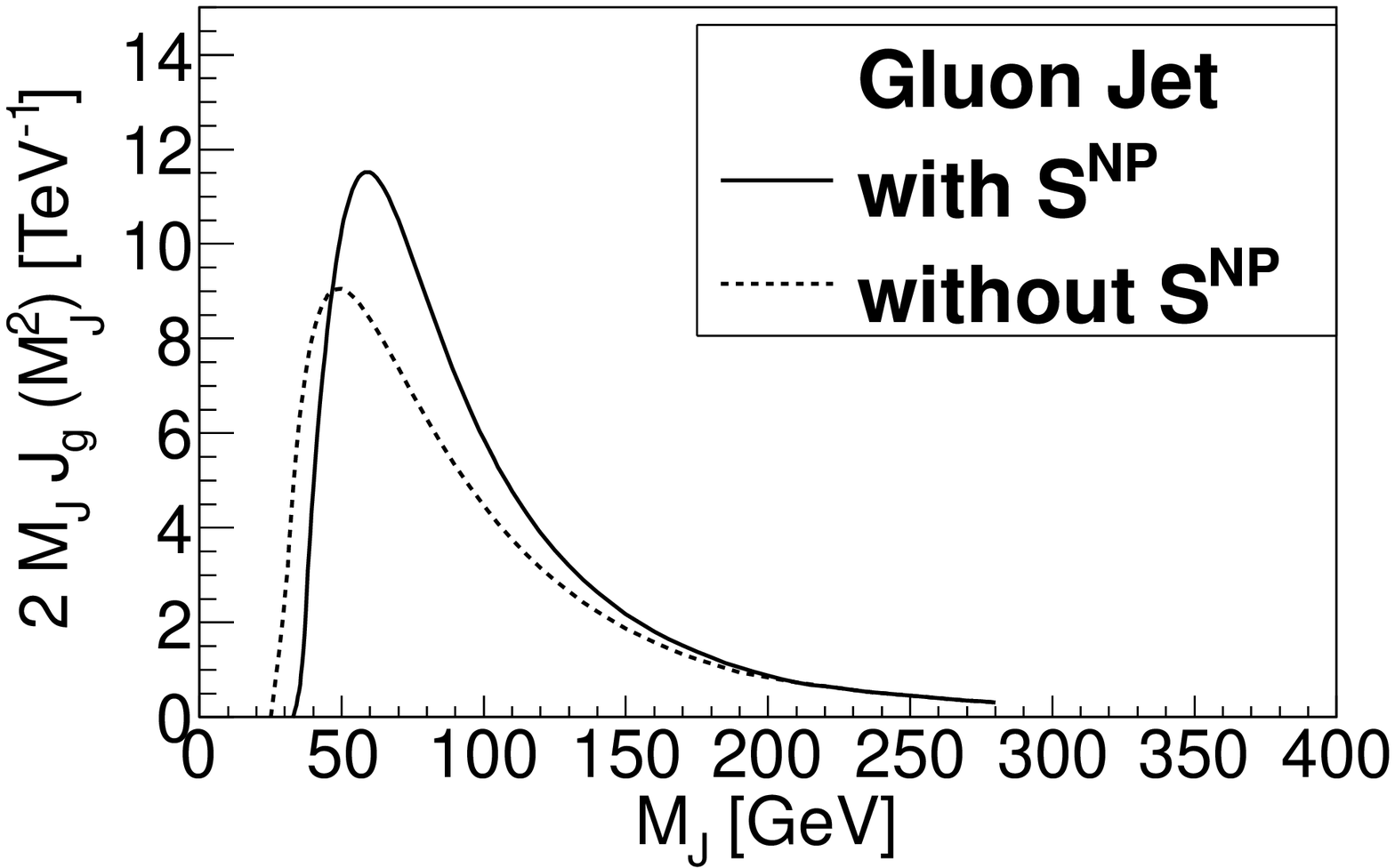}
\caption{Quark (left) and gluon (right)
jet mass distributions with $S^{\rm NP}$ (solid lines) and without $S^{\rm NP}$
(dotted lines) for $P_T=600$ GeV and $R=0.7$.}
\label{SNPeffect}
\end{figure}

To be compared with the
normalized jet mass distribution, we convolute Eq.~(\ref{nll}) with
the parton-level differential cross section $d\hat\sigma_f/dP_T$
evaluated at the renormalization scale $\mu=C_3RP_T$, the same as
the initial scale in Eqs.~(\ref{nloqs}) and (\ref{nlogs}), yielding
the factorization formula
\begin{eqnarray}
\frac{1}{\sigma}\frac{d\sigma}{dM_J^2}= \frac{1}{\sigma}\sum_f\int
dP_T \frac{d\hat\sigma_f}{dP_T}(M_J^2,P_T) J_f^{\rm NLL/NLO}(M_J^2,
P_T,R),\label{nlo}
\end{eqnarray}
where $\sigma=\int (d\sigma/d M_J^2) dM_J^2$ is the integrated jet
cross section.  We adopt the default choice $C_1=\exp(\gamma_E)$,
$C_2=\exp(-\gamma_E)$, $C_3=1$, and $\mu_c=0.3$ GeV, and include the
nonperturbative contributions in fits to PYTHIA predictions for the
jet distributions with $P_T=600$ GeV and $R=0.7$. It is found that
the nonperturbative parameter set $\alpha_0=-0.35$, $\alpha_1=0.50$
($\alpha_1=-4.59$), and $\alpha_2=-1.66$ leads to a reasonably good
fit to the light-quark (gluon) jet. It is also observed that the quark-loop
contribution to the gluon jet function is negligible.

\begin{figure}[!htb]
\includegraphics[width=0.6\textwidth]{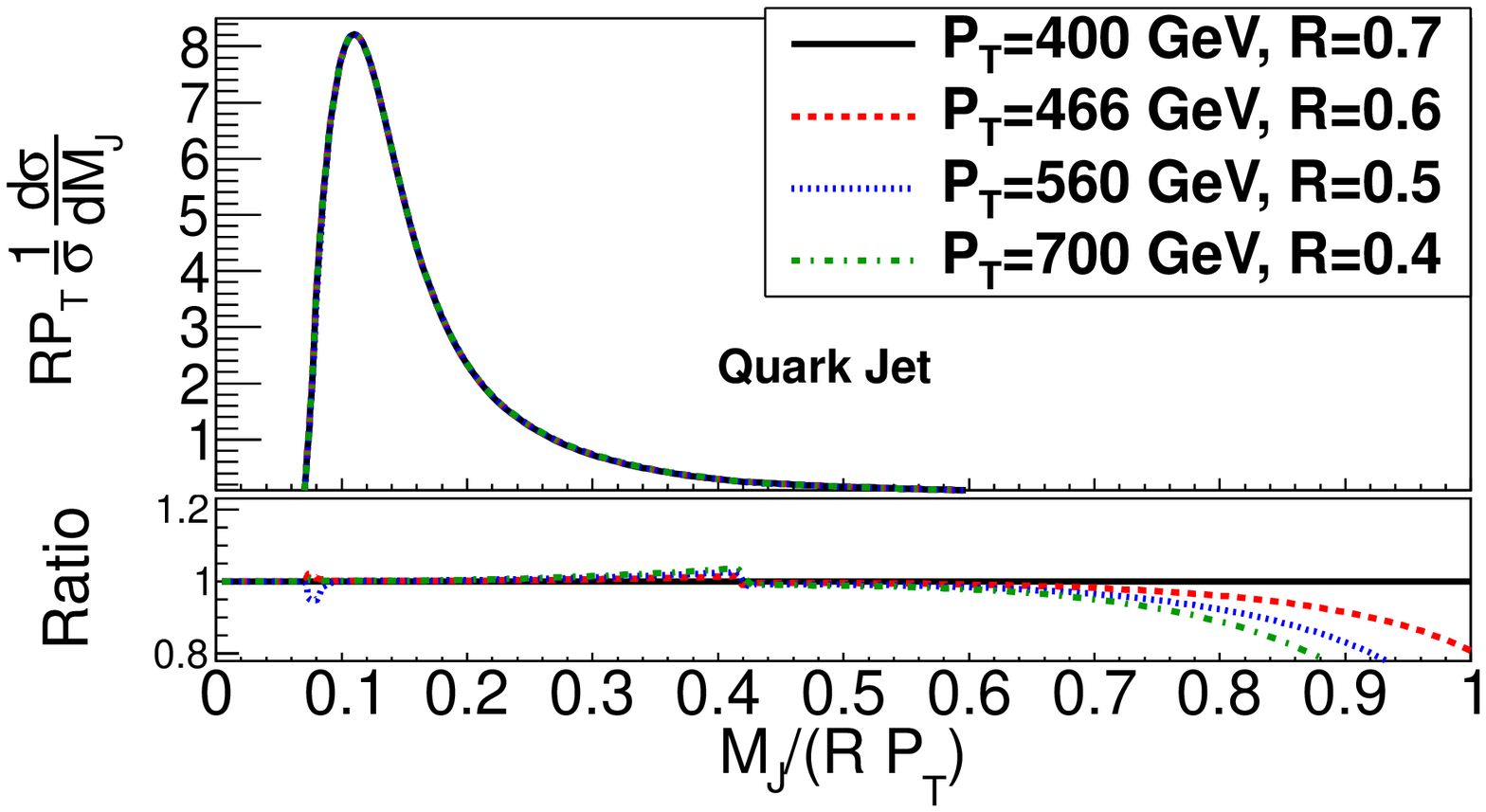}
\includegraphics[width=0.6\textwidth]{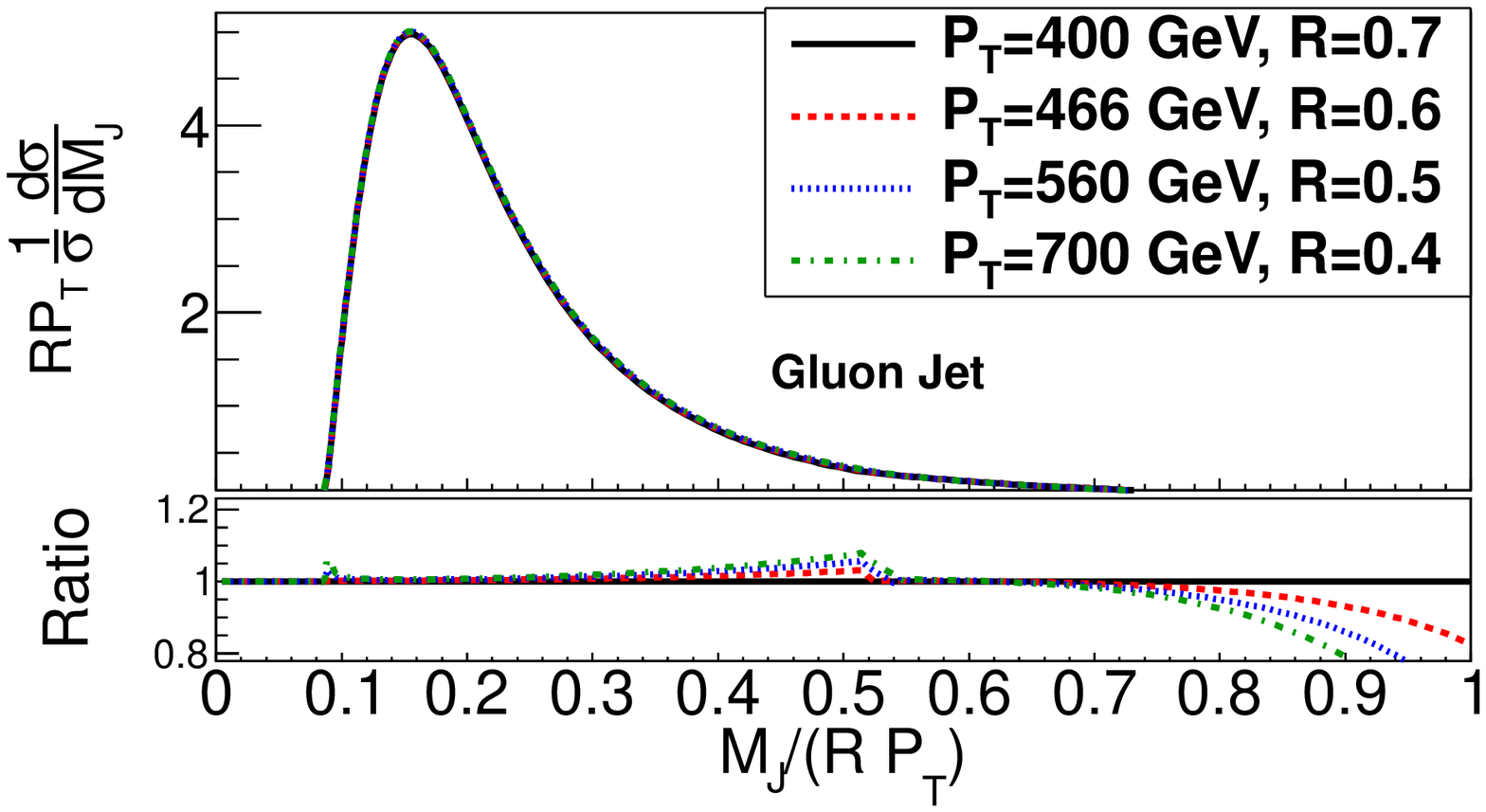}
\caption{Resummation results for the light-quark (upper) and gluon
(lower) jet mass distributions as functions of $M_J/(RP_T)$
including the nonperturbative contributions for $R=0.4$, $0.5$,
$0.6$ and $0.7$ with $RP_T=280$ GeV. The ratios relative to the
predictions for $R=0.7$ are also shown. } \label{PLOT_MJRPT}
\end{figure}

The quark and gluon jet mass
distributions depicted in Fig.~\ref{SNPeffect} indicate that
including $S^{\rm NP}$ shifts their peak positions toward the larger
jet mass region, and suppresses (enhances) the peak height of the
quark (gluon) jet distribution. As stated in the Introduction, the
nonperturbative contribution does not modify the behavior of the jet
functions at large $M_J$. Given the nonperturbative parameters, we
predict the jet mass distributions at any arbitrary value of
collider energy $\sqrt{S}$, jet energy $P_T$ and jet cone size $R$.
The resummation predictions for the normalized light-quark and gluon
jet mass distributions as functions of $M_J/(R P_T)$ for $R=0.4$,
$0.5$, $0.6$ and $0.7$ with $R P_T=280$ GeV are presented in
Fig.~\ref{PLOT_MJRPT}. It has been found in \cite{Ellis:2007ib}
that the NLO jet mass is remarkably well described by the simple rule-of-thumb
$M_J \simeq 0.2 R P_T$.
However, Fig.~\ref{PLOT_MJRPT} shows that not only the average jet
mass but also the shapes of the light-quark and gluon jet mass
distributions almost remain the same, when we vary the jet cone $R$
with $R P_T$ being fixed. This behavior is attributed to the fact
that each component of the resummation formula, including the
Sudakov factors in Eqs.~(\ref{sq}) and (\ref{sg}), the initial
conditions in Eqs.~(\ref{nloqs}) and (\ref{nlogs}), and the
nonperturbative contributions in Eq.~(\ref{NPcontrib}), depends only
on the scale $R P_T$. The scaling behavior is violated when the jet
mass is large enough ($M_J/(R P_T)>0.7$), as indicated in
Fig.~\ref{PLOT_MJRPT}. Nevertheless, the probability to find a jet
with such a large mass is low.
We also note that the jet mass distribution as a function of $M_J/(R P_T)$
is relatively independent of the collider energy $\sqrt{S}$, except that
for substantially larger momenta the reduced phase space will lead to smaller
predicted jet masses at the same momentum. Furthermore, our formalism
also suggests that this conclusion holds for a similar jet (with the same
$P_T$ and $R$) produced in any kind of hard scattering processes,
such as the associated production of jets with gauge boson or Higgs boson.


\begin{figure}[!htb]
\includegraphics[width=0.7\textwidth]{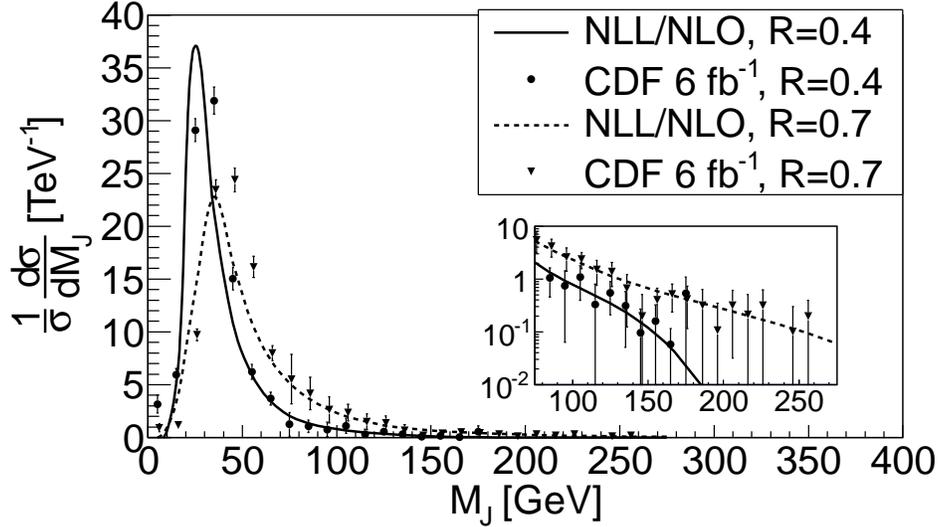}
\caption{Comparison of resummation predictions for the jet mass
distribution to Tevatron CDF data with the kinematic cuts $P_T>400$
GeV and $0.1<|Y|<0.7$ at $R=0.4$ and $R=0.7$. The inset shows the
detailed comparison in large jet mass region.} \label{CONVP1}
\end{figure}

\begin{figure}[!htb]
\includegraphics[width=0.7\textwidth]{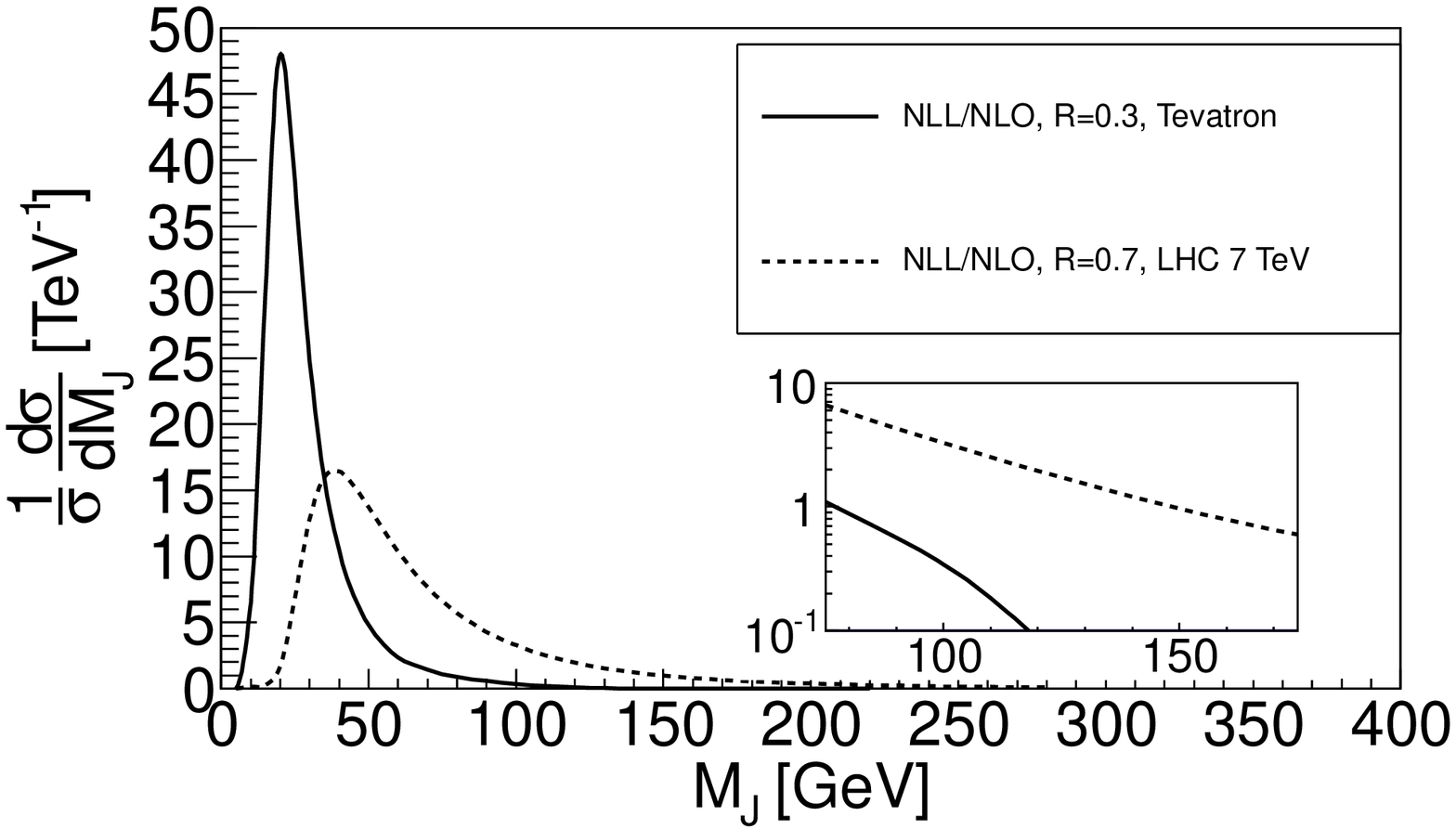}
\caption{Resummation predictions for the jet mass distribution for
Tevatron and LHC. The inset shows the detailed behaviors in large
jet mass region.} \label{CONVP2}
\end{figure}

Following Eq.~(\ref{nlo}), we convolute the light-quark and gluon
jet functions with the constituent cross sections of LO partonic
dijet processes at the Tevatron and the parton distribution
functions (PDF) CTEQ6L \cite{Pumplin:2002vw}. Here we have neglected
the soft gluon contribution \cite{Li:1996gi}, equivalent to the soft
function introduced in the Soft Collinear Effective Theory (SCET) \cite{Fleming:2007qr}, which couples the
light-particle jet and the partonic processes. The resummation
predictions for the jet mass distributions at $R=0.4$ and $R=0.7$
are compared to the Tevatron CDF data \cite{Aaltonen:2011pg} in
Fig.~\ref{CONVP1} with the kinematic cuts $P_T>400$ GeV and the
rapidity interval $0.1<|Y|<0.7$ . The above data were obtained using
the midpoint jet algorithm \cite{Blazey:2000qt}, and the data from
the anti-$k_t$ algorithm \cite{Cacciari:2008gp} do not vary much as
shown in \cite{Aaltonen:2011pg}. The consistency of the resummation
results with the CDF data is excellent at intermediate $M_J$.
The resummation formula describes the shapes and
the peak heights of the jet distributions in the
small $M_J$ region, but with the peak
positions being slightly lower than the CDF data. As indicated in
\cite{Aaltonen:2011pg}, the PDF uncertainties could induce large
variation in shapes of jet mass distributions around peak positions.
The difference from the data in Fig.~\ref{CONVP1} is within the
above variation. This is the first time that the pQCD factorization
theorem explains the observed jet mass distributions successfully.
Note that the jet mass distribution, which corresponds to the
angularity distribution with $a=0$ \cite{Ellis:2010rwa}, cannot be
well described in the SCET formalism. In Fig.~\ref{CONVP2} we
display the resummation predictions for the jet mass distributions
at the Tevatron with $R=0.3$ and at the LHC with $R=0.7$, which can
be tested by Tevatron data and LHC experiments.

\section{JET ENERGY PROFILES}

We define the jet energy functions $J^E_f(M_J^2,P_T,\nu^2,R,r)$
with $f=q(g)$ denoting the light-quark (gluon), which describe the energy
accumulation within the cone of size $r<R$. The definition is chosen, such that
$J^{E(0)}_f=P_T\delta(M_J^2)$ at LO. In this section we will study the energy profile
of a light-particle jet in the framework of QCD resummation at leading power of
$r$. The Feynman rules for $J^E_f$
are similar to those for the jet functions $J_f$ at each order of
$\alpha_s$, except that a sum of the step functions
$\sum_ik_i^0\Theta(r-\theta_i)$  is inserted, where $k_i^0$
($\theta_i$) is the energy (the angle with respect to the jet axis)
of the final-state particle $i$. For example, the jet energy
functions $J^E_f$ are expressed, at NLO, as
\begin{eqnarray}
J_q^{E(1)}(M_J^2,P_T,\nu^2,R,r,\mu^2)&=&\frac{(2\pi)^3}
{2\sqrt{2}(P_J^0)^2N_c}\sum_{\sigma,\lambda}
\int\frac{d^3p}{(2\pi)^3 2p^0}\frac{d^3k}{(2\pi)^3 2k^0}
[p^0\Theta(r-\theta_p)+k^0\Theta(r-\theta_k)]
\nonumber \\
&& \times{\rm Tr}\left\{\not\xi\langle 0|q(0)W_n^{(\bar
q)\dagger}(\infty,0)|p,\sigma;k,\lambda\rangle \langle
k,\lambda;p,\sigma|W_n^{(\bar q)}(\infty,0) \bar
q(0)|0\rangle\right\}\nonumber\\
& &\times\delta(M_J^2-(p+k)^2)\delta^{(2)}(\hat e-\hat
e_{\bf{p}+\bf{k}})\delta(P_J^0-p^0-k^0), \nonumber \\
J_g^{E(1)}(M_J^2,P_T,\nu^2,R,r,\mu^2)&=&\frac{(2\pi)^3}
{2(P_J^0)^3N_c}\sum_{\sigma,\lambda} \int\frac{d^3p}{(2\pi)^3
2p^0}\frac{d^3k}{(2\pi)^3 2k^0}
[p^0\Theta(r-\theta_p)+k^0\Theta(r-\theta_k)]
\nonumber \\
&& \times \langle 0|\xi_\sigma
F^{\sigma\nu}(0)W_n^{(g)\dagger}(\infty,0)
|p,\sigma;k,\lambda\rangle\langle
k,\lambda;p,\sigma|W_n^{(g)}(\infty,0)
F_\nu^\rho(0)\xi_\rho|0\rangle\nonumber\\
& &\times\delta(M_J^2-(p+k)^2)\delta^{(2)}(\hat e-\hat
e_{\bf{p}+\bf{k}})\delta(P_J^0-p^0-k^0),
\label{jetENLO1}
\end{eqnarray}
where the expansion of the Wilson links in $\alpha_s$ is understood.
As shown in the previous section, the quark-loop
contribution to the gluon jet function is not important, cf.
Eq.~(\ref{gloop}), with a proper choice of
the factorization scale $\mu$  in the resummation calculation.
Hence, the quark-loop contribution to the energy profile of the gluon jet
can also be ignored with an appropriate choice of $\mu$.

When $r$ approaches zero, the phase space of real radiation is
strongly constrained, so infrared enhancement does not cancel exactly
with that in virtual contribution and results in large logarithms,
e.g., $\alpha_s \ln^2 r$. An evolution equation for summing these
logarithms to all orders in $\alpha_s$ in the jet energy functions
can be constructed, whose derivation is similar to that for the jet
functions discussed in Sec.~II: the variation of the Wilson line
direction introduces the same special vertex in the differentiated
jet energy functions. The virtual gluons emitted from the special
vertex are factorized into the same hard kernel $G^{(1)}$ and the
same virtual soft kernel $K_v^{(1)}$. For example, their expressions
for the light-quark jet energy function $J_q^E$ are given by
Eqs.~(\ref{g2}) and (\ref{jv2}), respectively. For the real soft
gluon emitted from the special vertex, we split the sum of the step
functions into
\begin{eqnarray}
\sum_ik_i^0\Theta(r-\theta_i)=\sum_{i'} k_{i'}^0
\Theta(r-\theta_{i'})+l^0\Theta(r-\theta),\label{split}
\end{eqnarray}
in which $\sum_{i'}$ means a summation over final-state particles with
the real soft gluon being excluded. The first term in
Eq.~(\ref{split}) gives
\begin{eqnarray}
K_r^{(1)}\otimes J_q^E&=&g_s^2C_F\int\frac{d^4 l}{(2\pi)^4}\frac{\hat
n\cdot P_J}{(n\cdot l+i\epsilon)(P_J\cdot
l-i\epsilon)}2\pi\delta(l^2) \Theta\left(r-\frac{|{\bf l}|\sin\theta}{P_T}\right)
J_q^E(M_J^2-2P_J\cdot l,P_T,\nu^2,R,r). \label{ee1}
\end{eqnarray}
Because of the real soft gluon emission with the polar angle
$\theta$, the jet axis of the rest of particles, described by
$J_q^E$ on the right-hand side of the above expression, inclines by
an angle $|{\bf l}|\sin\theta/P_T$ with respect to the jet momentum
$P_J$. The step function in Eq.~(\ref{ee1}) imposes a phase-space
constraint on the real soft gluon emission, such that the jet axis
of the rest of particles cannot move outside of the jet cone $r$.
Applying the Mellin transformation with respect to $x\equiv
M_J^2/(RP_T)^2$, we have
\begin{eqnarray}
& &\int_0^1 dx (1-x)^{N-1}K_r^{(1)}\otimes
J_q^E ={\bar K}_r^{(1)}(N){\bar J}_q^E(N,P_T,\nu^2,R,r),
\end{eqnarray}
with the definition
\begin{eqnarray}
{\bar K}_r^{(1)}(N)&=& g_s^2C_F\int_0^1 dz(1-z)^{N-1}\int\frac{d^4
l}{(2\pi)^3}\frac{2P_J\cdot l n^2}{(n\cdot l+i\epsilon)^2(2P_J\cdot
l+\lambda^2)} \nonumber\\ & &\times
\delta(l^2)\delta\left(z-\frac{2|{\bf
l}|}{RP_T}(1-\cos\theta)\right) \Theta\left(r-\frac{|{\bf
l}|\sin\theta}{P_T}\right).\label{ee2}
\end{eqnarray}

The second term in Eq.~(\ref{split}) leads to
\begin{eqnarray}
K_e^{(1)}\otimes J_q&=&g_s^2C_F\int\frac{d^4 l}{(2\pi)^4}\frac{\hat
n\cdot P_Jl^0\Theta(r-\theta)}{(n\cdot l+i\epsilon)(P_J\cdot
l-i\epsilon)} 2\pi\delta(l^2)J_q(M_J^2-2P_J\cdot l,P_T,\nu^2,R),
\label{kr1}
\end{eqnarray}
whose Mellin transformation gives
\begin{eqnarray}
\int_0^1 dx (1-x)^{N-1} K_e^{(1)}\otimes J_q ={\bar
K}_e^{(1)}(N){\bar J}_q(N,P_T,\nu^2,R),
\end{eqnarray}
with the definition
\begin{eqnarray}
{\bar K}_e^{(1)}(N)= g_s^2C_F\int_0^1 dz(1-z)^{N-1}\int\frac{d^4
l}{(2\pi)^3}\frac{n^2l^0\Theta(r-\theta)}{(n\cdot l+i\epsilon)^2}
\delta(l^2)\delta\left(z-\frac{2|{\bf
l}|}{RP_T}(1-\cos\theta)\right) \Theta\left(\frac{P_T}{2}-|{\bf
l}|\right).\label{ke1}
\end{eqnarray}
Strictly speaking, the energy $|{\bf l}|$ of a real gluon cannot
approach infinity, so the step function at the end of the above
expression has been introduced. Working out the above integration,
we obtain
\begin{eqnarray}
{\bar K}_e^{(1)}(N)
&=& \frac{\alpha_s}{2\pi}C_F\frac{1}{N}\int
d\cos\theta\frac{n^2}{(n^0- n^x\cos\theta)^2}
\frac{RP_T}{2(1-\cos\theta)}
\left(1-\cos^N\theta\right)
\Theta(r-\theta),\label{ke3}
\end{eqnarray}
which is down by $1-\cos^N r$ and negligible in the small $r$
region. This result is attributed to the suppression of the second
term in Eq.~(\ref{split}) by soft $l$. Hence,
this piece will not be considered from now on.

The jet energy profiles are measured by summing over all jet
invariant masses in experiments. Therefore, we perform a
corresponding analysis with the $M_J^2$ dependence being integrated
out of the jet energy profiles, namely, by considering only the
$N=1$ moment. A straightforward computation leads Eq.~(\ref{ee2}) to
\begin{eqnarray}
{\bar K}_r^{(1)}(1)
&=&\frac{\alpha_s}{2\pi}C_F\ln\frac{\nu^2R^2P_T^4r^2}{\lambda^4},
\label{en3}
\end{eqnarray}
where the infrared regulator $\lambda^2$ will be taken to be zero
eventually, and $\nu^2$ is defined as in Eq.~(\ref{nu}). Using the
same counterterm, Eqs.~(\ref{jv2}) and (\ref{en3}) are combined to
form
\begin{eqnarray}
{\bar K}^{(1)}(1)&=&{\bar K}_v^{(1)}+{\bar K}_r^{(1)}(1)
=\frac{\alpha_s}{2\pi}C_F\ln \frac{\nu^2 P_T^2 r^2
C_1^2}{\mu^{\prime
2}}+\frac{\alpha_s}{2\pi}C_F\ln\frac{C_2^2}{C_1^2}.
\end{eqnarray}
which contains the large single logarithm $\ln r$.

Solving the RG equation for the kernels,
\begin{eqnarray}
\mu'\frac{d}{d\mu'} G=\lambda_K=-\mu'\frac{d}{d\mu'}{\bar K},
\end{eqnarray}
we derive
\begin{eqnarray}
& & {\bar K}\left(\frac{\nu P_T r C_1}{\mu'},\alpha_s(\mu^{\prime
2})\right) + G\left(\frac{\nu^2 C_2 R
P_T}{\mu'},\alpha_s(\mu^{\prime 2})\right)\nonumber\\ & &={\bar
K}\left(1,\alpha_s\left(\nu^2 P_T^2r^2C_1^2\right)\right) +
G\left(1,\alpha_s\left(\nu^4 C_2^2 R^2 P_T^2\right)\right)
-\frac{1}{2}\int_{C_1^2\nu^2 P_T^2 r^2}^{C_2^2 \nu^4 P_T^2
R^2}\frac{d\mu^{\prime 2}}{\mu^{\prime 2}}
\lambda_K(\alpha_s(\mu^{\prime 2})),\nonumber\\
& &=\frac{C_F}{2\pi}\ln\frac{C_2^2}{C_1^2}\alpha_s\left(\nu^2
P_T^2r^2C_1^2\right) +\frac{C_F}{2\pi}\alpha_s\left(\nu^4 C_2^2 R^2
P_T^2\right) -\frac{1}{2}\int_{C_1^2\nu^2 r^2}^{C_2^2 \nu^4 R^2}
\frac{d\omega}{\omega}\lambda_K(\alpha_s(\omega P_T^2)).
\end{eqnarray}
The light-quark jet energy function ${\bar J}_q^E$ then obeys a
differential equation similar to Eq.~(\ref{crd2}):
\begin{eqnarray}
& &-\frac{n^2}{v\cdot n}v_{\alpha}\frac{d}{dn_\alpha}{\bar
J}_q^E(1,P_T,\nu^2,R,r,\mu^2)=2\nu^2\frac{d}{d\nu^2}{\bar
J}_q^E(1,P_T,\nu^2,R,r,\mu^2)\nonumber\\
& &=2\left[{\bar K}\left(\frac{\nu P_T r
C_1}{\mu'},\alpha_s(\mu^{\prime 2})\right) + G\left(\frac{\nu^2 C_2
R P_T}{\mu'},\alpha_s(\mu^{\prime 2})\right)\right]{\bar
J}_q^E(1,P_T,\nu^2,R,r,\mu^2). \label{er2}
\end{eqnarray}
A similar equation also holds for describing the energy profile of the gluon 
jet. As solving these equations, we choose the factorization scale 
$\mu^2 \sim {\cal O}(r^2 P_T^2/(R^2 \nu^2))$,
so that the quark-loop contribution to the gluon jet energy profile can be
ignored, for the quark-loop contribution does not contain the large logarithm
$\ln (R^2/r^2)$ with this choice of the scale.

The strategy to solve the above equation is to evolve $\nu^2$
from the low value $\nu^2=\nu^2_{\rm in}\equiv C_1^2r^2/(C_2^2 R^2)$
to the large value $\nu^2=\nu^2_{\rm fi}\equiv 1$, which correspond
to the specific choices $n=n_{\rm in}\equiv
(1,(4C_2^2-r^2C_1^2)/(4C_2^2+r^2C_1^2),0,0)$ and $n=n_{\rm fi}\equiv
(1,(4-R^2)/(4+R^2),0,0)$, respectively. The solution of the above
equation is given by
\begin{eqnarray}
{\bar J}_q^E(1,P_T,\nu_{\rm fi}^2,R,r)&=&{\bar J}_q^E(1,P_T,\nu_{\rm
in}^2,R,r)\exp\left[ S_q(R,r) \right], \label{ep1}
\end{eqnarray}
with the Sudakov exponent
\begin{eqnarray}
S_q(R,r)&=& \int^C_{C\nu_{\rm in}^2}\frac{dy}{y} \left[
\frac{C_F}{2\pi}\ln\frac{C_2^2}{C_1^2}\alpha_s\left(y
P_T^2r^2C_1^2\right) +\frac{C_F}{2\pi}\alpha_s\left(y^2 C_2^2 R^2
P_T^2\right) -\frac{1}{2}\int_{C_1^2 y r^2}^{C_2^2 y^2
R^2}\frac{d\omega}{\omega}\lambda_K(\alpha_s(\omega P_T^2))
\right],\nonumber\\
&=&\int^C_{C\nu_{\rm in}^2}\frac{dy}{y} \left[
\frac{C_F}{\pi}\alpha_s\left(y^2 C_2^2 R^2
P_T^2\right)\left(\frac{1}{2}+\ln\frac{C_2}{C_1}\right)
-\frac{1}{2}\int_{y \nu_{\rm
in}^2}^{y^2}\frac{d\omega}{\omega}A_q(\alpha_s(\omega C_2^2 R^2
P_T^2)) \right],\label{sqr}
\end{eqnarray}
where the constant $C$ will be fixed below. The Sudakov
exponent $S_g(R,r)$ for the gluon jet is obtained by substituting
the color factor $C_A$ for $C_F$ in the above expression. The
resummation formulas are summarized as
\begin{eqnarray}
{\bar J}_f^E(1,P_T,\nu_{\rm fi}^2,R,r)&=&{\bar
J}_f^E(1,P_T,\nu_{\rm in}^2,R,r)\exp\left[ S_f(R,r) \right],
\label{sfr}
\end{eqnarray}
with the subscript $f=q$ or $g$. The ${\cal O}(1)$ constants
are chosen as $C_1=C_2=1$ and $C=\exp(5/2)$ ($C=\exp(17/6)$) in
order to reproduce the single logarithm $\alpha_s\ln r$ in the NLO
light-quark (gluon) jet energy function. The initial conditions
${\bar J}_f^E(1,P_T,\nu_{\rm in}^2,R,r)$ of the Sudakov evolution,
in the absence of the large logarithms and with the factorization
scale $\mu\sim {\cal O}(P_T)$, are calculated up to NLO in
Appendix D.

\begin{figure}[!htb]
\includegraphics[width=0.6\textwidth]{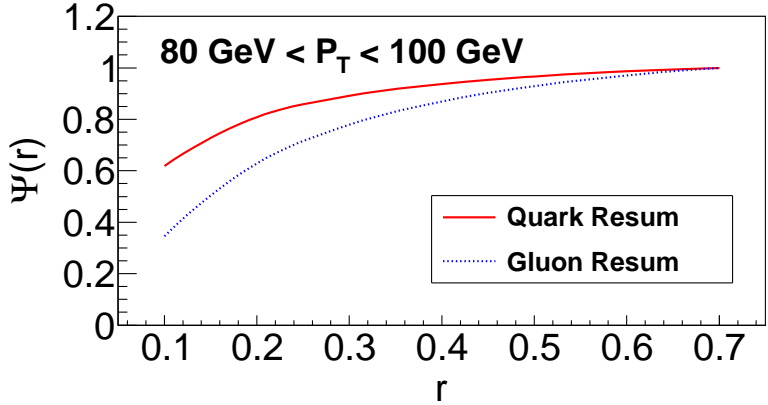}
\caption{Resummation predictions for the energy profiles of the
light-quark (solid curve) and gluon (dotted curve) jets with
$\sqrt{S}=7$ TeV and 80 GeV $< P_T<$ 100 GeV.} \label{COMP}
\end{figure}

Inserting the solutions in Eq.~(\ref{sfr}) into Eq.~(\ref{pro}), the
jet energy profile is written, in terms of the convolution with the
parton-level differential cross section, as
\begin{eqnarray}
\Psi(r)=\left[\sum_f \int \frac{dP_T}{P_T}\frac{d\hat\sigma_f}{dP_T}
\bar J_f^E(1, P_T, \nu_{\rm fi}^2, R,R)\right]^{-1}
\sum_f \int \frac{dP_T}{P_T}\frac{d\hat\sigma_f}{dP_T}
\bar J_f^E(1, P_T, \nu_{\rm fi}^2, R,r),
\label{JE2PSI}
\end{eqnarray}
which respects the normalization $\Psi(R)=1$, and vanishes as $r\to
0$. Note that a jet energy profile, with $N=1$, is not sensitive to
the nonperturbative contribution, so our predictions are free of the
nonperturbative parameter dependence, in contrast to the case of
describing the jet invariant mass distribution, cf. Sec.~II. We find
that the light-quark jet has a narrower energy profile than the
gluon jet, as exhibited in Fig.~\ref{COMP} for $\sqrt{S}=7$ TeV and
the interval $80$ GeV $< P_T<$ $100$ GeV of the jet transverse
momentum. The broader distribution of the gluon jet results from
stronger radiations caused by the larger color factor $C_A=3$,
compared to $C_F=4/3$ for a light-quark jet.

\begin{figure}[!htb]
\includegraphics[width=0.32\textwidth]{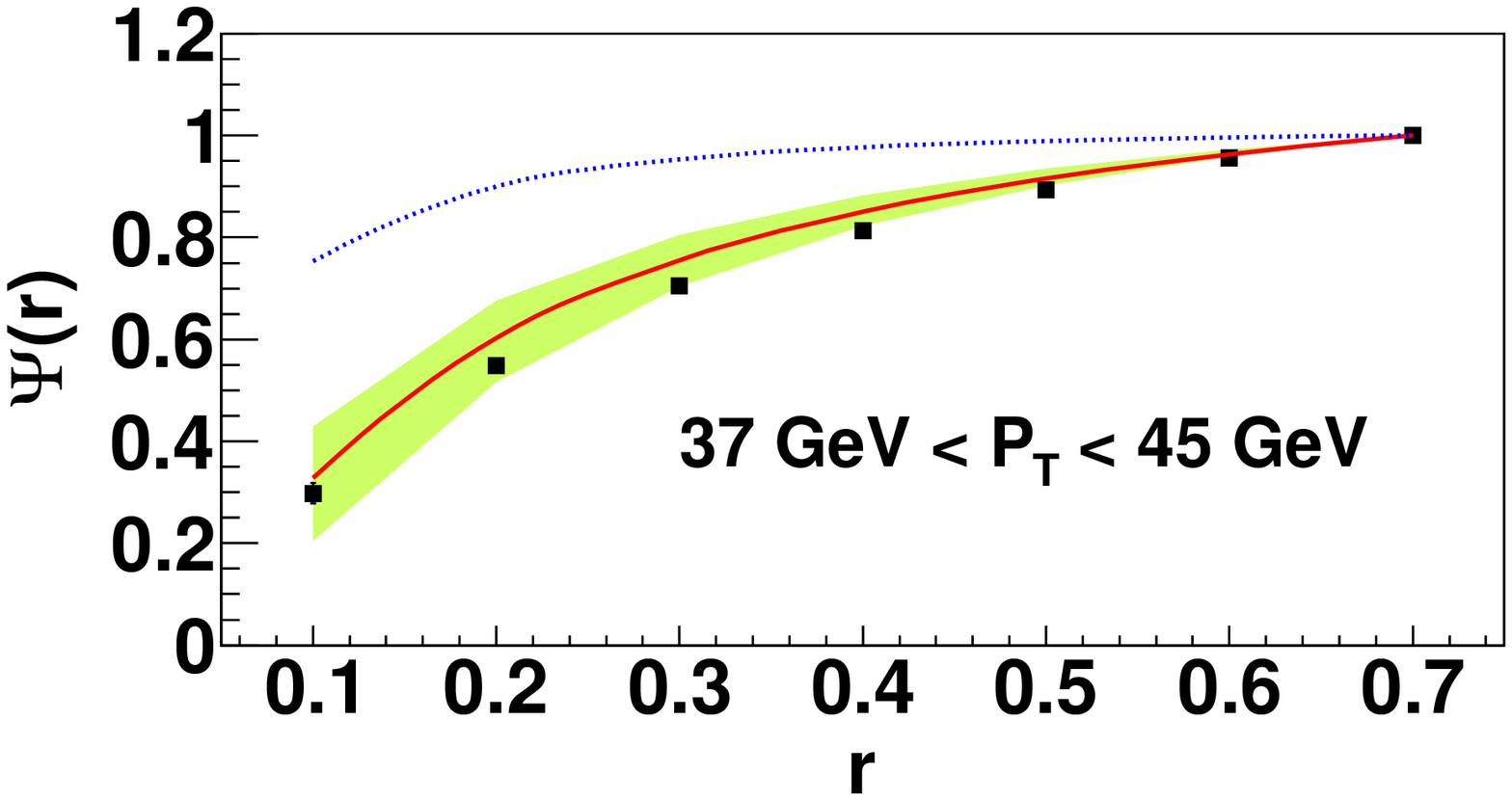}
\includegraphics[width=0.32\textwidth]{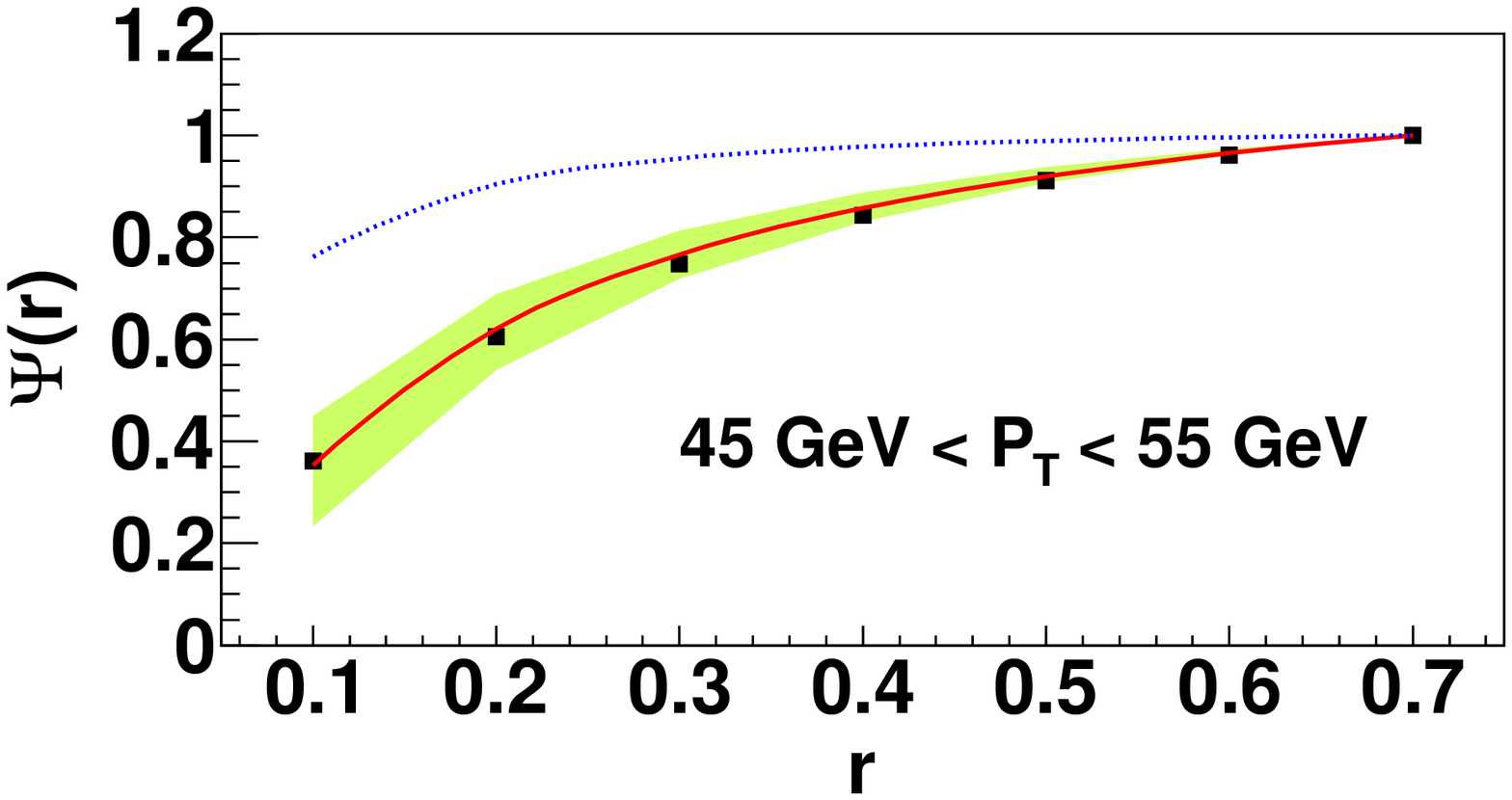}
\includegraphics[width=0.32\textwidth]{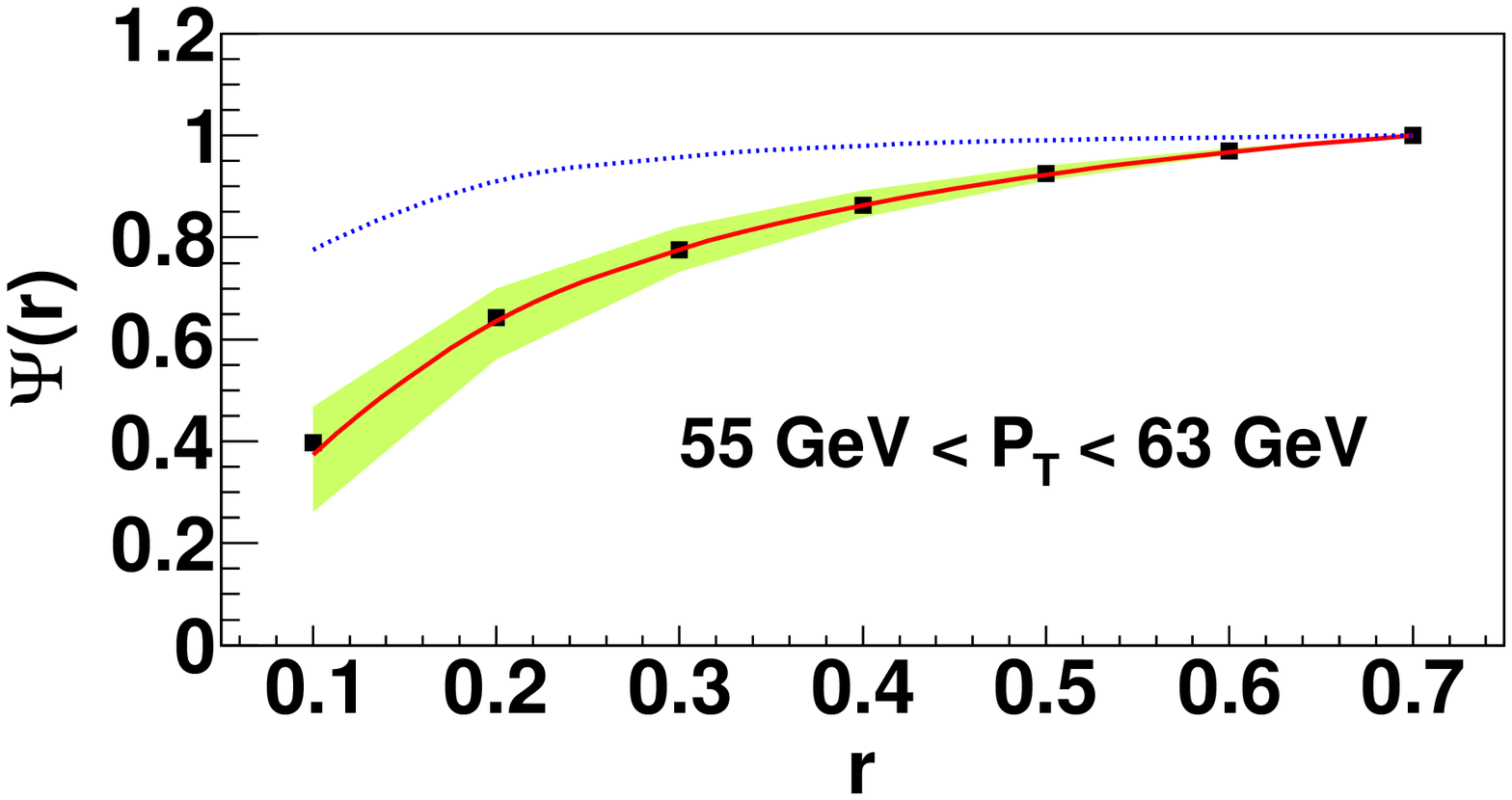}
\includegraphics[width=0.32\textwidth]{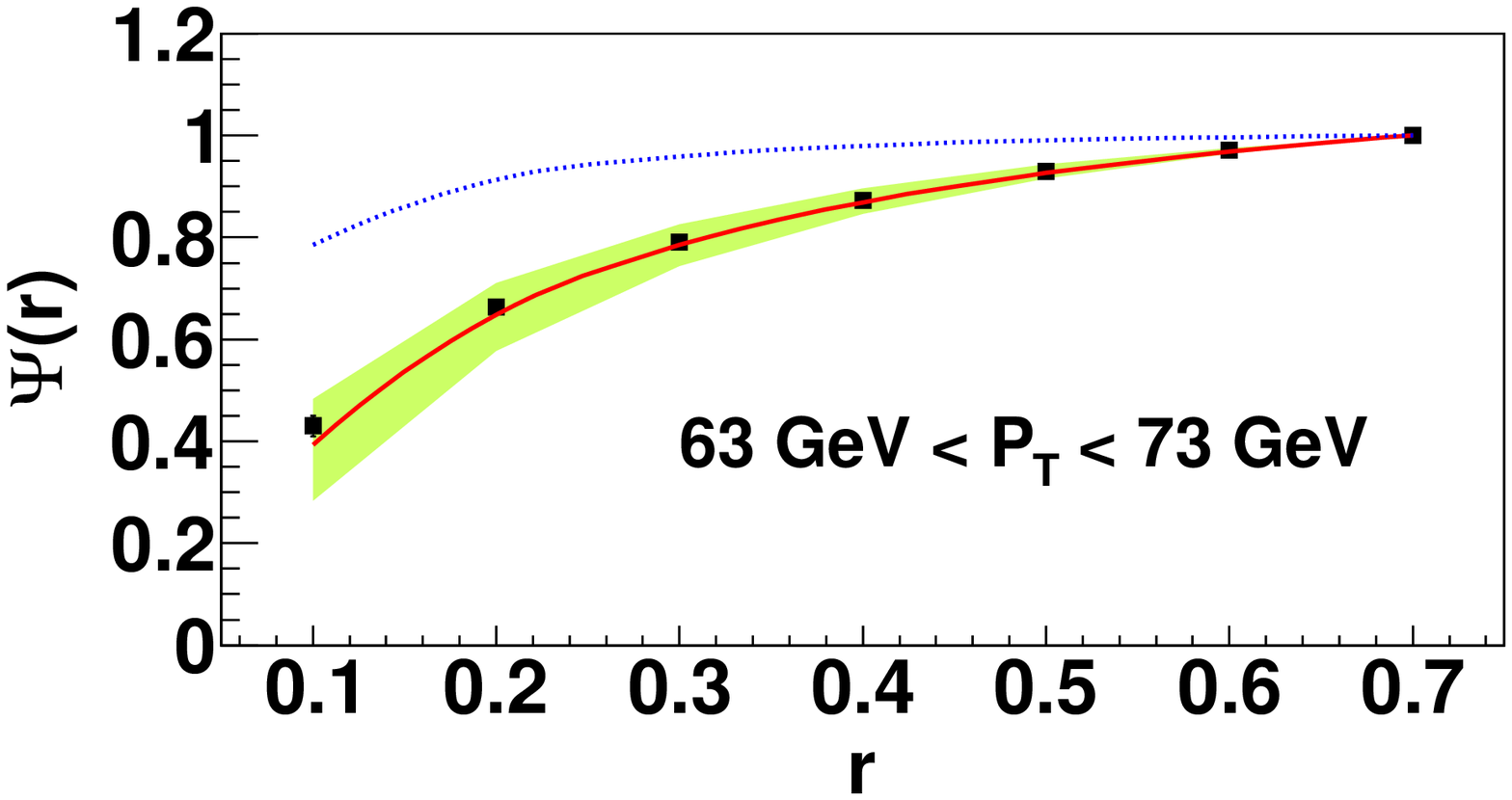}
\includegraphics[width=0.32\textwidth]{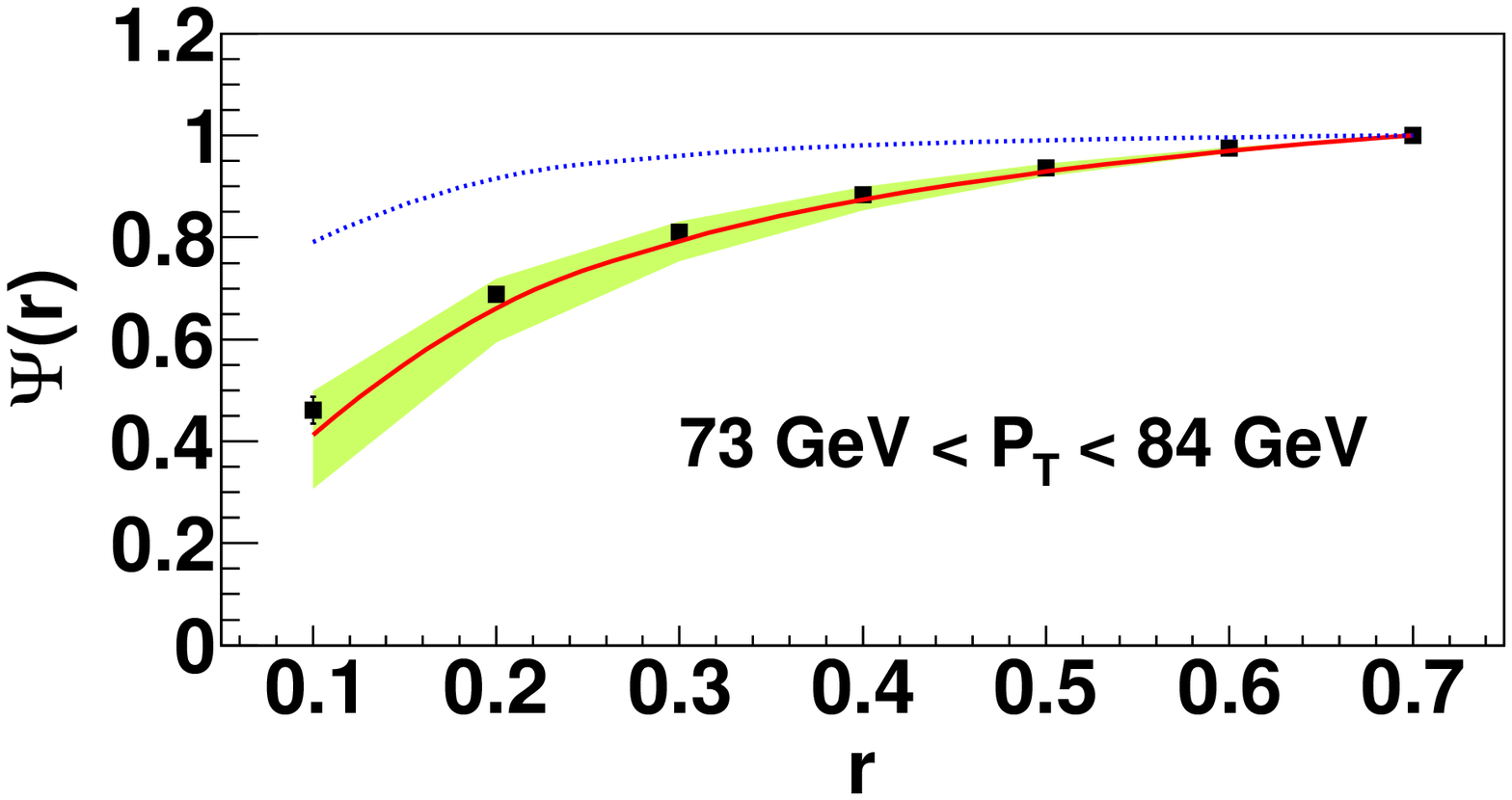}
\includegraphics[width=0.32\textwidth]{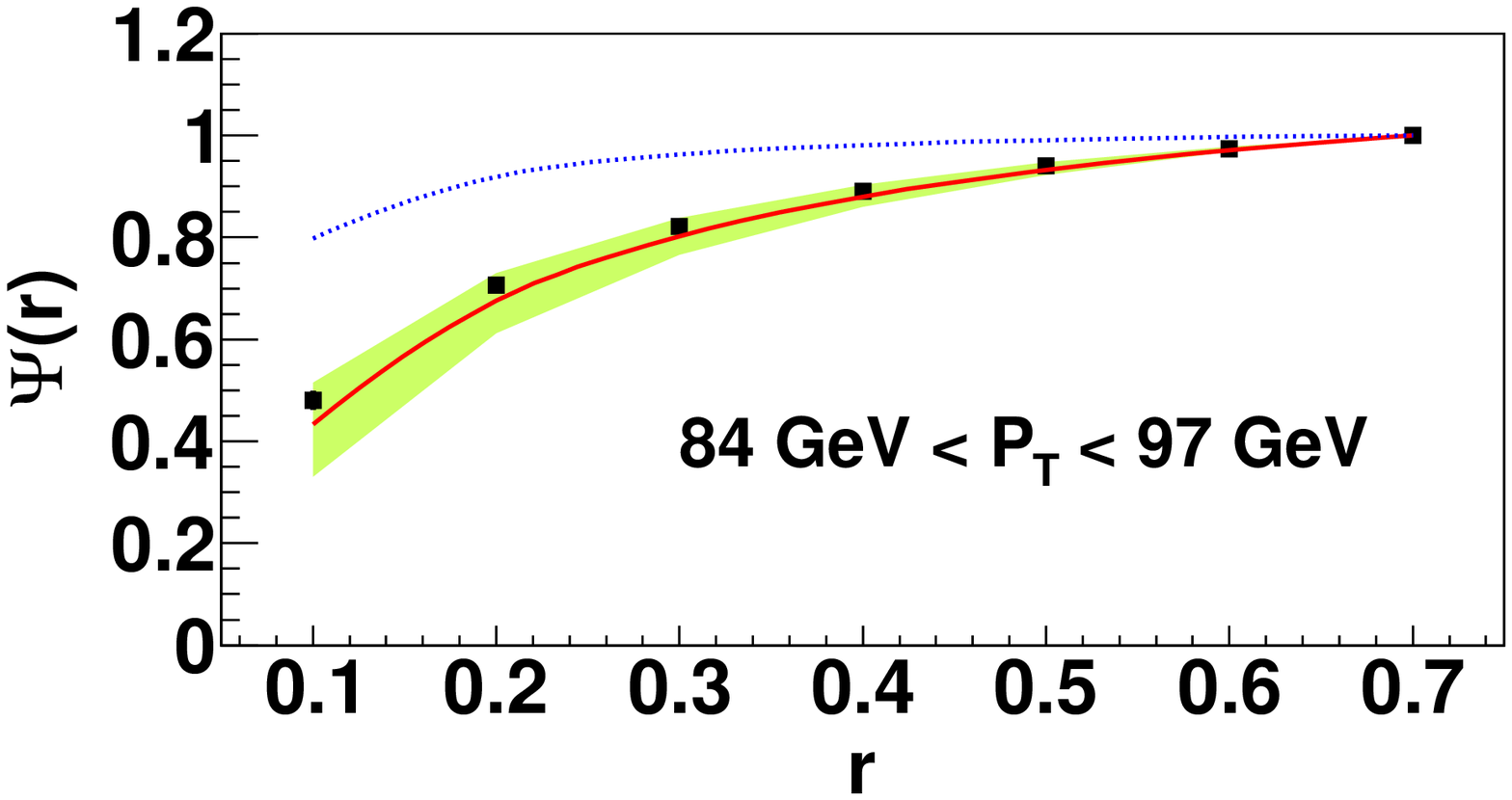}
\includegraphics[width=0.32\textwidth]{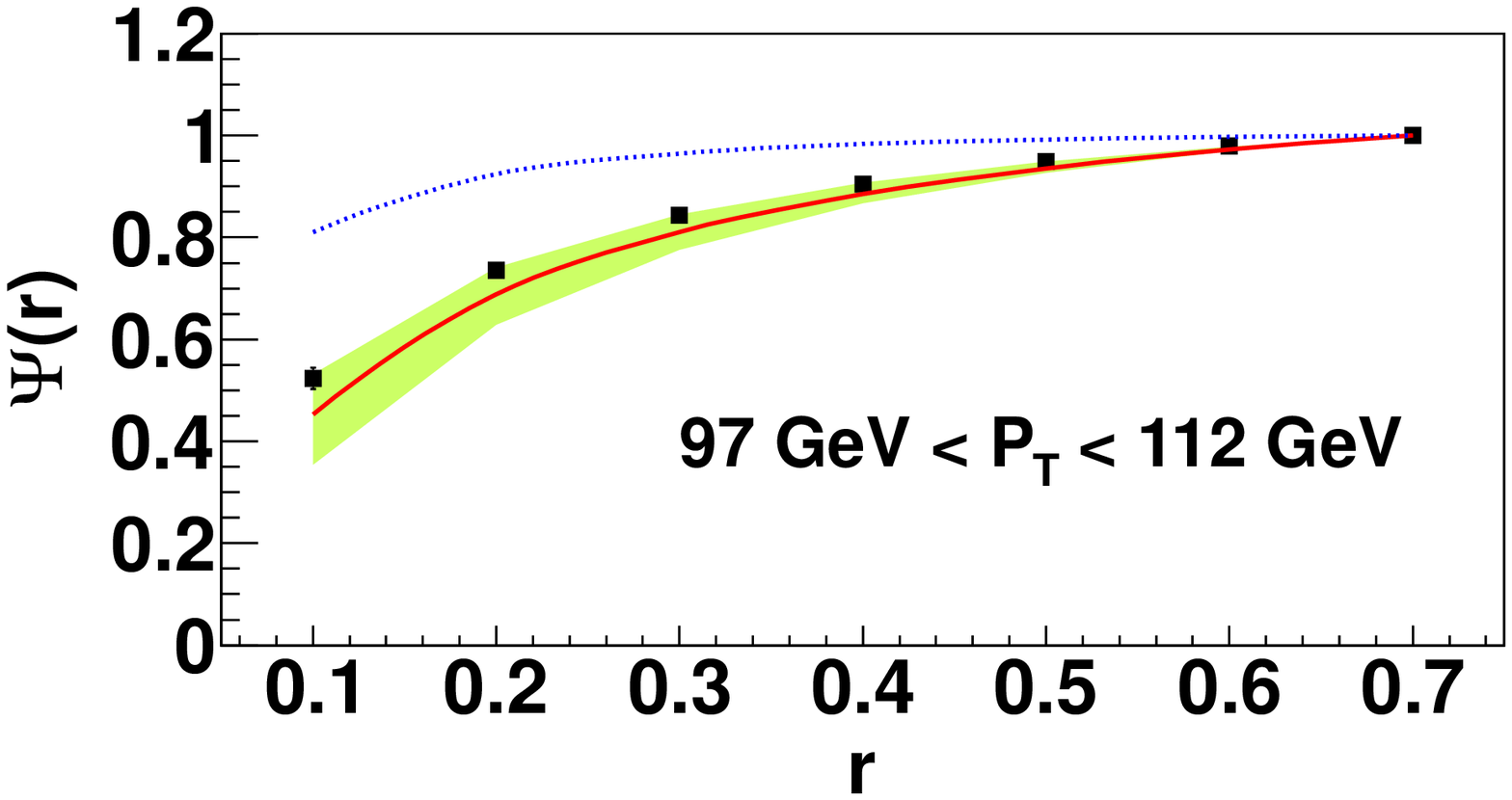}
\includegraphics[width=0.32\textwidth]{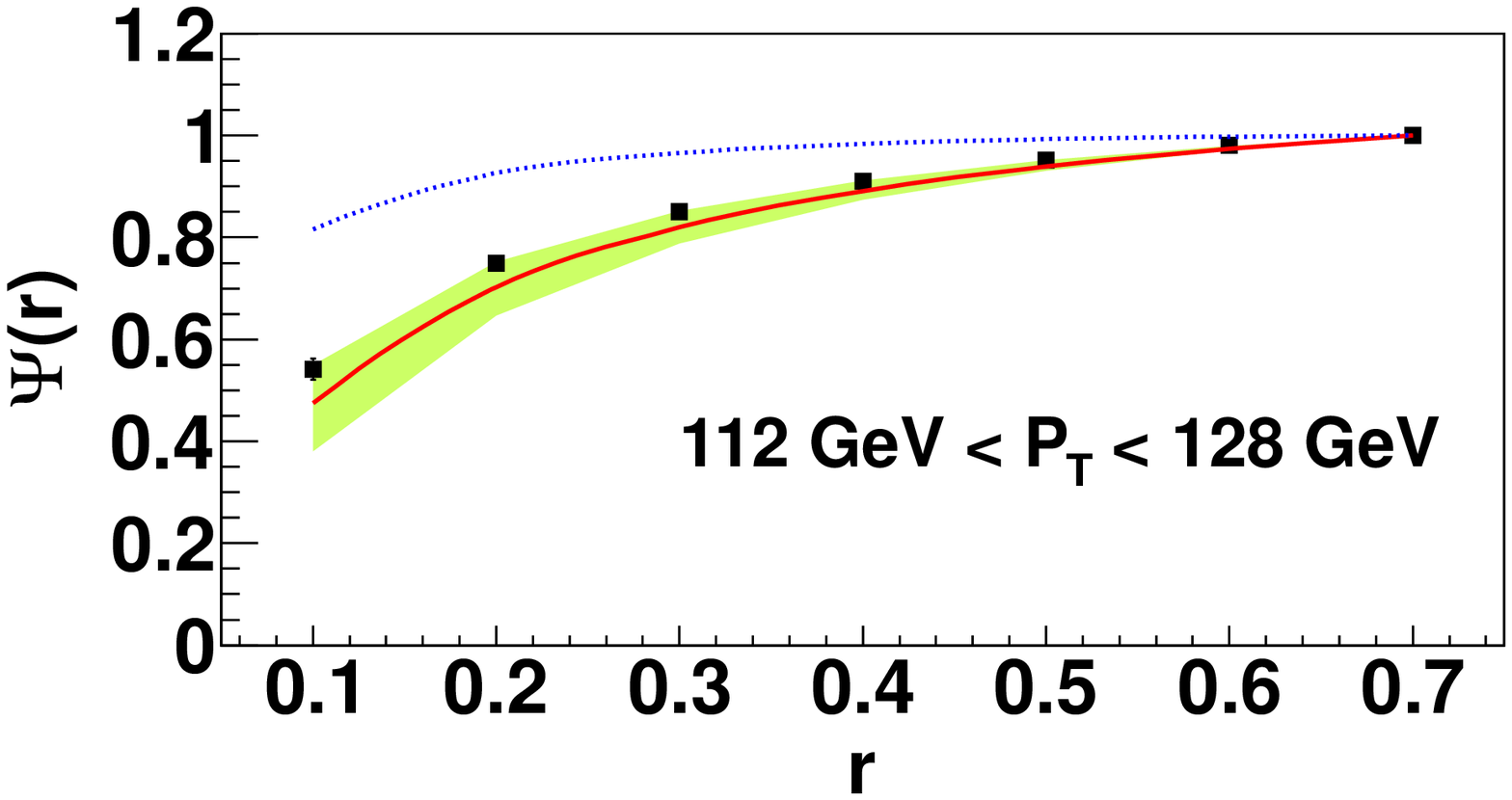}
\includegraphics[width=0.32\textwidth]{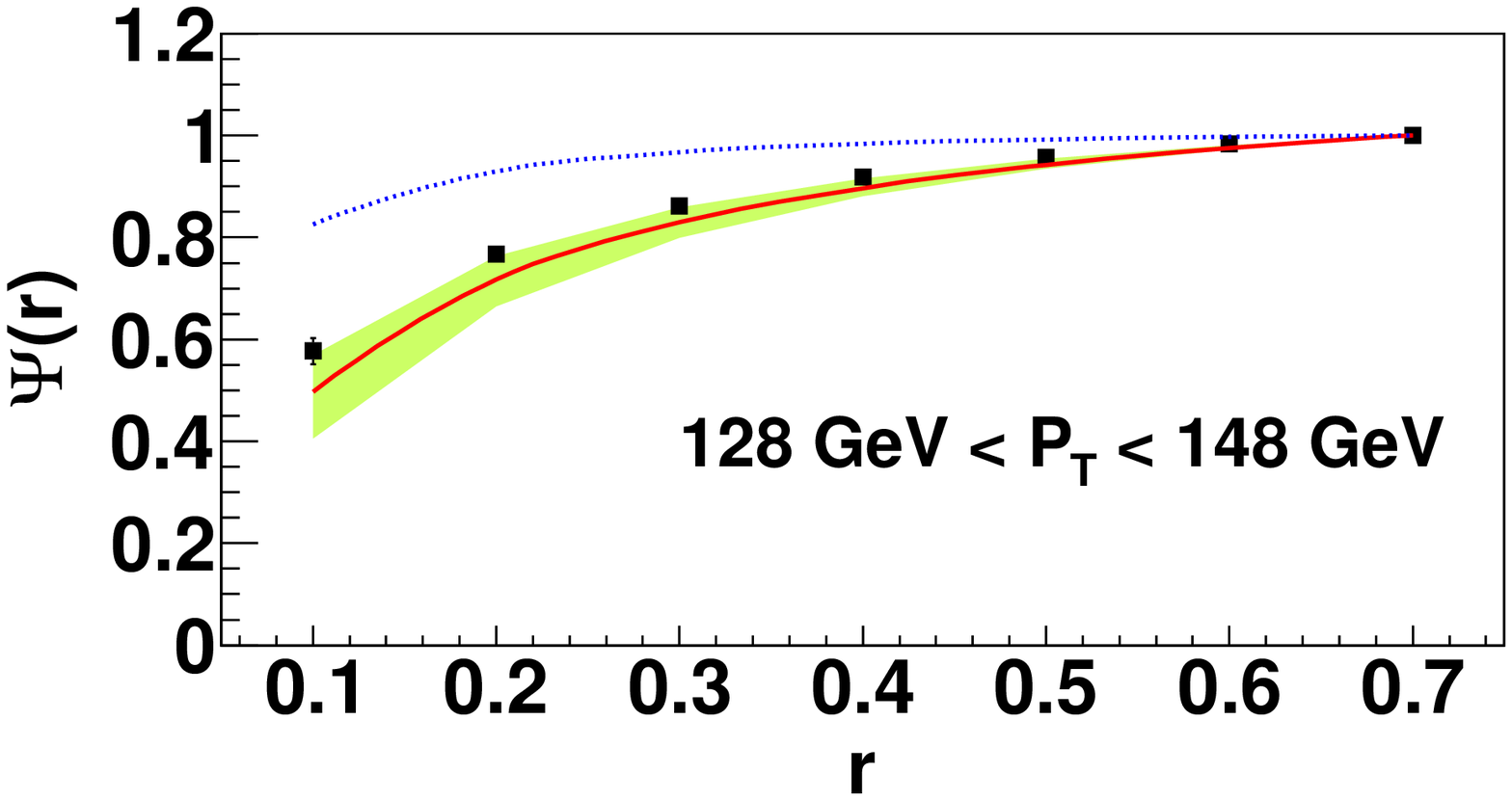}
\includegraphics[width=0.32\textwidth]{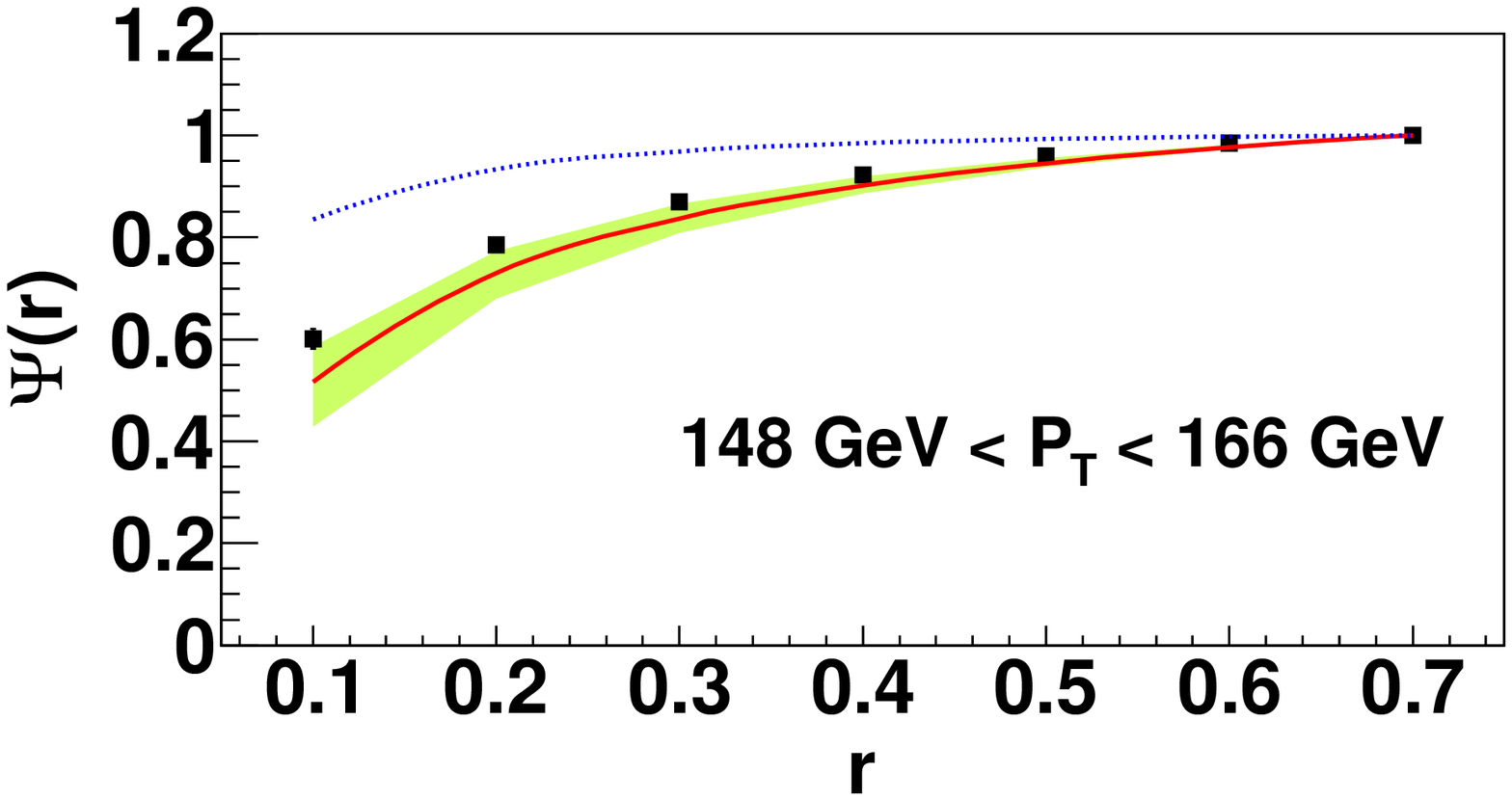}
\includegraphics[width=0.32\textwidth]{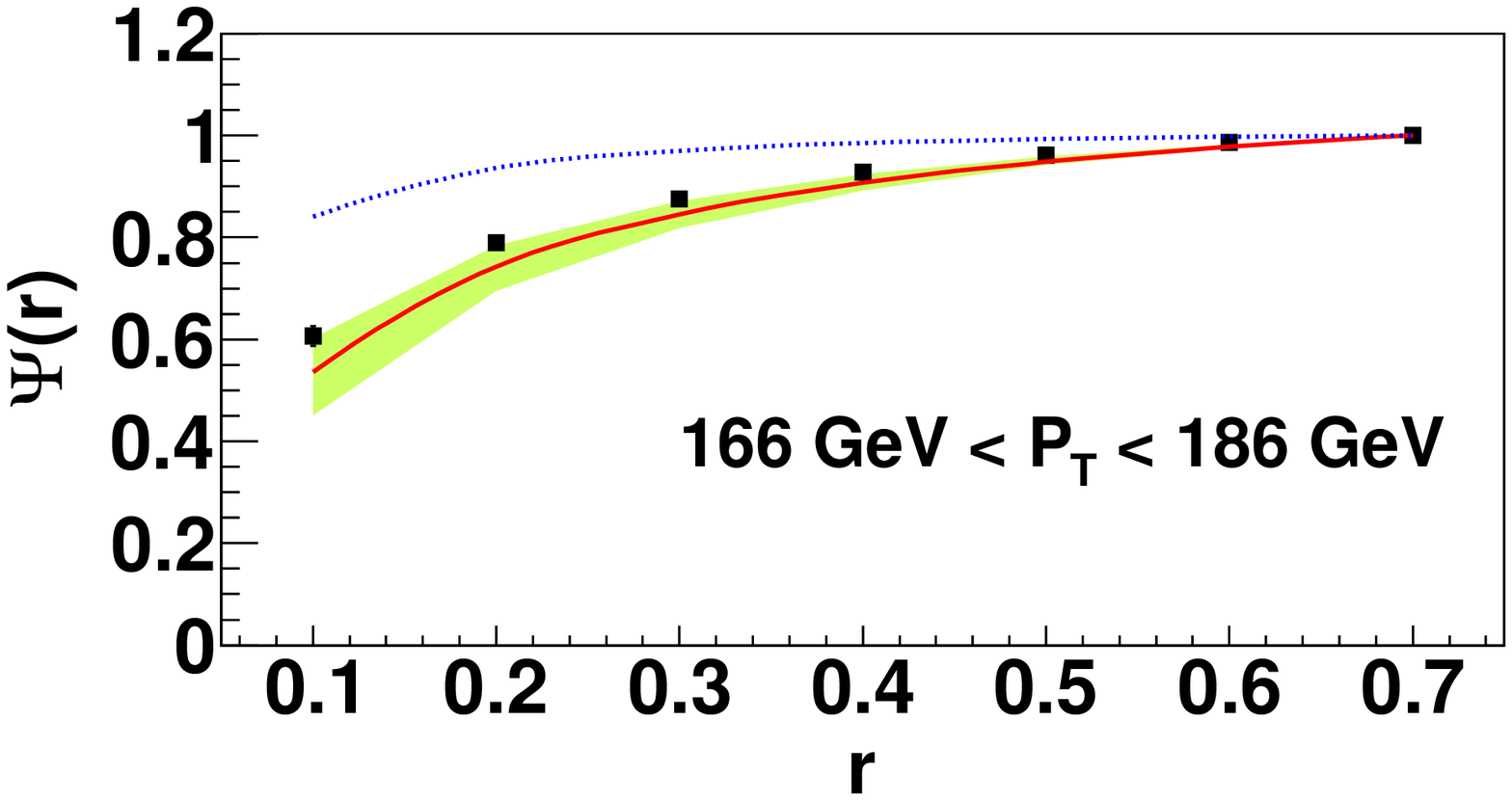}
\includegraphics[width=0.32\textwidth]{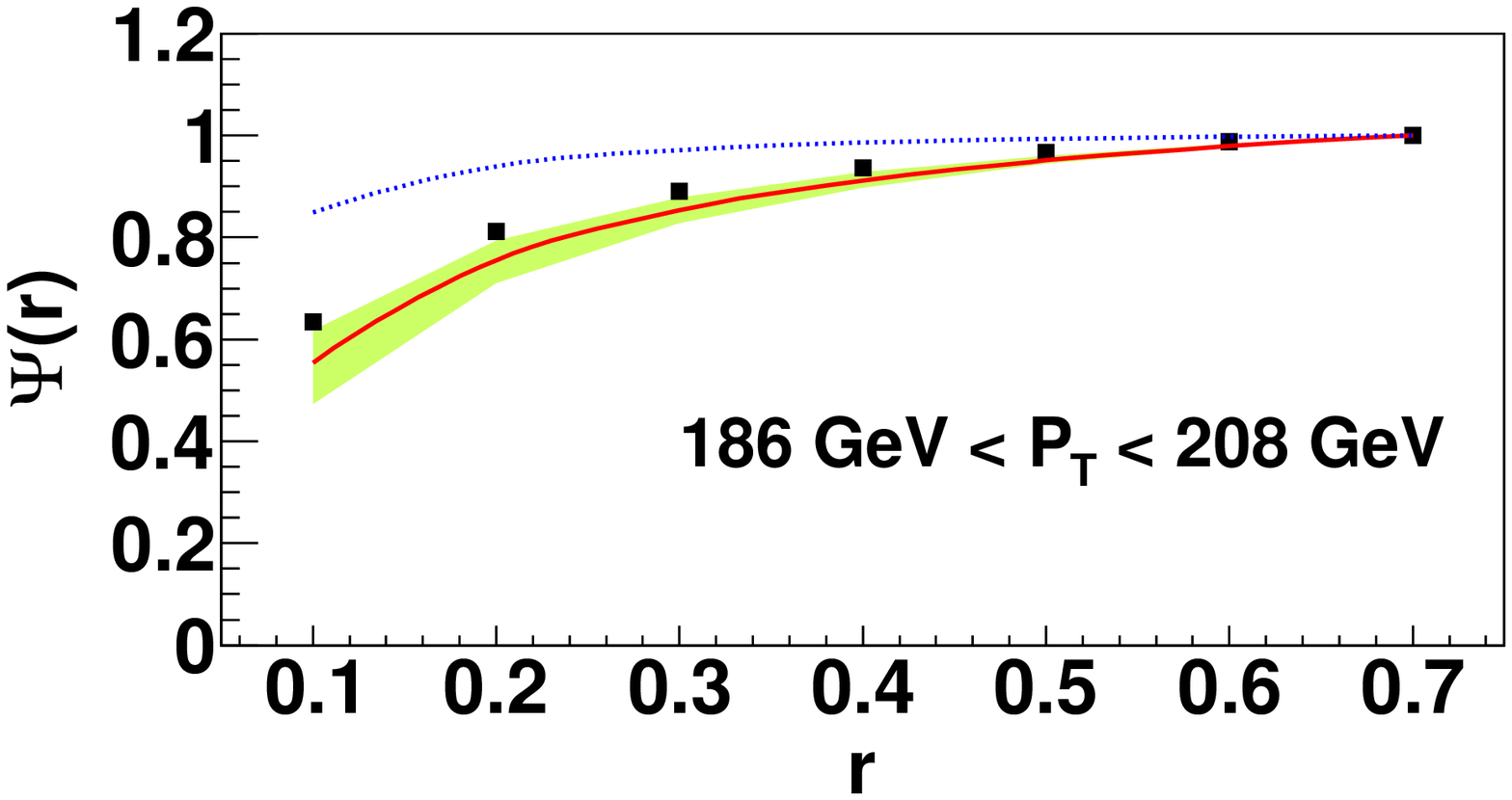}
\includegraphics[width=0.32\textwidth]{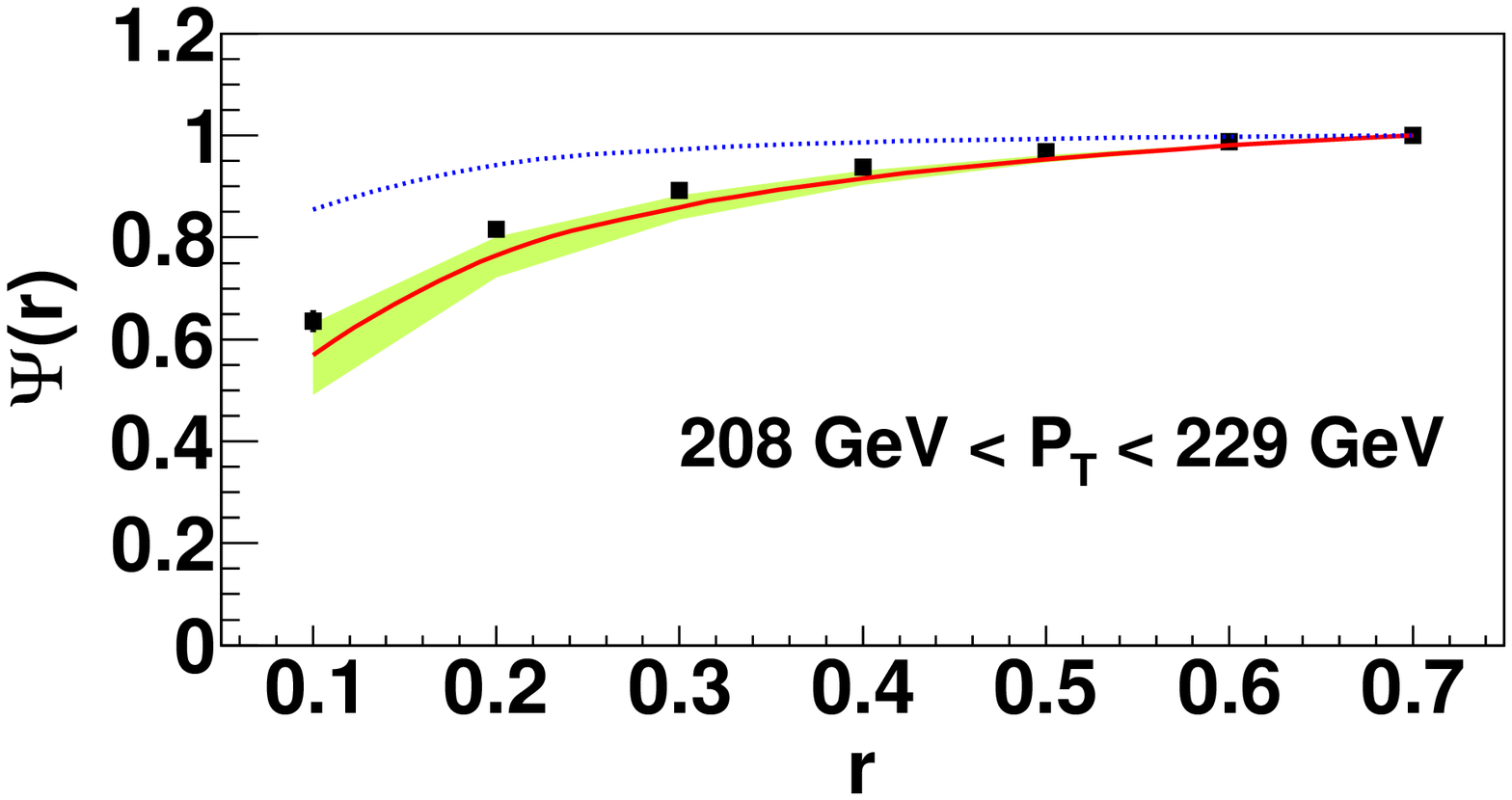}
\includegraphics[width=0.32\textwidth]{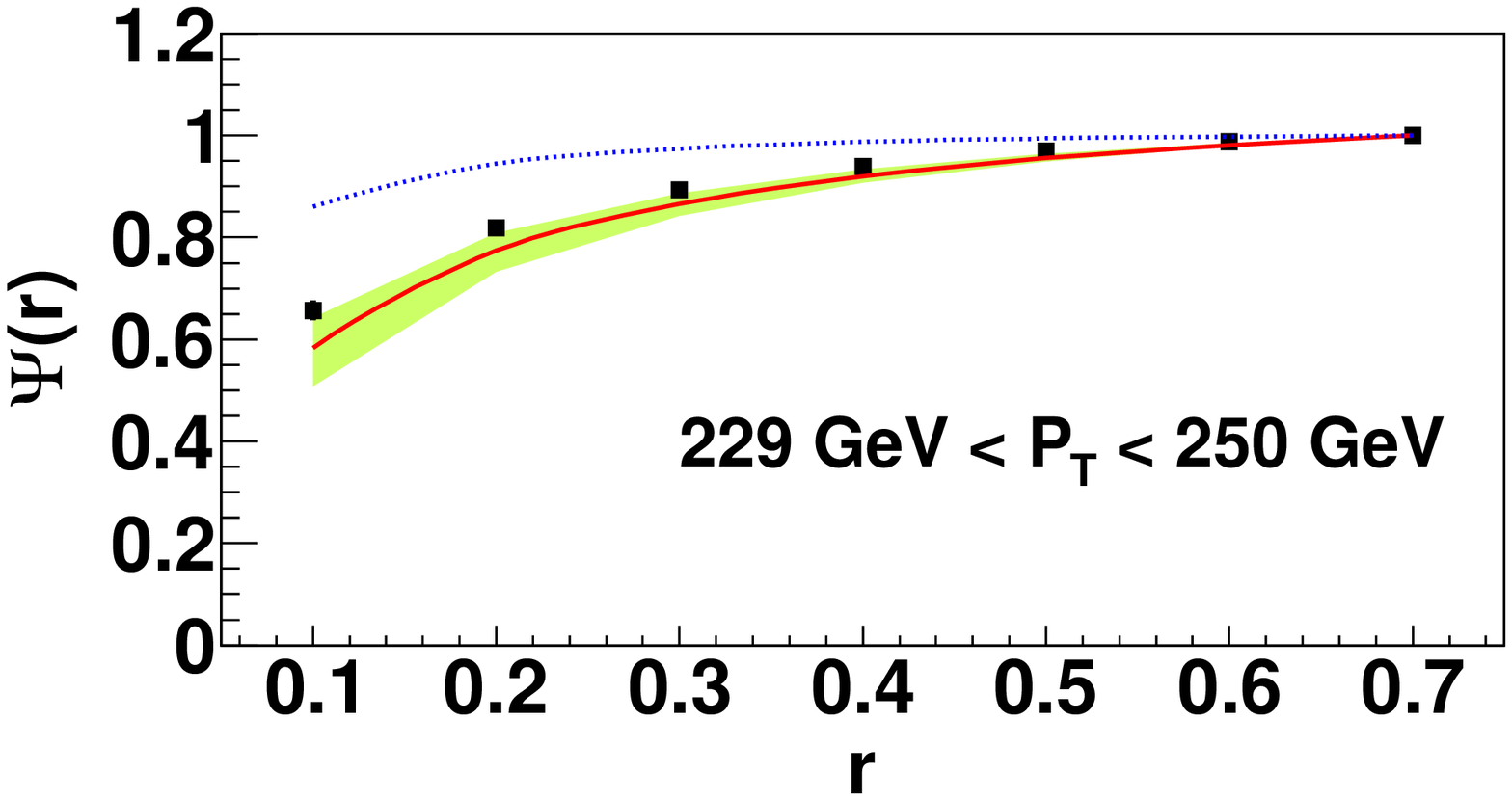}
\includegraphics[width=0.32\textwidth]{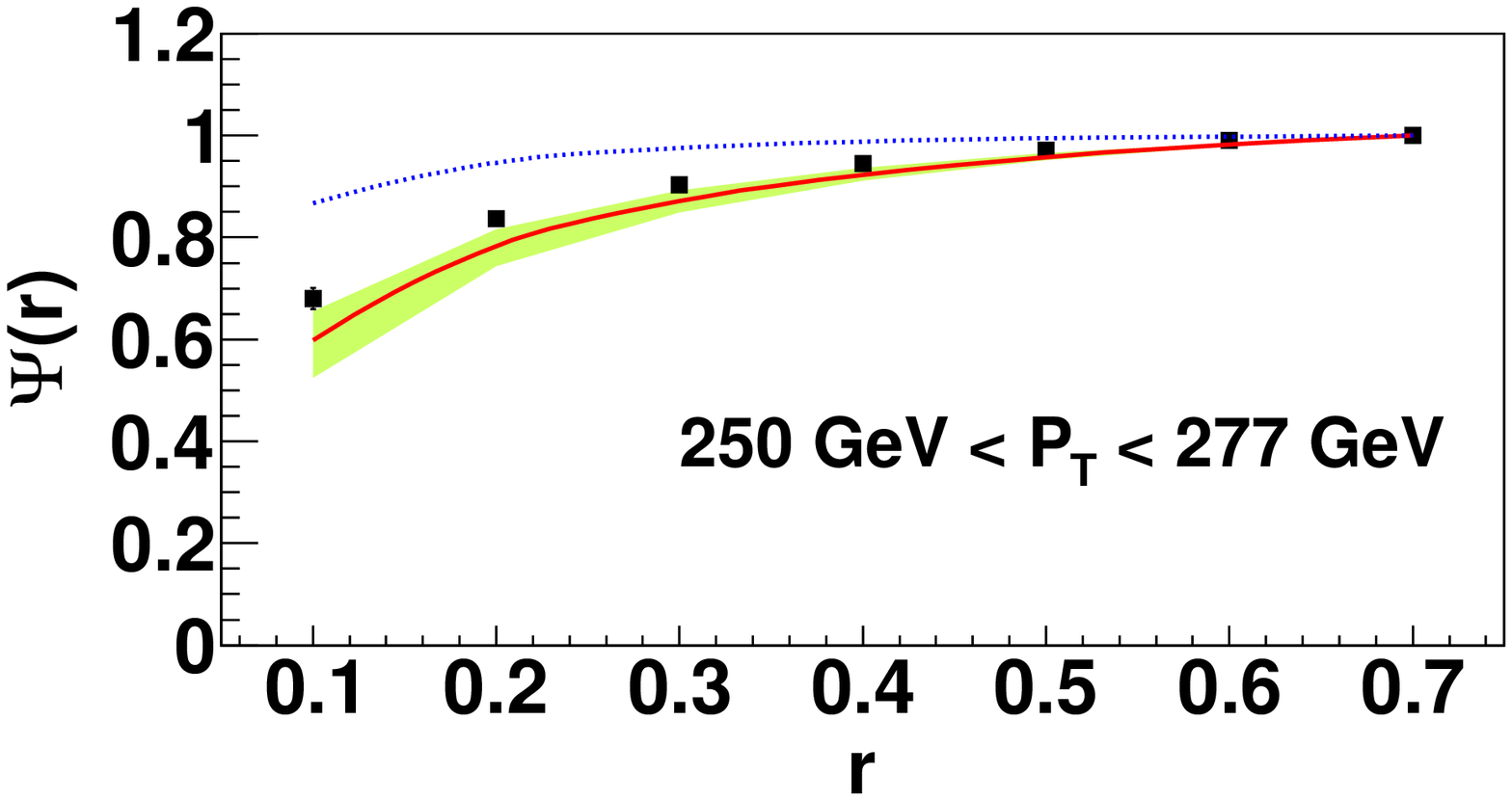}
\includegraphics[width=0.32\textwidth]{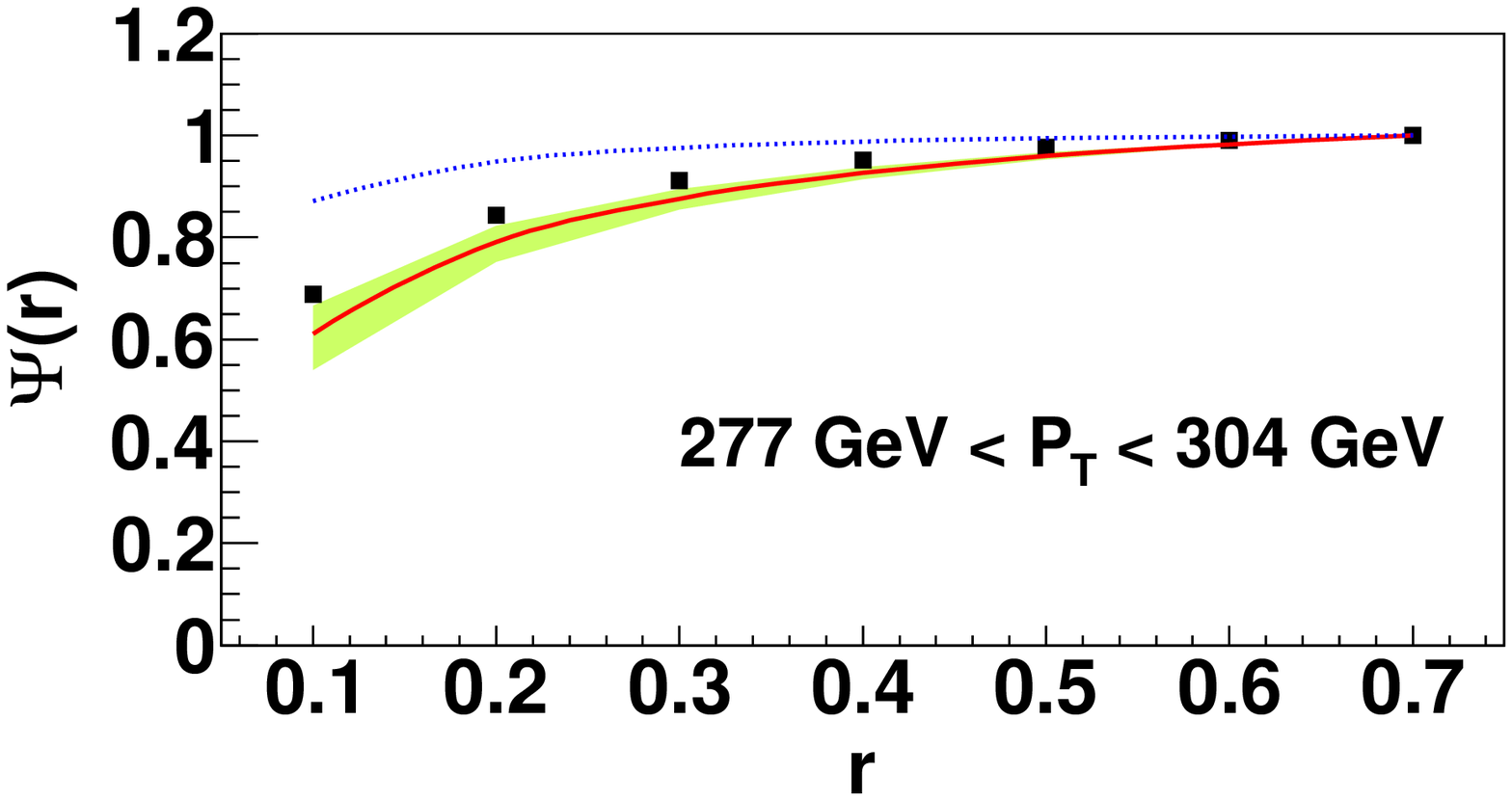}
\includegraphics[width=0.32\textwidth]{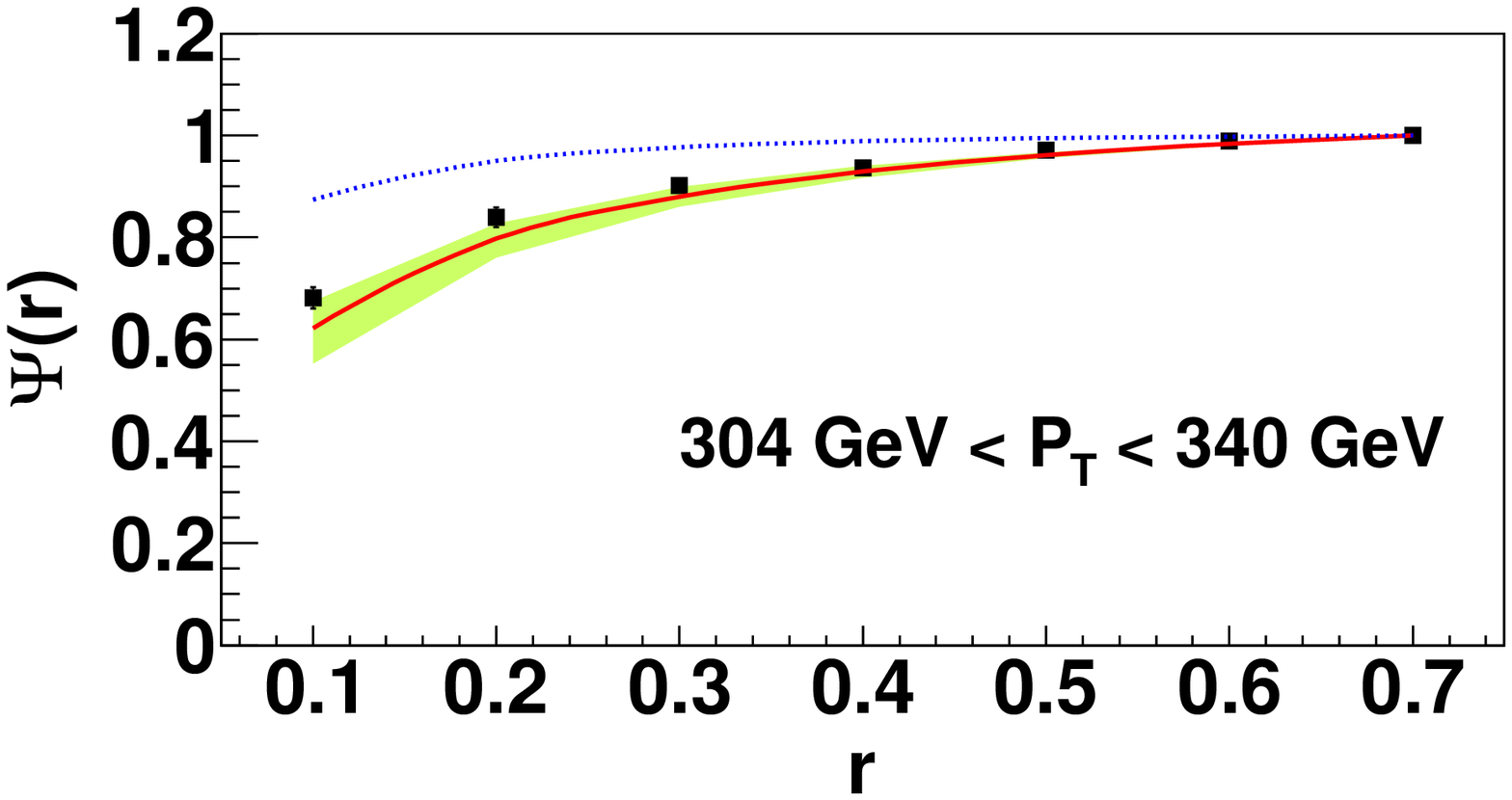}
\includegraphics[width=0.32\textwidth]{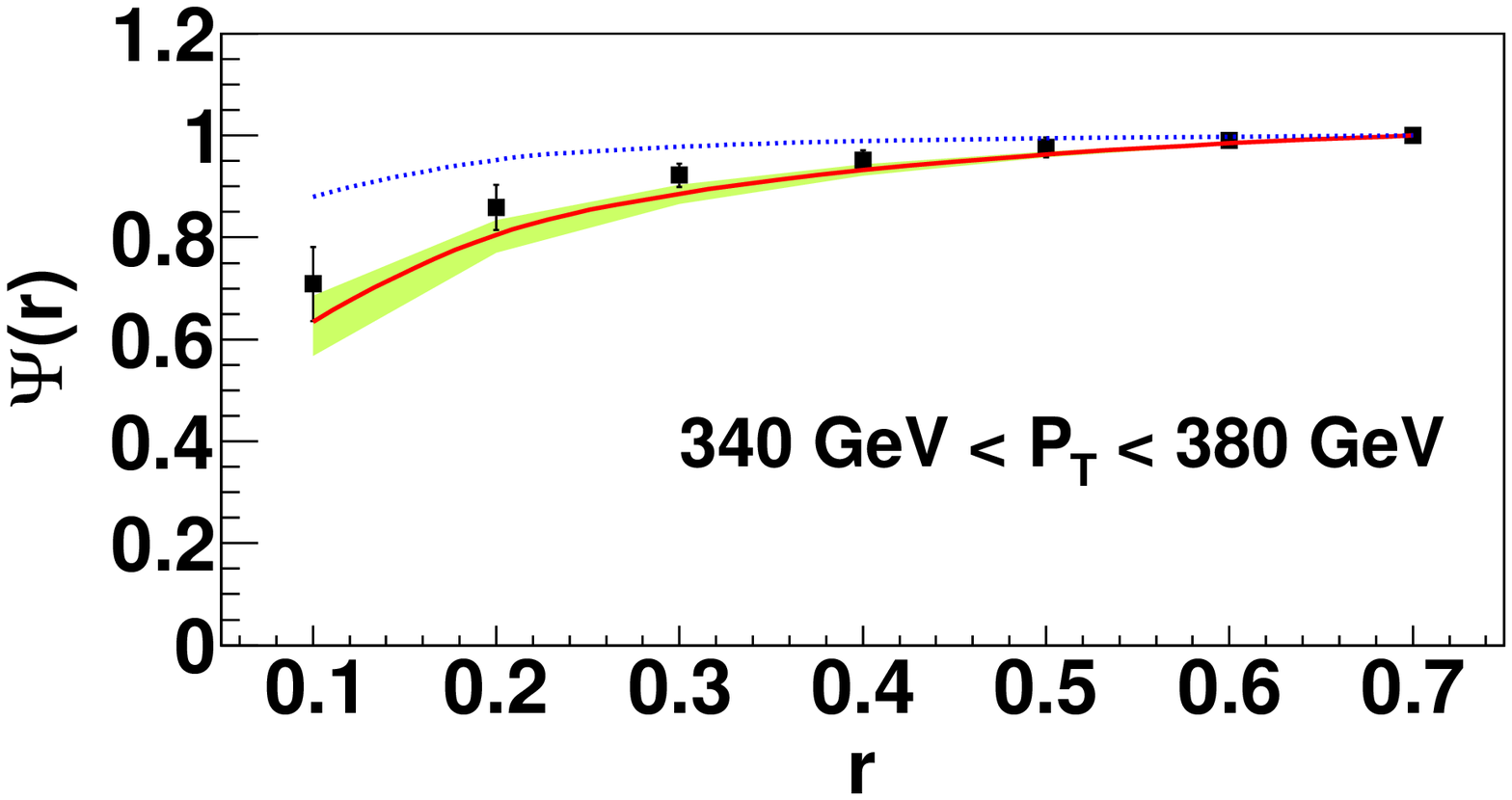}
\caption{Comparison of resummation predictions for the jet energy
profiles with $R=0.7$ to Tevatron CDF data in various $P_T$ intervals. The NLO
predictions denoted by the dotted curves are also displayed.}
\label{CDFJE}
\end{figure}

\begin{figure}[!htb]
\includegraphics[width=0.45\textwidth]{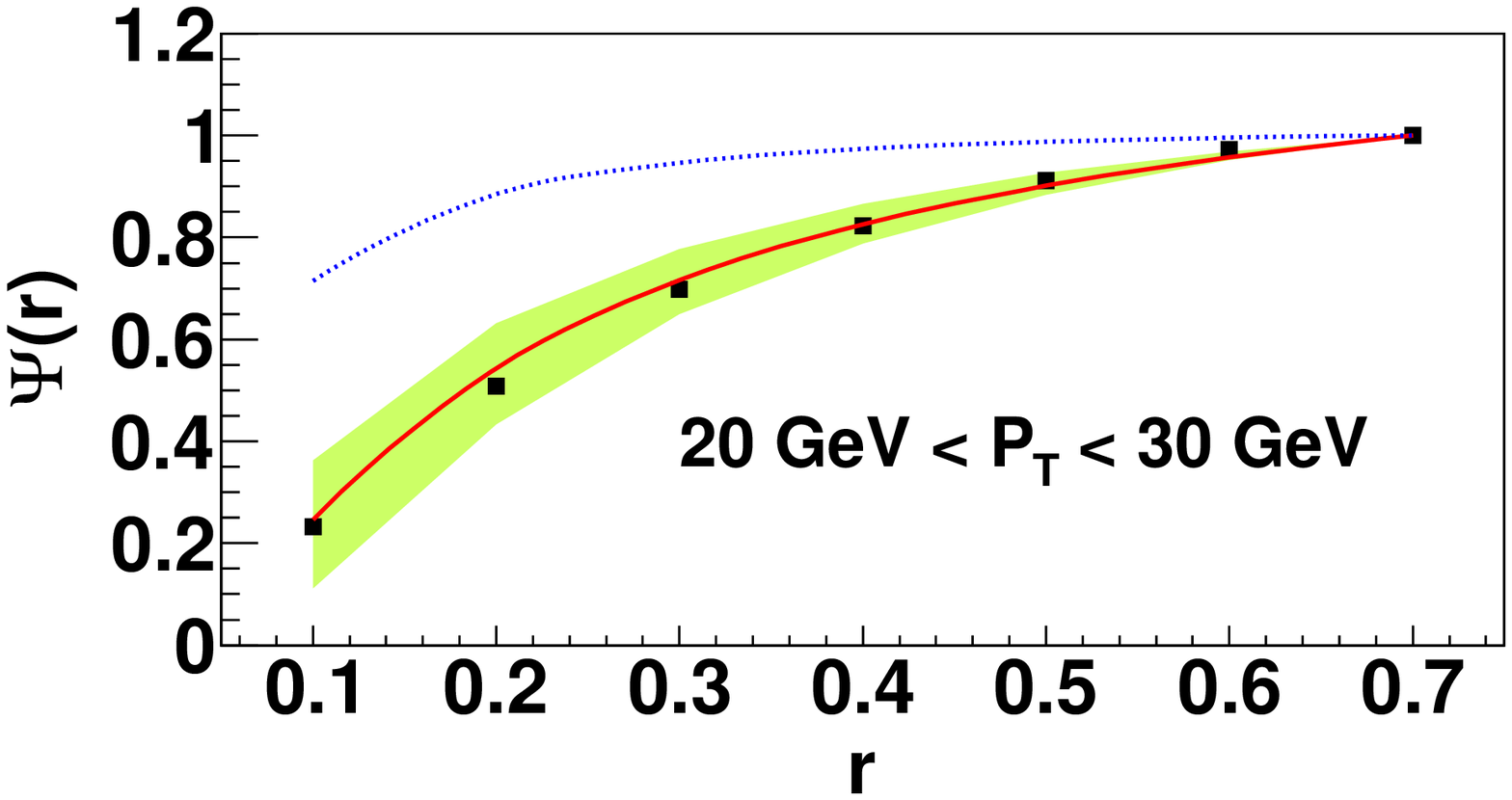}
\includegraphics[width=0.45\textwidth]{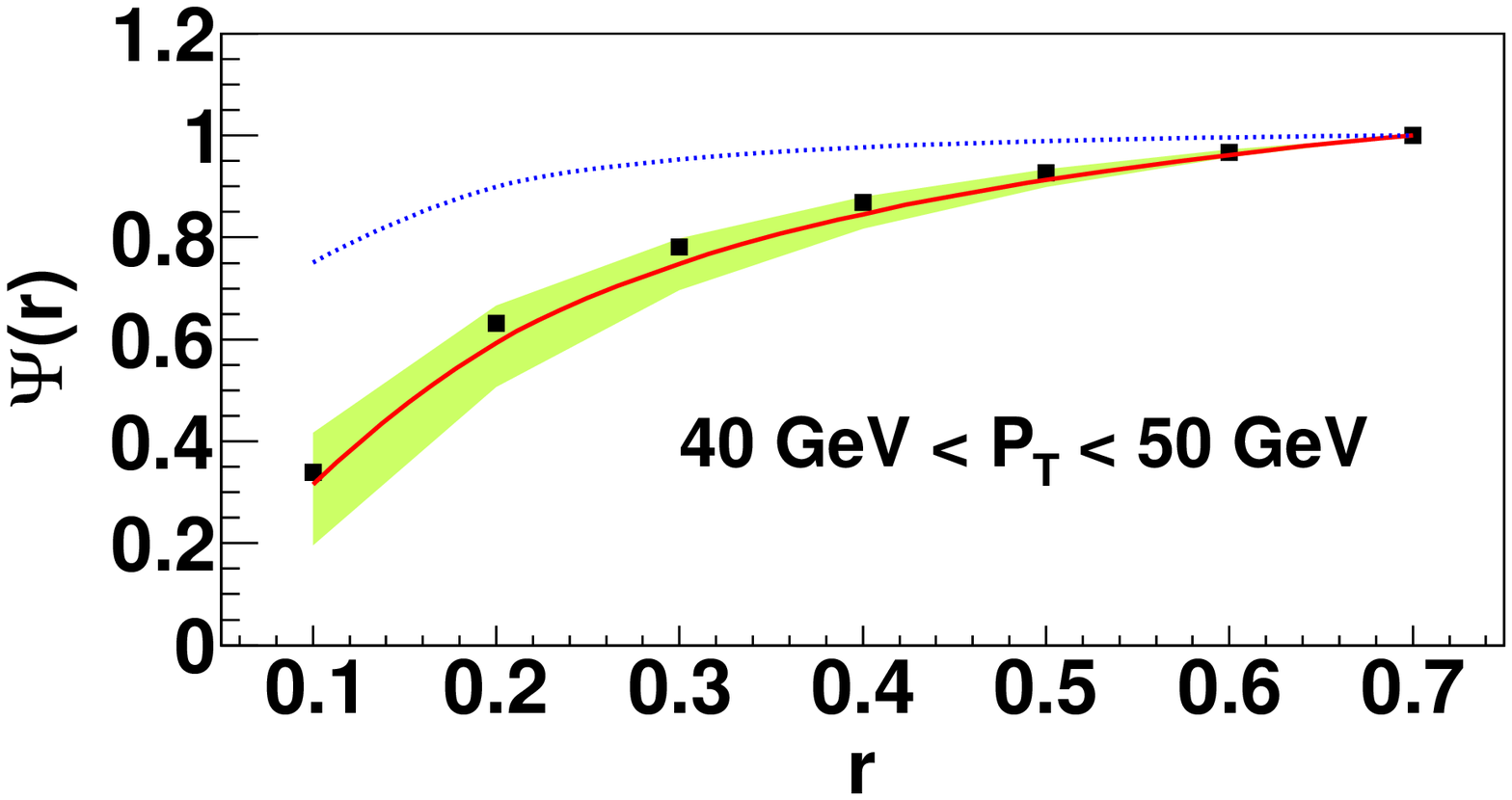}
\includegraphics[width=0.45\textwidth]{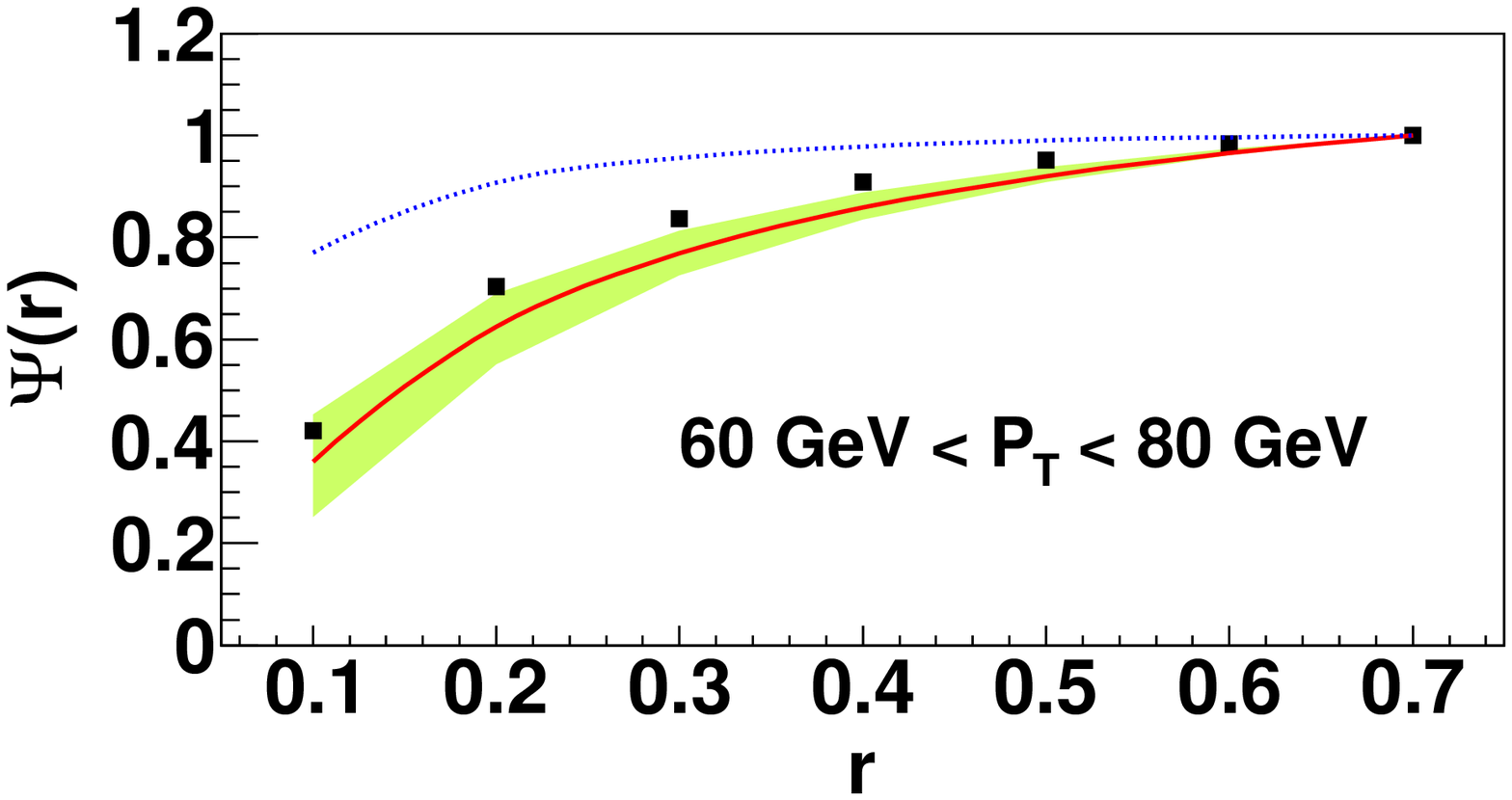}
\includegraphics[width=0.45\textwidth]{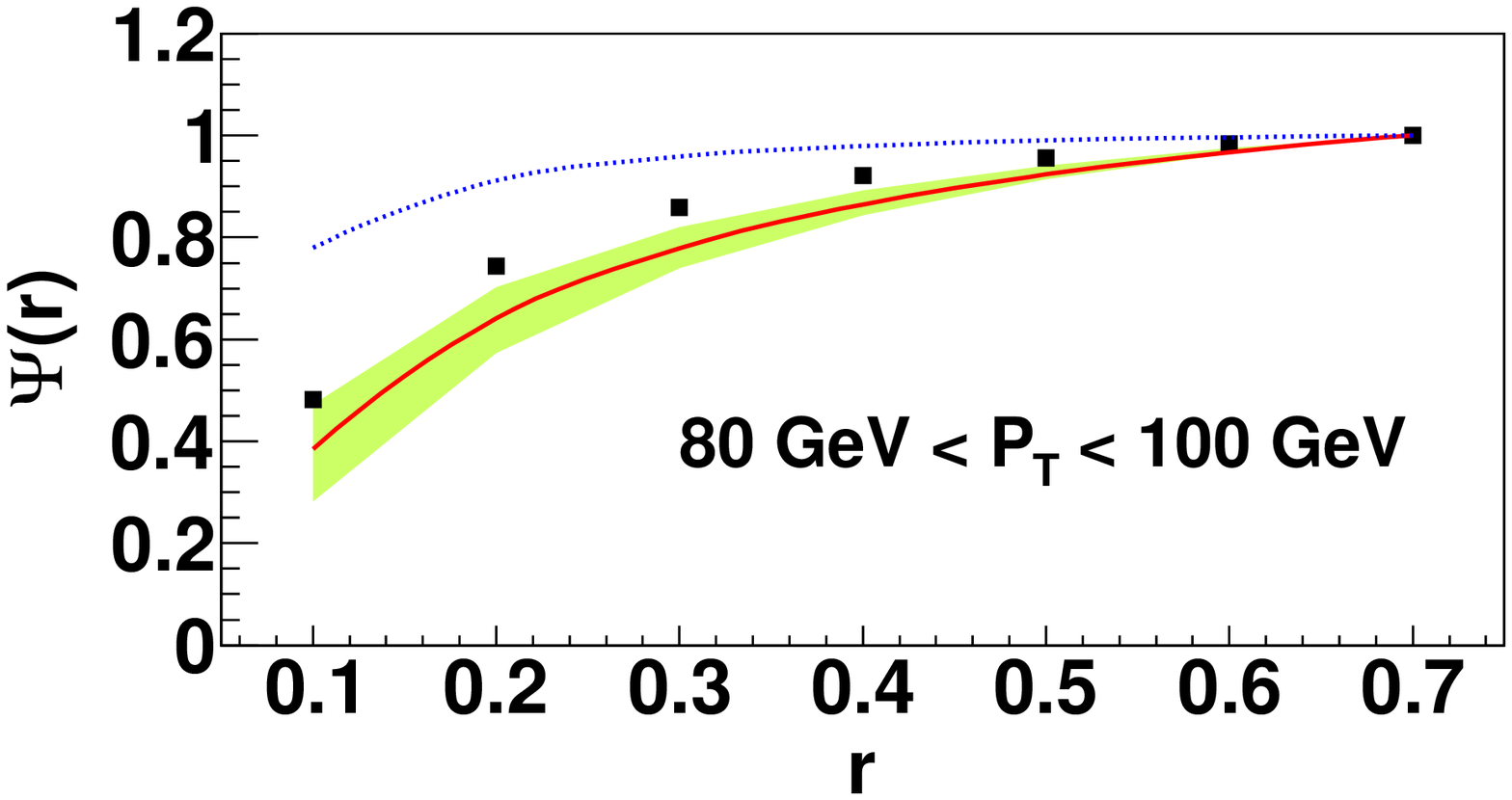}
\caption{Resummation predictions for the
jet energy profiles with $R=0.7$ compared to LHC CMS data
in various $P_T$ intervals. The NLO
predictions denoted by the dotted curves are also displayed.}
\label{CMSJE}
\end{figure}

We then convolute the light-quark and gluon jet energy functions
with the  constituent cross sections of the LO partonic subprocess
and CTEQ6L PDFs \cite{Pumplin:2002vw} at certain collider energy.
The predictions are directly compared with experiment data, such as
the Tevatron CDF data \cite{Acosta:2005ix} using the midpoint jet
algorithm \cite{Blazey:2000qt}, as shown in Fig.~\ref{CDFJE}. The
band represents the theoretical uncertainty caused by the variation
of the parameters from $C_1=C_2=\exp(\gamma_E)\approx 1.78$ to
$C_1=C_2=\exp(-\gamma_E)\approx 0.56$, which serves as an estimate
of the subleading logarithmic effect that is not included in our
formula. It is evident that the resummation predictions agree well
with the data in all $P_T$ intervals. Although there is slight
difference between the data and the central values of the
resummation predictions, the deviation is within the theoretical
uncertainty. The NLO predictions derived from $\bar
J_f^{E(1)}(1,P_T,\nu_{\rm fi}^2, R, r)$ are also displayed for
comparison, which obviously overshoot the data. The resummation
predictions for the jet energy profiles are compared with the LHC
CMS data at 7 TeV \cite{CMSJE} from the anti-kt jet algorithm
\cite{Cacciari:2008gp} in Fig.~\ref{CMSJE}, which are also
consistent with the data in various $P_T$ intervals. Since we can
separate the contributions from the light-quark jet and the gluon
jet, the comparison with the CDF and CMS data implies that
high-energy (low-energy) jets are mainly composed of the light-quark
(gluon) jets. It indicates that our resummation formula has captured
the dominant dynamics in a jet energy profile. Hence, a precise
measurement of the jet energy profile as a function of jet
transverse momentum can be used to experimentally test the
production mechanism of jets in association with other particles,
such as electroweak gauge bosons, top quarks and Higgs bosons.

A careful look at Figs.~\ref{CDFJE} and \ref{CMSJE} reveals that the
resummation predictions fall a bit below the data, as the jet
transverse momentum $P_T$ increases. One of the reasons for this deviation
may be traced back to the kinematic constraint for the real soft
gluon emitted from the special vertex in Eq.~(\ref{ee1}). This
constraint will include too much radiation outside the inner jet
cone $r$ into the estimate of the energy profile, especially when
the jet axis of the rest of particles moves toward the edge of the
inner jet cone. The extra radiation can be regarded as a power
correction to the energy profile in the small $r$ region, because
its effect is proportional to $r$. Since more radiation
will be included as $r$ increases, the
energy profile at large $r$ has been overestimated in our formalism.
The energy profile is normalized
to unity at $r=R$, so the overestimate actually causes suppression of
the distribution at small $r$, explaining the little falloff of the
resummation predictions in comparison with the data. When $P_T$
grows, the power correction in the small $r$ region is strengthened
due to the narrowness of the jet, explaining why the deviation
becomes more obvious at high $P_T$. The above reasoning suggests a
more restricted phase space for the real soft gluon in order to
reduce the power correction and to improve the consistency between
the predictions and the data. This subject will be investigated in a
future work. Besides, we note that the effects from hadronization
and underlying events on jet energy profiles have been estimated by
using the PYTHIA code and removed from the published Tevatron CDF
data \cite{Acosta:2005ix}. On the contrary, these effects have not
been removed in the published LHC CMS data \cite{CMSJE}.

\section{CONCLUSION}

We have developed a theoretical framework for studying jet physics
based on the QCD resummation technique in this paper. The evolution
equations for a light-quark jet function and for a gluon jet
function have been derived and numerically solved in the Mellin
($N$) space. The inverse Mellin transformation from the $N$ space to
the jet mass space was performed, which demands the inclusion of the
nonperturbative contribution in the large $N$ region, in order to
avoid the Landau pole, and to phenomenologically parameterize the
effects from hadronization and underlying events. It has been
observed that the nonperturbative contribution is crucial for
describing the jet mass distribution in the low invariant mass
region. The needed nonperturbative parameters were determined by
fits of the resummation formula including the nonperturbative
contribution to the PYTHIA predictions for the light-quark and gluon
jet distributions at certain jet momentum and cone size, which were
then employed to make predictions for other kinematic
configurations. The above complete resummation formula, convoluted
with the LO partonic hard scattering matrix elements and PDFs, have
led to the jet mass distributions in good agreement with the
Tevatron CDF data at different jet momenta and cone sizes. Our
solutions for the light-particle jet functions are ready to be
implemented into factorization formulas for jet production cross
sections from various processes.

We have also derived the evolution equations for the light-quark and
gluon jet energy functions. With the jet invariant mass being
integrated out, the evolution equations can be straightforwardly
solved in the Mellin space. The energy profiles were then predicted
by convoluting the solutions with LO partonic hard scattering and
PDFs. It has been checked that the resummation results for the
energy profiles associated with a light-quark jet and a gluon jet
agree with the PYTHIA simulations. We have demonstrated that the
resummation predictions for the jet energy profiles are consistent
with the Tevatron CDF data and the LHC CMS data within the
theoretical uncertainty, while the NLO predictions overshoot the
data. It should be emphasized that our formula for this jet
substructure is insensitive to the nonperturbative contribution, and
does not involve tunable parameters. Hence, the agreement with the
data is a highly nontrivial success of the perturbative QCD theory.
Besides, an improvement to reduce the power corrections to the
predicted energy profiles can be done and will be investigated in a
forthcoming paper.

Since final states observed in experiments are usually composed of
quark and gluon jets, jet substructures are sensitive to the ratios
between quark and gluon contributions in a given kinematic region.
It is also known that the components of the quark and gluon jets are
related to the initial-state PDFs. For example, the quark (gluon)
jet component can be related to the initial-state gluon (quark) PDF
in the $W$ boson and jet associated production. By analyzing the
ratio between the quark and gluon contributions to jet
substructures, we may extract additional information on the PDFs,
especially on the gluon PDF in the small $x$ region. On the other
hand, new physics beyond the SM introduces more hard subprocesses,
which may contribute differently to quark and gluon productions in
final states. Therefore, a jet substructure, e.g., the jet energy
profile, can be used to search indirectly for new physics in the
region, where PDFs are relatively stable, when both theoretical
predictions and experiment data become precise enough.

At last, we reiterate that our framework is ready for the extension
to the study of heavy-particle jets produced at the LHC, which
contain energetic light decay products. For instance, a boosted top
quark at the TeV scale will appear as an energetic jet, when it
decays through its hadronic modes. Likewise, a boosted $W$, $Z$, or
Higgs boson decaying into jet modes at the TeV scale will also
appear as an energetic jet. The heavy-particle jet function and
energy profile can be defined at a high energy scale in a similar
way in the factorization theorem as presented in this work. The
additional ingredient is the factorization of the light final states
from the heavy-particle jet at the lower heavy-particle mass scale,
for which the conventional heavy-quark expansion can be implemented.
The solutions for the light-particle jet functions and energy
profiles established in this work will serve as the inputs of this
factorization formula for the heavy-particle jet. The above
illustrations manifest potential and broad applications of our
formalism to jet physics.

\acknowledgments{

This work was supported by National Science Council of R.O.C. under
Grant No. NSC 98-2112-M-001-015-MY3; by the U.S. National Science
Foundation under Grand No. PHY-0855561. CPY and ZL thank the
hospitality of Academia Sinica and National Center for Theoretical
Sciences in Taiwan, where part of this work was done. We thank Pekka
Sinervo and Raz Alon for providing CDF jet mass distribution data.

}

\appendix

\section{NLO JET FUNCTIONS}

In this Appendix we calculate the NLO light-quark and gluon jet
functions by expanding Eq.~(\ref{jet1}) to ${\mathcal O}(\alpha_s)$,
and demonstrate the cancellation of infrared divergences between the
virtual and real corrections in the Mellin space. After regularizing
the UV divergence in the $\overline{\text{MS}}$ scheme, the NLO
virtual correction to the light-quark and gluon jet functions are
given by
\begin{eqnarray}
J_q^{(1)V}
&=& \frac{\alpha_s(\mu^2) C_F}{\pi}\left[
-\frac{1}{2}\ln^2\frac{4P_T^2(1-n_x)}{\lambda^2(1+n_x)}
+\frac{3}{4}\ln\frac{4P_T^2(1-n_x)}{\lambda^2(1+n_x)}
\right. \nonumber \\ &&\left.
+\frac{1}{4}\ln\frac{\mu^2}{R^2P_T^2}+\frac{1}{2}\gamma_E-\frac{\pi^2}{3}-\frac{9}{8}
\right]\delta(M_J^2),\\
J_g^{(1)V}&=& \frac{\alpha_s(\mu^2) C_A}{\pi}\left[
-\frac{1}{2}\ln^2\frac{4P_T^2(1-n_x)}{\lambda^2(1+n_x)}
+\frac{11}{12}\ln\frac{4P_T^2(1-n_x)}{\lambda^2(1+n_x)} \right.
\nonumber \\ &&\left.
+\frac{5}{12}\left(\ln\frac{\mu^2}{R^2P_T^2}-\gamma_E\right)
-\frac{\pi^2}{3}+\frac{1}{2}(\ln 2-3)+\frac{1}{36}
\right]\delta(M_J^2),
\end{eqnarray}
respectively, where $\lambda^2$ is an infrared regulator, and the Wilson line
direction has been chosen as $n=(1,n_x,0,0)$ for convenience. The quark-loop
contributions to the gluon jet function will be elaborated at the end of
this Appendix.

The explicit expressions for the NLO real corrections to the
light-quark and gluon jet functions are written as
\begin{eqnarray}
J_q^{(1)R}&=&\frac{\alpha_s(\mu^2) C_F \beta(1+\beta)}{8\pi
M_J^2(\beta-n_x)^2}\left\{ \frac{(\beta-\cos
R)(\beta-n_x)[\beta(2n_x -1)+n_x-2]}{1-\beta\cos R} \right.\nonumber
\\ &&\left. +(1+\beta)^2(1-n_x)^2\ln\frac{(1+\beta^2)(1+n_x\cos
R)-2\beta(n_x+\cos R)}{(1-\beta^2)(1-\cos Rn_x)}
\right\},\\
J_g^{(1)R}&=&\frac{\alpha_s(\mu^2) C_A \beta(1+\beta)^2}{96\pi
M_J^2(\beta-n_x)^3}\left\{ \frac{(\beta-\cos
R)(\beta-n_x)}{(\beta\cos R-1)^3} \right.\nonumber \\ &&\left.
\times [\beta(\beta^3-3\beta+18+4(\beta^2-9\beta-3)\cos
R+(7+18\beta-3\beta^2)\beta\cos^2 R) \right.\nonumber \\ &&\left.
+n_x^2(7\beta^2+18\beta-3-4\beta(3\beta^2+9\beta-1)\cos R+
(6\beta^4+18\beta^3-3\beta^2+1)\cos^2 R) \right.\nonumber\\ &&\left.
-2\beta n_x((9\beta^3+3\beta^2+9\beta+1)\cos^2
R-2(9\beta^2+4\beta+9)\cos R+\beta^2+9\beta+3) -18n_x+6]
\right.\nonumber \\ &&\left.
+3(1+\beta)^3(1-n_x)^3\ln\frac{(1+\beta^2)(1+\cos Rn_x)-2\beta(\cos
R+n_x)}{(1-\beta^2)(1-n_x\cos R)} \right\},
\end{eqnarray}
respectively, where the polar angle of the radiated particle momentum
has been constrained to be within the cone size $R$. In the $M_J\to
0$ limit and without restricting the phase space of the soft
radiation, i.e., with $R\to\pi$, the large logarithms in the above
expressions are collected into
\begin{eqnarray}
J_q^{(1)R,\rm asym}&=&\frac{\alpha_s(\mu^2) C_F}{\pi M_J^2}\left[
\ln\frac{4(1-n_x) P_T^2}{(1+n_x) M_J^2}-\frac{3}{4}\right],\\
J_g^{(1)R,\rm asym}&=&\frac{\alpha_s(\mu^2) C_A}{\pi M_J^2}\left[
\ln\frac{4(1-n_x) P_T^2}{(1+n_x) M_J^2}-\frac{11}{12}\right].
\end{eqnarray}
This isolation of the $R$-independent soft contributions at NLO has
followed the treatment of the evolution kernel from the real soft
gluon emission in Eq.~(\ref{nr2}).

Combining the NLO real and virtual corrections to the light-quark
jet function in the Mellin space, we arrive at an infrared finite
expression
\begin{eqnarray}
\int_0^1 dx (1-x)^{N-1} (J_q^{(1)V}+J_q^{(1)R,\rm asym})
&=&\frac{\alpha_s(\mu^2) C_F}{\pi R^2 P_T^2}\left[
-\frac{1}{2}\ln^2(\nu^2\bar N)
+\frac{3}{4}\ln(\nu^2\bar N)\right.\nonumber \\
&&\left.+\frac{1}{4}\ln\frac{\mu^2}{R^2P_T^2}
+\frac{1}{2}\gamma_E-\frac{\pi^2}{4}-\frac{9}{8}
\right],
\end{eqnarray}
in which the infrared regulator $\lambda^2$ has disappeared. Those
$N$-dependent terms suppressed by $1/\bar N$ have been dropped,
whose effect is expected to be minor. Similarly, the NLO gluon jet
function is given, in the Mellin space, by
\begin{eqnarray}
\int_0^1 dx (1-x)^{N-1} (J_g^{(1)V}+J_g^{(1)R,\rm asym})
&=&\frac{\alpha_s(\mu^2) C_A}{\pi R^2 P_T^2}\left[
-\frac{1}{2}\ln^2(\nu^2\bar N) +\frac{11}{12}\ln(\nu^2\bar N)\right. \nonumber \\
&&\left. +\frac{5}{12}\left(\ln\frac{\mu^2}{R^2P_T^2}
-\gamma_E\right)-\frac{\pi^2}{4}+\frac{1}{2}(\ln 2-3)+\frac{1}{36}
\right].
\end{eqnarray}

Applying the derivative $\nu^2 d/d\nu^2$ in Eq.~(\ref{cr2}) to the above
expressions, it is easy to see that the double logarithms
reduce to single logarithms, which contribute to the kernel $G+K$
in Eq.~(\ref{crd2}). Since the double logarithms are $\mu^\prime$-independent,
$G+K$ is $\mu^\prime$-independent, and satisfies the RG equations in Eq.~(\ref{rg}).
Choosing the renormalization scale $\mu^2=C_3^{\prime 2} R^2P_T^2/(\bar N\nu^2)$,
the above NLO jet functions become
\begin{eqnarray}
&& \int_0^1 dx (1-x)^{N-1} (J_q^{(1)V}+J_q^{(1)R,\rm asym})
\nonumber \\
&&=\frac{C_F}{\pi R^2 P_T^2}\alpha_s\left(\frac{C_3^{\prime 2} R^2P_T^2}{\bar N\nu^2}\right) \left[
-\frac{1}{2}\ln^2(\nu^2\bar N)+\frac{1}{2}\ln(\nu^2\bar N)
+\frac{1}{4}\ln C_3^{\prime 2}
+\frac{1}{2}\gamma_E-\frac{\pi^2}{4}-\frac{9}{8}
\right],\label{a12}\\
&& \int_0^1 dx (1-x)^{N-1} (J_g^{(1)V}+J_g^{(1)R,\rm asym})
\nonumber \\
&&=\frac{ C_A}{\pi R^2 P_T^2}\alpha_s\left(\frac{C_3^{\prime 2}
R^2P_T^2}{\bar N\nu^2}\right) \left[ -\frac{1}{2}\ln^2(\nu^2\bar
N)+\frac{1}{2}\ln(\nu^2\bar N) +\frac{5}{12}\ln C_3^{\prime 2}
-\frac{5}{12}\gamma_E-\frac{\pi^2}{4}+\frac{1}{2}(\ln
2-3)+\frac{1}{36} \right].\label{a13}
\end{eqnarray}

The choice of $\mu$ depends on $\nu^2$ in the way that we
have $\mu\sim {\mathcal O}(RP_T)$ as $\nu^2=\nu^2_{\rm in}\equiv
C_1/(C_2\bar N)$ for the initial conditions, which then do not
contain the large logarithms $\ln\bar N$. The NLO initial conditions of
the Sudakov evolution
\begin{eqnarray}
&& \int_0^1 dx (1-x)^{N-1} (J_q^{(1)V}+J_q^{(1)R,\rm asym})_{\rm initial}
\nonumber \\ &&=
\frac{C_F}{\pi R^2 P_T^2}\alpha_s\left(C_3^2 R^2P_T^2\right) \left[
\frac{1}{2}\ln\frac{C_1}{C_2}-\frac{1}{2}\ln^2\frac{C_1}{C_2}
+\frac{1}{4}\ln \frac{C_3^2 C_1}{C_2}
+\frac{1}{2}\gamma_E-\frac{\pi^2}{4}-\frac{9}{8}
\right], \label{init}\\
&& \int_0^1 dx (1-x)^{N-1} (J_g^{(1)V}+J_g^{(1)R,\rm asym})_{\rm initial}
\nonumber \\ &&=
\frac{C_A}{\pi R^2 P_T^2}\alpha_s\left(C_3^2 R^2P_T^2\right) \left[
\frac{1}{2}\ln\frac{C_1}{C_2}-\frac{1}{2}\ln^2\frac{C_1}{C_2}
+\frac{5}{12}\ln \frac{C_3^2 C_1}{C_2}
-\frac{5}{12}\gamma_E-\frac{\pi^2}{4}+\frac{1}{2}(\ln 2-3)+\frac{1}{36}
\right],\label{initial}
\end{eqnarray}
are derived from Eqs.~(\ref{a12}) and (\ref{a13}), respectively,
with $C_3^2=C_3^{\prime 2} C_2/C_1$.
The original definitions of the jet functions in Eq.~(\ref{jet1})
involve the Wilson links on the light cone along the vector $\xi$.
Setting $\nu^2=\nu_{\rm fi}^2\equiv 1$, Eqs.~(\ref{a12}) and
(\ref{a13}) reproduce the $\ln\bar N$ terms in these original
definitions at NLO, leading to the final conditions
\begin{eqnarray}
&& \int_0^1 dx (1-x)^{N-1} (J_q^{(1)V}+J_q^{(1)R,\rm asym})_{\rm final}
\nonumber \\ &&=
\frac{C_F}{\pi R^2 P_T^2}\alpha_s\left(\frac{C_3^2 C_1 R^2P_T^2}{C_2 \bar N}\right) \left[
-\frac{1}{2}\ln^2\bar N+\frac{1}{2}\ln\bar N
+\frac{1}{4}\ln \frac{C_3^2 C_1}{C_2}
+\frac{1}{2}\gamma_E-\frac{\pi^2}{4}-\frac{9}{8}
\right],\label{final1} \\
&& \int_0^1 dx (1-x)^{N-1} (J_g^{(1)V}+J_g^{(1)R,\rm asym})_{\rm
final} \nonumber \\ &&= \frac{C_A}{\pi R^2
P_T^2}\alpha_s\left(\frac{C_3^2 C_1 R^2P_T^2}{C_2 \bar N}\right)
\left[ -\frac{1}{2}\ln^2\bar N+\frac{1}{2}\ln\bar N +\frac{5}{12}\ln
\frac{C_3^2 C_1}{C_2}
-\frac{5}{12}\gamma_E-\frac{\pi^2}{4}+\frac{1}{2}(\ln
2-3)+\frac{1}{36} \right].\label{final2}
\end{eqnarray}
It is seen that as the integration
variable $\nu^2$ in Eq.~(\ref{cr2}) varies from $\nu_{\rm in}^2$ to
$\nu_{\rm fi}^2$, the scale $\mu^2$ varies from  ${\mathcal O}(R^2P_T^2)$
to ${\mathcal O}(R^2P_T^2/{\bar N})$. The latter describes the
soft and collinear radiations in the jet mass distribution appropriately,
because they mainly occur at a lower scale.

The NLO terms in the expansion of the Sudakov exponent contain
\begin{eqnarray}
\left.\exp[S_f(N)]\right|_{\alpha_s}=\frac{C_f}{\pi}\alpha_s
\left(- \frac{1}{2}\ln^2\bar N
+\frac{1}{2}\ln\bar N
+ \frac{1}{2}\ln\frac{C_2}{C_1}
+\frac{1}{2}\ln^2\frac{C_2}{C_1} \right) ,
\end{eqnarray}
where $C_f=C_F$ or $C_A$, for $S_q$ or $S_g$, respectively.
Combining the above expansion with Eqs.~(\ref{init}) and
(\ref{initial}), it is straightforward to show that the resummed jet
functions in Eqs.~(\ref{qloop}) and (\ref{gloop}) indeed agree with
the final conditions in Eqs.~(\ref{final1}) and (\ref{final2}) at
NLO, respectively. That is, our resummation formalism is matched to
the NLO jet functions with $\mu^2\sim {\mathcal O}(R^2P_T^2/{\bar
N})$, implying that the single logarithm introduced by our choice of
$\mu^2$ has been also summed into the Sudakov factor. The all-order
summation of this single logarithm corresponds to the RG evolution
in $\mu^2$ from $\mu^2=C_3^2 R^2P_T^2$ to $\mu^2=(C_3^2 C_1
R^2P_T^2)/(C_2 \bar N)$.

At last, we discuss the treatment of the virtual and real quark-loop
contributions to the gluon jet function
\begin{eqnarray}
J_{g\to q\bar q}^{(1)V} & = & -\frac{\alpha_s(\mu^2) n_f
C_F}{3\pi}\left(\ln\frac{\mu^2}{\lambda^2}
-\frac{1}{3}\right)\delta(M_J^2),\label{jvg}\\
J_{g\to q\bar q}^{(1)R} & = & \frac{\alpha_s(\mu^2) n_f C_F
\beta^3(1+\beta)^2(\beta-\cos R)}{48\pi M_J^2(1-\beta\cos R)^3}
\left[\beta^2(1+3\cos^2R)-8\beta\cos R+3+\cos^2R\right],\label{jrg}
\end{eqnarray}
respectively.
In the $M_J\to 0$ and $R\to\pi$ limits, Eq.~(\ref{jrg}) gives
\begin{eqnarray}
J_{g\to q\bar q}^{(1)R,\rm asym} &=& \frac{\alpha_s(\mu^2) n_f C_F}{3\pi M_J^2},
\end{eqnarray}
and the infrared finite expression
\begin{eqnarray}
\int_0^1 dx (1-x)^{N-1} (J_{g\to q \bar q}^{(1)V}+J_{g\to q \bar
q}^{(1)R,\rm asym})& = & -\frac{\alpha_s(\mu^2) n_f C_F}{3\pi R^2
P_T^2} \left( \ln\bar N -\frac{1}{3}
+\ln\frac{\mu^2}{R^2P_T^2}\right).
\end{eqnarray}
With our choice of $\mu^2$, the final condition from the quark-loop
contributions is written as
\begin{eqnarray}
\int_0^1 dx (1-x)^{N-1} (J_{g\to q \bar q}^{(1)V}+J_{g\to q \bar
q}^{(1)R,\rm asym})_{\rm final}=\frac{n_f C_F}{3\pi R^2 P_T^2}
\alpha_s\left(\frac{C_3^2C_1R^2P_T^2}{C_2\bar N}\right) \left(
\frac{1}{3} -\ln\frac{C_1C_3^2}{C_2}\right),
\end{eqnarray}
which has been added into Eq.~(\ref{gloop}).
The absence of the logarithm $\ln\bar N$ implies that the quark-loop
contribution is not important, as verified in the numerical analysis.

The initial conditions of the jet functions,
namely, the prefactors of $S_f$ in Eqs.~(\ref{nloqs}) and
(\ref{nlogs}) are evaluated at the hard scale $\mu\sim {\mathcal O}(RP_T)$.
After applying the inverse Mellin transformation to obtain the jet
functions $J_f^{\rm NLL/NLO}(M_J^2,P_T,R)$, which is
inserted into Eq.~(\ref{nlo}) to obtain theoretical predictions,
the hard scale $\mu\sim {\mathcal O}(RP_T)$ remains. Since
$J_f^{(1)V}+J_f^{(1)R,\rm asym}$ was organized in the resummation formalism,
the regular piece $J_f^{(1)R}-J_f^{(1)R,\rm asym}$ has to be added back
in order to reproduce the complete NLO corrections to the jet functions.
This piece is also evaluated at the hard scale $\mu\sim {\mathcal O}(RP_T)$ in
Eq.~(\ref{nll}). Similarly, the regular piece of the quark-loop contribution,
$J_{g\to q\bar q}^{(1)R}-J_{g\to q\bar q}^{(1)R, \rm asym}$, should be
included too, which has been combined into $J_g^{(1)R}-J_g^{(1)R,\rm asym}$
on the right-hand side of Eq.~(\ref{nll}).

\section{INVERSE MELLIN TRANSFORMATION}

\begin{figure}[t]
\begin{center}
\includegraphics[height=6cm]{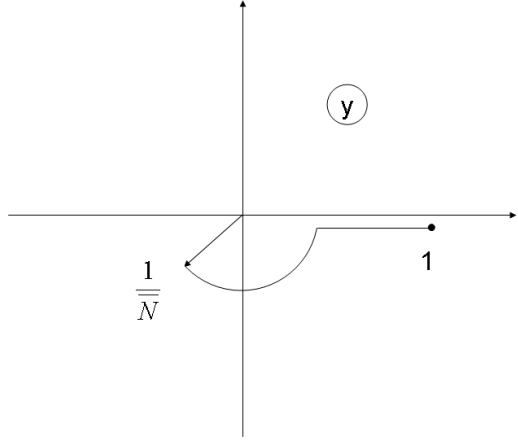}
\caption{Contour for the integration variable $y$ in
Eqs.~(\ref{sq}) and (\ref{sg}).}
\label{contourY}
\end{center}
\end{figure}

Because the evolution equations were solved in the Mellin space,
we need to perform the inverse Mellin transformation to get the solutions
in the space of the jet invariant mass. As stated in Sec.~III,
the argument $\mu^2$ of $\alpha_s(\mu^2)$ in the Sudakov
integrals should be treated as a complex number in the inverse Mellin
transformation. Besides, the argument becomes very small (lower than
the QCD scale $\Lambda_{\text{QCD}}$) in the large $N$ region, and the
running coupling constant
suffers the Landau pole problem \cite{Vogt:2000ci,Amsler:2008zz}.
To avoid the Landau pole, we introduce a critical scale $\mu_c$,
below which the running coupling constant is
frozen to a constant value $\alpha_s(\mu_c^2)$. To be precise,
the following prescription is proposed
\begin{eqnarray}
\alpha_s(\mu^2)=\left\{
\begin{array}{ll}
\alpha_s(\mu_c^2\exp[2{\rm Arg}(\mu)]), & |\mu|<\mu_c
\\
\alpha_s(\mu^2), & |\mu|>\mu_c
\end{array}
\right..
\end{eqnarray}
We have adopted the perturbative expansion of $\alpha_s$ in the numerical analysis
\begin{eqnarray}
\alpha_s(Q^2)=\frac{\alpha_s(\mu^2)}{X}
\left\{1-\frac{\alpha_s(\mu^2)}{2\pi}\frac{\beta_1}{\beta_0}\frac{\ln X}{X}\right\},
\end{eqnarray}
with
\begin{eqnarray}
X&=&1+\frac{\alpha_s(\mu^2)}{4\pi}\beta_0\ln\frac{Q^2}{\mu^2},\nonumber\\
\beta_0&=&11-\frac{2}{3}n_f,\;\;\;\;\beta_1=51-\frac{19}{3}n_f.
\end{eqnarray}

The variable $N$, appearing in the lower bound of $y$ in Eqs.~(\ref{sq}) and
(\ref{sg}), should be also treated as a complex number in the
inverse Mellin transformation. The
contour in the complex $y$ plane is depicted in Fig.~\ref{contourY},
according to which an integral over $y$ is handled in the following way,
\begin{eqnarray}
\int^{C_2}_{C_1/\bar N} dy F(y)&=&\int^{C_2}_{C_1/|\bar N|}
dy_1 F(y_1)+\int^{C_1/|\bar N|}_{C_1/\bar N} dy_2 F(y_2),
\nonumber \\
&=&
\int_0^1(C_2-C_1/|\bar N|)dt ~F(y_1)
-\int_0^1 y_2 {\rm i}{\rm Arg}(1/\bar N)dt ~F(y_2),
\end{eqnarray}
with the variable changes $y_1\equiv C_1/|\bar N|+(C_2-C_1/|\bar N|)t$
and $y_2\equiv C_1/|\bar N| \exp({\rm i} {\rm Arg}(1/\bar N)(1-t))$.

\begin{figure}[!hbt]
\includegraphics[width=0.3\textwidth]{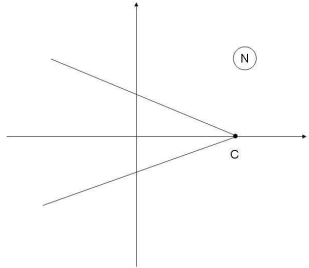}
\caption{Conventional contour of $N$ adopted in inverse Mellin transformation.}
\label{Ncontour1}
\end{figure}

\begin{figure}[!hbt]
\includegraphics[width=0.3\textwidth]{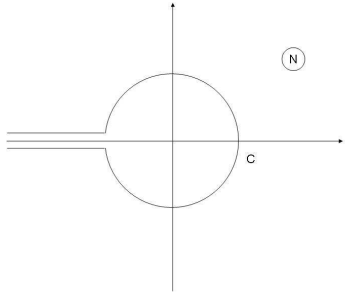}
\caption{Contour of $N$ adopted in our inverse Mellin transformation.}
\label{Ncontour2}
\end{figure}

The inverse Mellin transformation for the jet function is defined as
\begin{eqnarray}
J(M_J^2, P_T, \nu_{fi}^2, R, \mu^2)=\frac{1}{2\pi i}\int_{\rm C} dN (1-x)^{-N}
\bar J(N, P_T, \nu_{fi}^2, R, \mu^2),
\end{eqnarray}
with $\rm C$ labelling a contour of $N$. The
conventional contour of $N$ shown in Fig.~\ref{Ncontour1} is not suitable
for a numerical approach using a grid file, since different jet masses
require different parameterizations of this contour in order to
get enough information in the large $N$ region. Instead,
we choose the contour depicted in Fig.~\ref{Ncontour2}.
The inverse Mellin transformation along the upper-half part of this contour
is written as
\begin{eqnarray}
\frac{1}{2\pi i}\int_C dN(1-x)^{-N} F(N)
&=&\frac{1}{2\pi i}\int_0^{\pi-\epsilon/c} N_1 id\phi(1-x)^{-N_1} F(N_1\equiv c e^{i\phi})
\nonumber \\ &&
+\frac{1}{2\pi i}\int_{c+i\epsilon}^{-\infty+i\epsilon}dN_2 (1-x)^{-N_2} F(N_2),
\nonumber \\
&=&\frac{1}{2\pi i}\int_0^1 N_1 i(\pi-\epsilon/c)dt(1-x)^{-N_1}
F\left[N_1\equiv c\exp(i(\pi-\epsilon/c)t)\right]
\nonumber \\ &&
+\frac{1}{2\pi i}\int_0^1 L\frac{-1}{(1-t)^2}dt(1-x)^{-N_2}
F\left[N_2\equiv -c+i\epsilon+L\frac{-t}{1-t}\right].\label{upper}
\end{eqnarray}
The expression associated with the lower-half contour can be obtained
by taking the complex conjugate of Eq.~(\ref{upper}). The parameters
involved in the integral variables $N_1$ and $N_2$ are set to $c=5$, $L=10$,
and $\epsilon=10^{-6}$ in our numerical analysis.

\section{NLO JET ENERGY PROFILES}

In this Appendix we calculate the NLO light-quark and gluon jet energy
functions defined in Eq.~(\ref{jetENLO1}). Due to their lengthy expressions, we
focus only on the logarithmic terms below.
The NLO virtual corrections with the factorization scale $\mu=P_T$
and the real corrections to the Mellin-transformed
jet energy functions are given by
\begin{eqnarray}
\bar J_q^{E(1),\rm V}
&=& \frac{C_F \alpha_s}{\pi P_T}\left[
-\frac{1}{2}\ln^2\frac{4P_T^2(1-n_x)}{\lambda^2(1+n_x)}
+\frac{3}{4}\ln\frac{4P_T^2(1-n_x)}{\lambda^2(1+n_x)}
\right],\\
\bar J_g^{E(1),\rm V}
&=& \frac{C_A \alpha_s}{\pi P_T}\left[
-\frac{1}{2}\ln^2\frac{4P_T^2(1-n_x)}{\lambda^2(1+n_x)}
+\frac{11}{12}\ln\frac{4P_T^2(1-n_x)}{\lambda^2(1+n_x)}
\right],
\end{eqnarray}
and
\begin{eqnarray}
\bar J_q^{E(1),\rm R}&=&
\frac{C_F\alpha_s}{\pi P_T}\left[
\frac{1}{2}\ln^2\frac{\lambda^2}{P_T^2}
-\left(\ln\frac{4(1-n_x)}{(1+n_x) r^2}-\frac{3}{4}\right)
\ln\frac{\lambda^2}{P_T^2}
\right.\nonumber \\&&\left.
+\frac{1}{4}\ln^2\frac{4(1-n_x)}{(1+n_x)}-\frac{3}{2}\ln\frac{4(1-n_x)}{(1+n_x)}
-\frac{1}{4}\ln^2 r^2+\frac{1}{2}\ln r^2\ln\frac{4(1-n_x)}{(1+n_x)}+\frac{3}{4}\ln r^2
\right].\\
\bar J_g^{E(1),\rm R}&=&
\frac{C_A\alpha_s}{\pi P_T}\left[
\frac{1}{2}\ln^2\frac{\lambda^2}{P_T^2}
-\left(\ln\frac{4(1-n_x)}{(1+n_x) r^2}-\frac{11}{12}\right)
\ln\frac{\lambda^2}{P_T^2}
\right.\nonumber \\&&\left.
+\frac{1}{4}\ln^2\frac{4(1-n_x)}{(1+n_x)}-\frac{11}{6}\ln\frac{4(1-n_x)}{(1+n_x)}
-\frac{1}{4}\ln^2 r^2+\frac{1}{2}\ln r^2\ln\frac{4(1-n_x)}{(1+n_x)}+\frac{11}{12}\ln r^2
\right],
\end{eqnarray}
respectively. Combining the virtual and real corrections, we derive the infrared finite NLO expressions
\begin{eqnarray}
\bar J_q^{E(1),\rm V}+\bar J_q^{E(1),\rm R}=
\frac{\alpha_s C_F}{P_T \pi} \left[-\frac{1}{4}\ln^2\frac{4(1-n_x)}{r^2(1+n_x)}
-\frac{3}{4}\ln\frac{4(1-n_x)}{r^2(1+n_x)} \right],\label{efq}\\
\bar J_g^{E(1),\rm V}+\bar J_g^{E(1),\rm R}=
\frac{\alpha_s C_A}{P_T \pi} \left[-\frac{1}{4}\ln^2\frac{4(1-n_x)}{r^2(1+n_x)}
-\frac{11}{12}\ln\frac{4(1-n_x)}{r^2(1+n_x)} \right],\label{efg}
\end{eqnarray}
in the $r\to 0$ limit, where the infrared regulator $\lambda^2$ has disappeared.

The singular NLO terms of the resumed jet energy
functions in Eq.~(\ref{ep1}) are given by
\begin{eqnarray}
{\bar J}_f^{E(1)}(1,P_T,\nu_{\rm fi}^2,R,r)
=\frac{C_f\alpha_s}{\pi P_T}\left[-\frac{1}{4}\ln^2\frac{R^2}{r^2}+\frac{1}{2}(1-\ln C)\ln\frac{R^2}{r^2}
+\frac{1}{4}\ln^2\frac{C_1^2}{C_2^2}-\frac{1}{2}(1-\ln C)\ln\frac{C_1^2}{C_2^2}
\right].
\label{asy}
\end{eqnarray}
Substituting the vector $n_{\rm fi}\equiv (1,(4-R^2)/(4+R^2),0,0)$ for
$n$ in Eqs.~(\ref{efq}) and (\ref{efg}) to obtain the final conditions, and
choosing the ${\mathcal O}(1)$ constant $C=\exp(5/2)$ ($C=\exp(17/6)$)
for the light-quark (gluon) jet in Eq.~(\ref{asy}), we
observe the consistency between Eqs.~(\ref{efq}) and (\ref{efg}), and Eq.~(\ref{asy}).
That is, the resummation formula in Eq.~(\ref{ep1}) has collected the important
logarithms in the NLO jet energy functions. The complete expressions for the NLO
initial conditions corresponding to the choice
$n_{\rm in}=(1,(4 C_2^2 - r^2 C_1^2)/(4 C_2^2 + r^2 C_1^2),0,0)$,
and their convolution formulas with the LO partonic
subprocesses and PDFs, can be downloaded from the web site
http://hep.pa.msu.edu/people/yuan/public\_codes/JETENPRO/code\_energy\_convolute\_public.zip.

\bibliography{lightjet1}

\end{document}